\documentclass[prl,twocolumn,showpacs,preprintnumbers,superscriptaddress,longbibliography]{revtex4-2}

\usepackage[utf8]{inputenc}
\usepackage{amsmath}
\usepackage{amssymb}
\usepackage{mathtools}
\usepackage{graphicx}
\usepackage[dvipsnames]{xcolor}
\usepackage{hyperref}
\usepackage{mathrsfs}
\definecolor{dark-red}{rgb}{0.4,0.15,0.15}
\definecolor{dark-blue}{rgb}{0.15,0.15,0.4}
\definecolor{medium-blue}{rgb}{0,0,0.5}
\hypersetup{
	colorlinks,
	linkcolor={dark-blue},
	citecolor={dark-blue},
	urlcolor={medium-blue}
}

\begin{document}

\title{Symmetric Mass Generation in a Bilayer Honeycomb Lattice with \texorpdfstring{$\mathrm{SU}(2)\times\mathrm{SU}(2)\times\mathrm{SU}(2)/\mathbb{Z}_2$}{SU(2)xSU(2)xSU(2)/Z_2} Symmetry}

\author{Cheng-Hao He}
\affiliation{Key Laboratory of Artificial Structures and Quantum Control (Ministry of Education), School of Physics and Astronomy, Shanghai Jiao Tong University, Shanghai 200240, China}
\author{Yi-Zhuang You}
\email{yzyou@physics.ucsd.edu}
\affiliation{Department of Physics, University of California, San Diego, CA 92093, USA}
\author{Xiao Yan Xu}
\email{xiaoyanxu@sjtu.edu.cn}
\affiliation{Key Laboratory of Artificial Structures and Quantum Control (Ministry of Education), School of Physics and Astronomy, Shanghai Jiao Tong University, Shanghai 200240, China}
\affiliation{Hefei National Laboratory, Hefei 230088, China}
\date{\today}

\begin{abstract}
A central question beyond the Landau paradigm is the non-perturbative critical theory of the symmetric mass generation (SMG) transition, where strong interactions gap Dirac fermions in (2+1) dimensions without triggering spontaneous symmetry breaking or topological order. While previous studies have already provided evidence for direct SMG transitions in (2+1) dimensions, the fermion scaling dimension---the key observable for distinguishing candidate critical theories---has not been determined in a controlled unbiased way. In this Letter, using large-scale determinant quantum Monte Carlo (DQMC) simulations of a bilayer honeycomb lattice model with $\mathrm{SU}(2)\times\mathrm{SU}(2)\times\mathrm{SU}(2)/\mathbb{Z}_2$ symmetry, we establish a direct continuous transition by observing the simultaneous opening of single-particle and bosonic gaps at a critical coupling $J_c \approx 2.6$ with correlation length exponent $\nu = 1.14(2)$, while an exhaustive search over all 19 symmetry-inequivalent fermion bilinear order parameters confirms the absence of any symmetry breaking. We further obtain the first controlled unbiased estimate of the fermion anomalous dimension, $\eta_\psi = 0.071(1)$, which deviates significantly from the large-$N$ prediction ($\eta_\psi \approx 0.595$) and variational Monte Carlo estimates ($\eta_\psi \approx 0.62$), thereby placing direct quantitative constraints on SMG criticality. By contrasting with a related $\mathrm{Spin}(5)\times\mathrm{U}(1)/\mathbb{Z}_2$ model that develops an intermediate excitonic phase, we show that pure non-Abelian symmetry plays a decisive role in stabilizing the direct SMG transition.
\end{abstract}

\maketitle
{\it Introduction}\,---\,
In both quantum field theory and condensed matter physics, mass generation is conventionally associated with symmetry breaking, as exemplified by the Higgs mechanism~\cite{Anderson1963,Englert1964,Higgs1964,Guralnik1964} and by symmetry-breaking transitions in lattice systems such as the sublattice-symmetry-broken gap in hexagonal boron nitride.

Recent theoretical developments have revealed the possibility of generating mass gaps without breaking any symmetries or developing topological order, a phenomenon known as symmetric mass generation (SMG)~\cite{wang2022symmetric,You2018,tong2022comments,Zeng2022,Lu2023,xu2021green,You2018from}. The idea of gapping fermions through strong interactions without symmetry breaking has roots in early lattice studies of Yukawa models and mirror-fermion decoupling~\cite{Eichten1986,Lee1990,lee1990lattice,bock1990phase,Bock1990,Abada1991,Hasenfratz1991,Banks1992,Golterman1993}, where symmetric phases with massive fermions were identified at strong bare coupling, though the connection to anomaly cancellation and universal low-energy physics was clarified only later. This mechanism represents a significant departure from the Landau paradigm, which classifies phases based on their symmetry-breaking patterns.

The theoretical foundation for SMG stems from anomaly cancellation in fermionic systems~\cite{You2014}. For SMG to occur, the system must have enough fermionic degrees of freedom to cancel all potential quantum anomalies, allowing for a fully gapped state without symmetry breaking. The minimal number of Dirac fermions required is 4 in (0+1)D and (1+1)D, and 8 in (2+1)D and (3+1)D~\cite{wang2022symmetric,You2018,tong2022comments}.

Various models exhibiting SMG have been proposed across different dimensions. In (0+1)D, the Majorana fermion model studied by Fidkowski and Kitaev~\cite{Fidkowski2010,Fidkowski2011} demonstrated how interactions can modify the classification of topological phases. In (1+1)D, the 3-4-5-0 chiral fermion model~\cite{Wang2013} provided a concrete lattice framework for regularizing chiral matter. For (2+1)D, several models have been investigated, including bilayer honeycomb lattice models~\cite{Slagle2015,He2016,Hou2023,Liu2024} and staggered fermion models~\cite{Catterall2016,Ayyar2015,Ayyar2016,maiti2026phase}. In (3+1)D, chiral fermion models have been explored for their potential to exhibit SMG~\cite{tong2022comments,Razamat2021}.

Despite these theoretical advances, unbiased numerical evidence for SMG in (2+1)D has remained elusive. In particular, an unbiased determination of the fermion scaling dimension has remained out of reach. High symmetry is a key ingredient, as it introduces large quantum fluctuations that can prevent symmetry breaking, but it alone does not guarantee a direct transition to a gapped phase. Indeed, recent variational Monte Carlo (VMC) studies of bilayer honeycomb models~\cite{Hou2023} could not fully settle the question due to the uncontrolled approximations inherent in the variational approach. Rigorous, sign-problem-free quantum Monte Carlo calculations are essential to determine whether a given model truly exhibits SMG or instead develops intermediate symmetry-breaking phases.

\begin{figure}[t]
		\centering
		\includegraphics[width=1\linewidth]{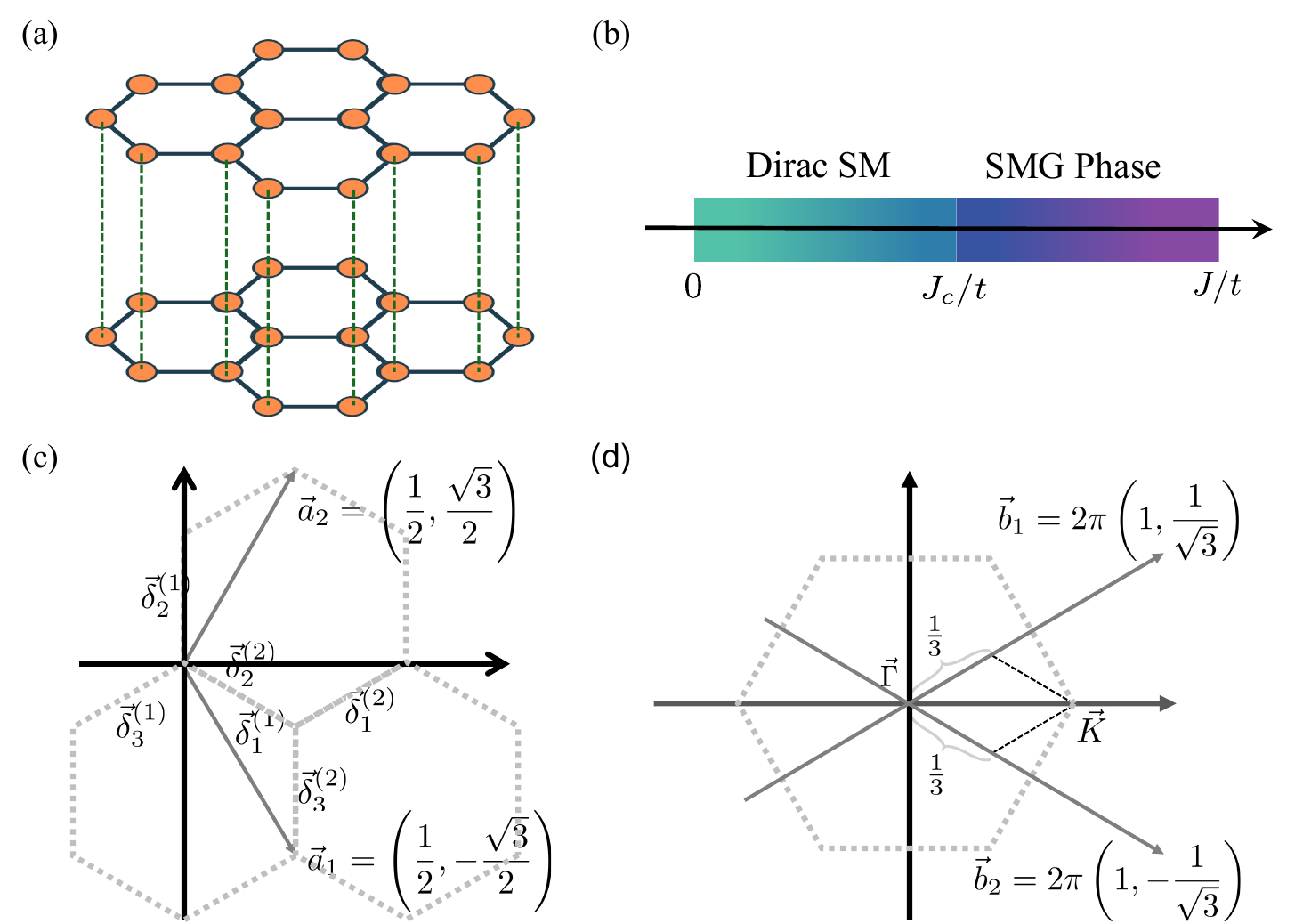}
		\caption{(a) Bilayer honeycomb lattice. (b) Phase diagram of the bilayer honeycomb lattice model. A direct SMG transition occurs between the gapless Dirac semimetal phase and the gapped SMG phase without any symmetry breaking. (c) Real-space and (d) momentum-space coordinates of the honeycomb lattice.}
		\label{fig:model_and_phase_diagram}
\end{figure}

In this Letter, we employ large-scale determinant quantum Monte Carlo (DQMC) simulations~\cite{blankenbeclerMonteCarloCalculations1981,scalapinoMonteCarloCalculations1981,assaadWorldlineDeterminantalQuantum2008,Ann.Phys.1986Sugiyama,Europhys.Lett.1989Sorella,Phys.Rev.B1989White,wang_resolving_2026} on the bilayer honeycomb lattice model proposed by Hou and You~\cite{Hou2023} [Fig.~\ref{fig:model_and_phase_diagram}(a)], with $\mathrm{SU}(2)\times\mathrm{SU}(2)\times\mathrm{SU}(2)/\mathbb{Z}_2$ (abbreviated as $\mathrm{SU}(2)^3/\mathbb{Z}_2$ in the following) symmetry. Our results provide the first numerically exact confirmation of a direct, continuous SMG transition in (2+1) dimensions and yield a fermion anomalous dimension $\eta_\psi = 0.071(1)$ that differs significantly from both VMC and large-$N$ predictions, pointing to a distinct universality class. By contrasting with a related $\mathrm{Spin}(5)\times\mathrm{U}(1)/\mathbb{Z}_2$ model that develops an intermediate excitonic phase (see the Supplemental Material (SM)~\cite{suppl}), we demonstrate the essential role of pure non-Abelian symmetry in enforcing the direct SMG transition. The symmetry analysis and classification of all fermion bilinear order parameters used here are developed systematically in a companion paper~\cite{He2026b}.

{\it Bilayer honeycomb lattice model}\,---\,
The bilayer honeycomb lattice model, proposed by Hou and You~\cite{Hou2023}, consists of two layers of honeycomb lattices with inter-layer interactions preserving a symmetry algebra $\mathfrak{su}(2)\oplus\mathfrak{su}(2)\oplus\mathfrak{su}(2)$ (abbreviated as $3\,\mathfrak{su}(2)$ in the following). The Hamiltonian reads
\begin{equation}
H = H_0 + H_{\text{int}}
\end{equation}
where $H_0$ represents the non-interacting part describing the hopping of fermions within each layer:
\begin{equation}
H_0 = -t \sum_{\langle i,j \rangle, \alpha, \sigma} (c^\dagger_{i\alpha\sigma} c_{j\alpha\sigma} + \text{h.c.})
\end{equation}
Here, $c^\dagger_{i\alpha\sigma}$ creates a fermion at site $i$ in layer $\alpha \in \{1,2\}$ with spin $\sigma \in \{\uparrow, \downarrow\}$, $t$ is the hopping amplitude, and the sum runs over nearest-neighbor bonds $\langle i,j \rangle$ within each layer. The interaction term $H_{\text{int}}$ includes inter-layer antiferromagnetic interactions:
\begin{equation}
H_{\text{int}} = J \sum_{i} \vec{S}_{i1} \cdot \vec{S}_{i2}
\end{equation}
where $\vec{S}_{i\alpha} = \frac{1}{2} \sum_{\sigma, \sigma'} c^\dagger_{i\alpha\sigma} \vec{\sigma}_{\sigma\sigma'} c_{i\alpha\sigma'}$ is the spin operator at site $i$ in layer $\alpha$, $\vec{\sigma}$ represents the Pauli matrices, and $J$ is the interaction strength. Here, we set $t=1$ and consider the half-filling condition. This interaction term couples the spins between the two layers at each site, promoting the formation of singlet states across the layers. In the large-$J$ limit, this layer singlet state is expected, and it is gapped and does not break any symmetry, making it an SMG phase.

\begin{figure}[t]
		\centering
		\includegraphics[width=1\linewidth]{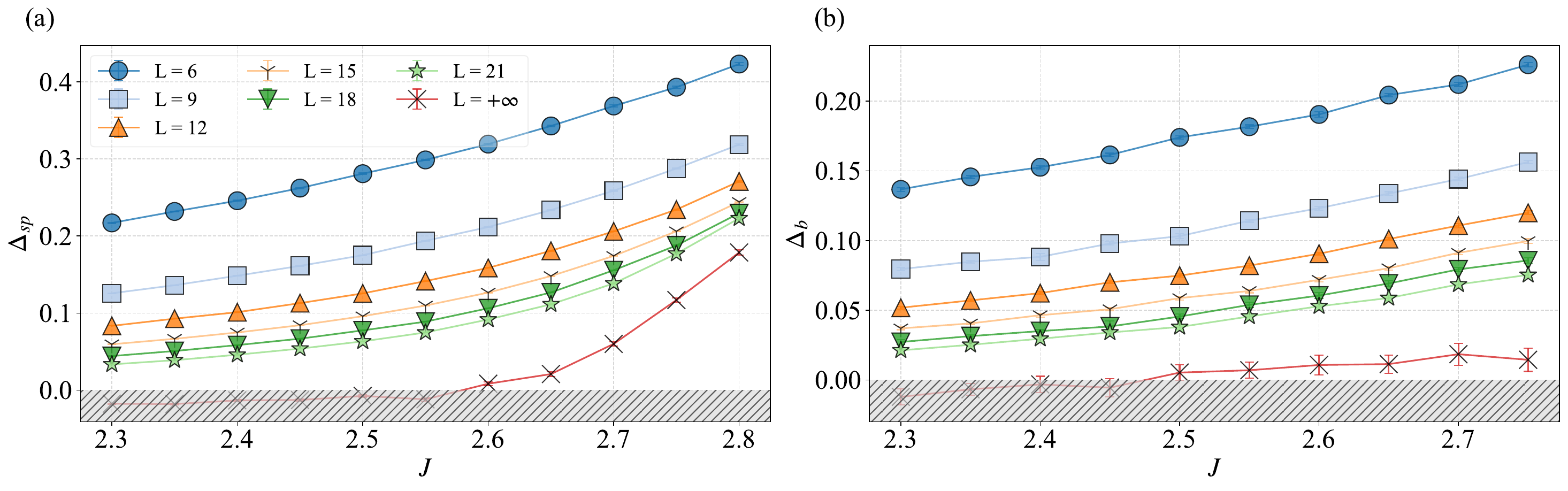}
		\caption{(a) Single-particle gap $\Delta_{sp}$ and (b) bosonic gap $\Delta_{b}$ as a function of $J$, extrapolated using the power-law form $\Delta=a\left(\frac{1}{L}\right)^{b}+c$ to the thermodynamic limit $L\to \infty$. $\Delta_{sp}$ and $\Delta_{b}$ open around $J=2.60$ and $2.55$, indicating a direct transition from the gapless phase to the gapped SMG phase.}
		\label{fig:gap}
\end{figure}

The model possesses a rich symmetry structure. First, it enjoys a total-spin $\mathrm{SU}(2)$ symmetry, reflecting the conservation of total spin, whose generators are 
$\vec{S}=\frac{1}{2}\sum_{i}\sum_{\alpha=1,2}
\left(c_{i,\alpha,\uparrow}^{\dagger},\,c_{i,\alpha,\downarrow}^{\dagger}\right)
\vec{\sigma}
\begin{pmatrix}c_{i,\alpha,\uparrow}\\c_{i,\alpha,\downarrow}\end{pmatrix}$. Second, each layer $\alpha=1,2$ independently carries a pseudo-spin $\mathrm{SU}(2)$ symmetry, generated by 
$K_{\alpha}^{z}=\frac{1}{2}\sum_{i}
\left(c_{i,\alpha,\uparrow}^{\dagger},\,c_{i,\alpha,\downarrow}\right)
\sigma^{z}
\begin{pmatrix}c_{i,\alpha,\uparrow}\\c_{i,\alpha,\downarrow}^{\dagger}\end{pmatrix}$
and 
$K_{\alpha}^{x,y}=\frac{1}{2}\sum_{i}\left(-1\right)^{i}
\left(c_{i,\alpha,\uparrow}^{\dagger},\,c_{i,\alpha,\downarrow}\right)
\sigma^{x,y}
\begin{pmatrix}c_{i,\alpha,\uparrow}\\c_{i,\alpha,\downarrow}^{\dagger}\end{pmatrix}$, where $(-1)^{i}=\pm 1$ for sites on the A/B sublattice. The symmetry algebra is $\mathfrak{su}(2)_S\oplus\mathfrak{su}(2)_{K_1}\oplus\mathfrak{su}(2)_{K_2}$. The global form of the symmetry group acting faithfully on the Fock space is $[\mathrm{SU}(2)_S\times\mathrm{SU}(2)_{K_1}\times\mathrm{SU}(2)_{K_2}]/\mathbb{Z}_2$, where the $\mathbb{Z}_2$ is the diagonal center generated by $2\pi$ rotation $\mathcal{C}_S\,\mathcal{C}_{K_1}\,\mathcal{C}_{K_2}$ which acts trivially on the Hilbert space, with $\mathcal{C}_S=e^{i2\pi S^z_\text{tot}}$ and $\mathcal{C}_{K_\alpha}=e^{i2\pi K^z_{\alpha,\text{tot}}}$ being the center elements of the three $\mathrm{SU}(2)$ factors. The high symmetry of the model is crucial for enabling SMG, as it introduces strong quantum fluctuations that can prevent symmetry breaking. 

At half-filling, the non-interacting part of the Hamiltonian describes two copies of graphene with Dirac points at the corners of the Brillouin zone. The low-energy excitations are eight massless Dirac fermions, consistent with the requirements for SMG in (2+1) dimensions. As the interaction strength $J$ increases, the model is expected to undergo a phase transition from a gapless Dirac semimetal to a gapped phase. The key question is whether this transition occurs directly, without intermediate symmetry-breaking phases, as would be expected for SMG. To investigate the phase diagram of the bilayer honeycomb lattice model and search for evidence of SMG, we employ the DQMC method, which is free from the sign problem here. We use a projection parameter $2\Theta = L + 30$ and Trotter step $\Delta\tau = 0.1$; details of the algorithm and convergence checks are provided in the SM~\cite{suppl}.

{\it Results}\,---\,
To characterize the nature of the phase transition driven by the coupling $J$, we first examine the evolution of the energy gaps in both the fermionic and bosonic sectors. We compute the imaginary-time single-particle Green's function as
\begin{equation}
	G\left(\vec{k},\tau\right)=\frac{1}{L^{4}}\sum_{j,l}\sum_{\alpha,\sigma}\left\langle c_{j,\alpha,\sigma}\left(\tau\right)c_{l,\alpha,\sigma}^{\dagger}\left(0\right)\right\rangle e^{\mathrm{i}\vec{k}\cdot\left(\vec{r}_{j}-\vec{r}_{l}\right)},
\end{equation}
where $L$ is the linear system size. The single-particle gap $\Delta_{sp}$ can be extracted at the Dirac point $\vec{K}$ from $G\left(\vec{K},\tau\right)\sim e^{-\Delta_{sp}\tau}$ in the large-$\tau$ limit. Simultaneously, to probe the collective excitations, we calculate the imaginary-time bosonic correlation functions for various symmetry channels. Among these, the lowest-energy bosonic excitation corresponds to a four-component field that captures inter-layer particle-hole and particle-particle fluctuations:
\begin{equation}
		\phi_{j}=\sum_{\sigma}\begin{pmatrix}\left(-1\right)^{j}\text{i}\left(c_{j,1,\sigma}^{\dag}c_{j,2,\sigma}-c_{j,2,\sigma}^{\dag}c_{j,1,\sigma}\right)\\
\left(-1\right)^{j}\left(c_{j,1,\sigma}^{\dag}c_{j,2,\sigma}+c_{j,2,\sigma}^{\dag}c_{j,1,\sigma}\right)\\
\left(-1\right)^{\sigma}\text{i}\left(c_{j,1,\sigma}^{\dag}c_{j,2,\bar{\sigma}}^{\dag}+c_{j,1,\sigma}c_{j,2,\bar{\sigma}}\right)\\
\left(-1\right)^{\sigma}\left(c_{j,1,\sigma}^{\dag}c_{j,2,\bar{\sigma}}^{\dag}-c_{j,1,\sigma}c_{j,2,\bar{\sigma}}\right)
\end{pmatrix},
\end{equation}
where $\left(-1\right)^{\sigma}=\pm$ for $\sigma=\uparrow/\downarrow$, $\bar{\sigma}$ represents the opposite spin of $\sigma$, and $\left(-1\right)^{j}=\pm$ for sublattices $j=A/B$. The correlation function associated with this channel exhibits the slowest decay in imaginary time, identifying it as the lowest-energy bosonic excitation. Moreover, $\phi_j$ is an irreducible representation of the global symmetry and transforms as an $\mathrm{SO}(4)$ vector, so it can serve as an order parameter for detecting symmetry breaking. This channel contains two equivalent orders: inter-layer exciton condensation (EC) order $\phi_{EC}\left(j\right)=\left(-1\right)^{j}\sum_{\sigma}\begin{pmatrix}\text{i}\left(c_{j,1,\sigma}^{\dag}c_{j,2,\sigma}-c_{j,2,\sigma}^{\dag}c_{j,1,\sigma}\right)\\c_{j,1,\sigma}^{\dag}c_{j,2,\sigma}+c_{j,2,\sigma}^{\dag}c_{j,1,\sigma}\end{pmatrix}$ and inter-layer singlet superconductivity (SC) order $\phi_{SC}\left(j\right)=\sum_{\sigma}\left(-1\right)^{\sigma}\begin{pmatrix}\text{i}\left(c_{j,1,\sigma}^{\dag}c_{j,2,\bar{\sigma}}^{\dag}+c_{j,1,\sigma}c_{j,2,\bar{\sigma}}\right)\\c_{j,1,\sigma}^{\dag}c_{j,2,\bar{\sigma}}^{\dag}-c_{j,1,\sigma}c_{j,2,\bar{\sigma}}\end{pmatrix}$. The bosonic correlation function is defined as
\begin{equation}
	P\left(\vec{k},\tau\right)=\frac{1}{L^{4}}\sum_{j,l}\left\langle \phi_{j}\left(\tau\right)\cdot\phi_{l}\left(0\right)\right\rangle e^{\mathrm{i}\vec{k}\cdot\left(\vec{r}_{j}-\vec{r}_{l}\right)},
\end{equation}
and the bosonic gap $\Delta_b$ can be obtained in the same way at the $\vec{\Gamma}$ point. By performing simulations on various lattice sizes $L$, we extrapolate these gaps to the thermodynamic limit. As shown in Fig.~\ref{fig:gap}, our data reveal that both $\Delta_{sp}$ and $\Delta_b$ remain vanishingly small in the weak coupling regime but open simultaneously as $J$ exceeds a critical threshold, suggesting a direct transition from a semimetallic to a gapped insulating phase.

\begin{figure}[t]
		\centering
		\includegraphics[width=1\linewidth]{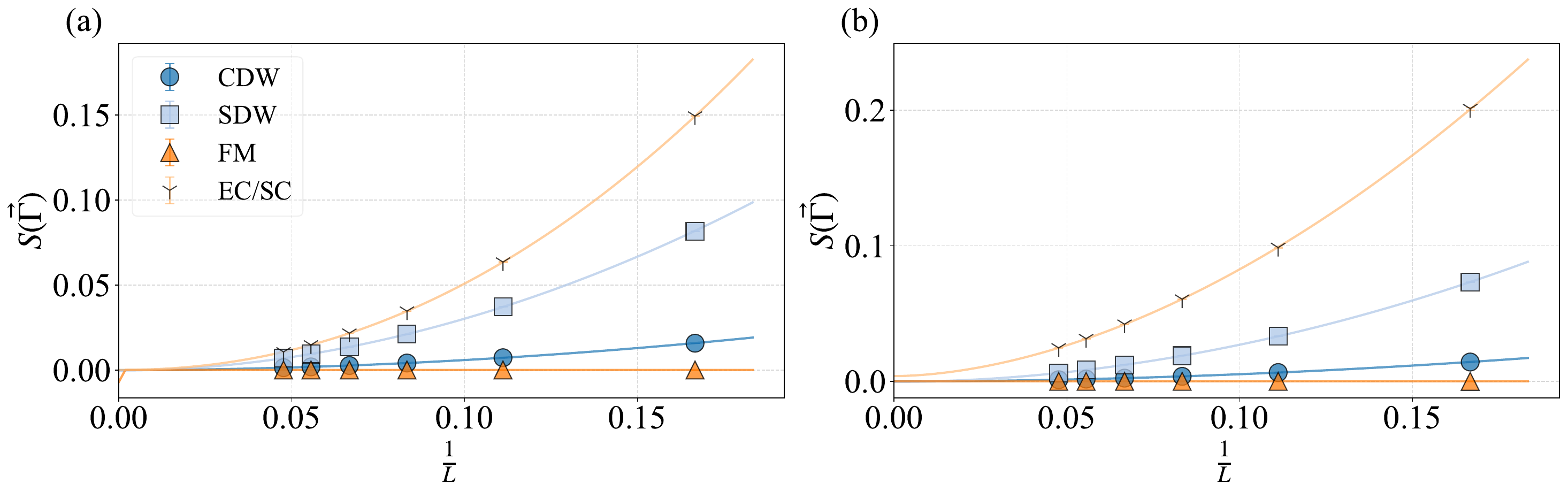}
		\caption{Structure factors of ferromagnetic (FM), charge density wave (CDW), spin density wave (SDW), and EC/SC order parameters as a function of $1/L$ for different $J$: (a) Dirac semimetal phase ($J=2.4$) and (b) SMG phase ($J=2.8$). All order parameters extrapolate to zero in the thermodynamic limit using power-law fits, indicating the absence of spontaneous symmetry breaking in the gapped phase.}
		\label{fig:order_parameters}
\end{figure}

A hallmark of the SMG mechanism is the absence of spontaneous symmetry breaking (SSB) in the gapped phase. To verify this, we performed an exhaustive group-theoretical classification of all possible fermion bilinear order parameters locally within the unit cell. Using the framework developed in the companion paper~\cite{He2026b}, the 119 nontrivial Majorana bilinears decompose into 19 irreducible multiplets under the full symmetry group, which is a $\mathbb{Z}_2^{s}\times\mathbb{Z}_2^{l}$ extension of $\mathrm{SU}(2)^3/\mathbb{Z}_2$, where $\mathbb{Z}_2^{s}$ and $\mathbb{Z}_2^{l}$ denote sublattice exchange and layer exchange, respectively (see the SM~\cite{suppl} for the complete list). We computed the structure factors for each multiplet across the transition. Remarkably, finite-size scaling analysis confirms that all 19 order parameters extrapolate to zero in the thermodynamic limit for all values of $J$ (for more details, see the SM~\cite{suppl}). Fig.~\ref{fig:order_parameters} shows that conventional ferromagnetic (FM), charge density wave (CDW), spin density wave (SDW), and EC/SC order are explicitly ruled out as representative examples. The absence of any SSB supports the realization of a featureless, symmetric insulator.

\begin{figure}[t]
		\centering
		\includegraphics[width=1\linewidth]{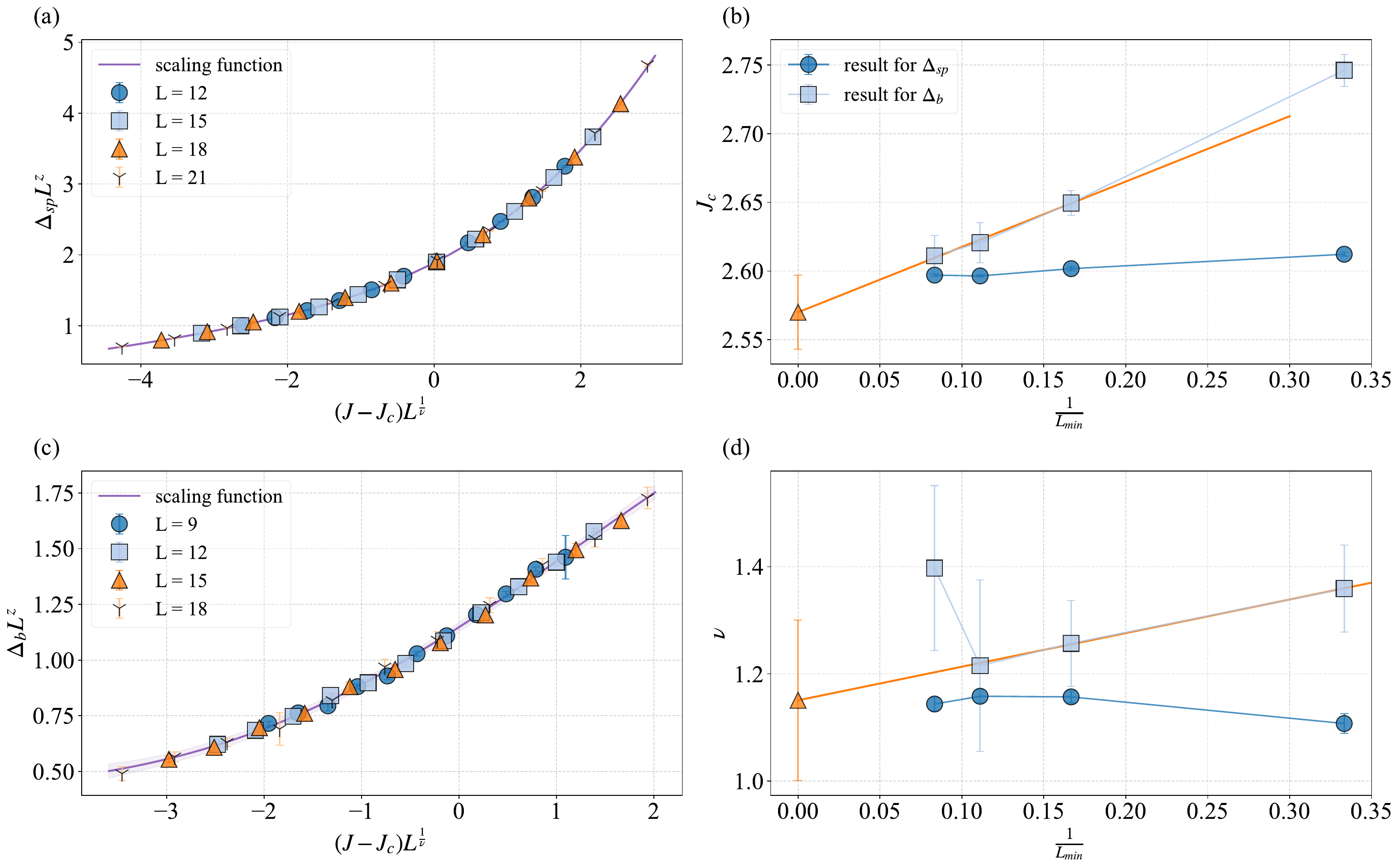}
		\caption{Finite-size scaling analysis for $\Delta_{sp}$ and $\Delta_b$ with a fitting window of four sizes. $L_{min}$ represents the minimum system size included in the fitting. (a) and (c) show representative data collapses for $\Delta_{sp}$ ($L_{min}=12$) and $\Delta_b$ ($L_{min}=9$), respectively. (b) and (d) show the evolution of the critical point $J_c$ and correlation length exponent $\nu$ as a function of $1/L_{min}$. The $\Delta_{sp}$ analysis converges, yielding $J_c=2.597(1)$ and $\nu=1.14(2)$. The $\Delta_b$ analysis does not fully converge; in particular, $\nu$ shows non-monotonic behavior due to large statistical noise at large sizes and imaginary times. Linear extrapolation gives $J_c=2.57(3)$ and $\nu=1.15(15)$.}
		\label{fig:gap_data_collapse}
\end{figure}

We determine the critical properties of this transition through a rigorous finite-size scaling analysis \cite{harada_bayesian_2011,harada_kernel_2015} of the gaps. Assuming the gap satisfies the standard scaling relation
\begin{equation}
		\Delta\left(L\right)\sim L^{-z}f\left(L^{1/\nu}(J-J_{c})\right),
\end{equation}
where the dynamical exponent is fixed to $z=1$, we collapse the data for different system sizes onto a single universal curve shown in Fig.~\ref{fig:gap_data_collapse}. The analysis of $\Delta_{sp}$ yields a critical point $J_c = 2.597(1)$ and a correlation length exponent $\nu = 1.14(2)$. A parallel analysis of $\Delta_{b}$ yields consistent results, with $J_c = 2.57(3)$ and $\nu = 1.15(15)$, albeit with slightly larger statistical uncertainties due to the higher-order Green's functions involved. The consistency of the critical points and exponents within error bars strongly supports a single, direct continuous phase transition. Further evidence against a first-order transition is provided by the smooth behavior of the energy derivative $\partial E_0 / \partial J$ across the transition (see the SM~\cite{suppl}), as well as the large value of $\nu = 1.14(2)$, far exceeding the first-order expectation $\nu = 1/(d+z) = 1/3$.

\begin{figure}[t]
		\centering
		\includegraphics[width=1\linewidth]{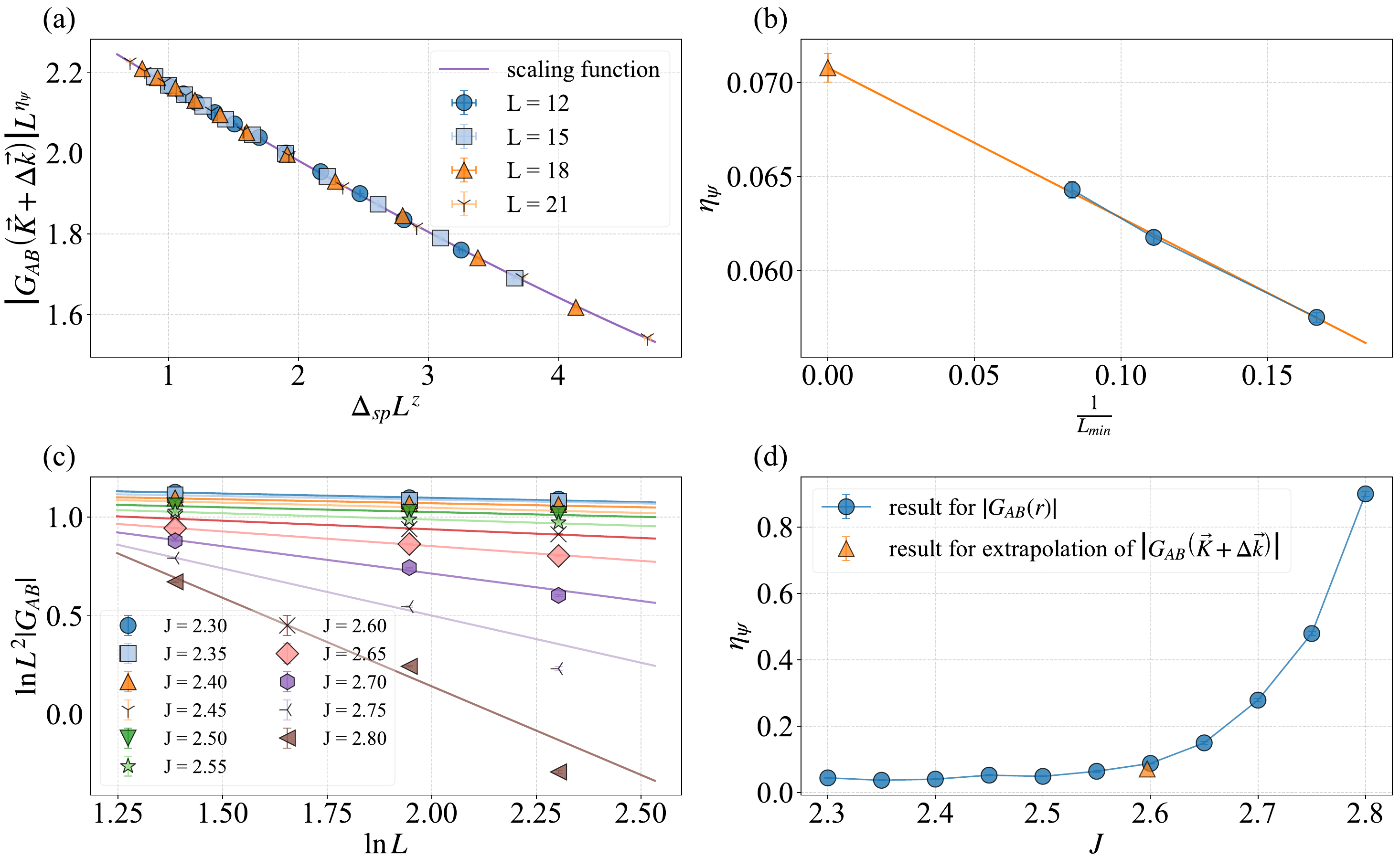}
		\caption{Scaling analysis of the fermion anomalous dimension $\eta_\psi$ from the off-diagonal Green's function in (a) momentum space and (c) real space. (b) shows the evolution of $\eta_\psi$ as a function of $1/L_{min}$ in momentum space; linear extrapolation gives $\eta_\psi = 0.071(1)$. (d) shows how $\eta_\psi$ obtained in real space varies with $J$ and gives $\eta_\psi=0.064(3)$ at $J=2.55$ and $\eta_\psi=0.088(3)$ at $J=2.60$, which are close to the result in momentum space.}
		\label{fig:fermion_anomalous_dimension}
\end{figure}

We probe the nature of the critical point by extracting the fermion anomalous dimension $\eta_\psi$. This is derived from the scaling behavior of the off-diagonal Green's function $|G_{AB}(\vec{K}+\Delta\vec{k})|$ near the Dirac points. To minimize finite-size corrections, we employ a moving window analysis, performing fits over ranges of system sizes with minimum linear size $L_{min} \in \{6, 9, 12\}$. By monitoring the evolution of the extracted exponent as a function of $1/L_{min}$ and extrapolating to the infinite-size limit, we obtain a precise estimate of $\eta_\psi = 0.071(1)$ (see Fig.~\ref{fig:fermion_anomalous_dimension}(a) and (b)).

To corroborate this result, we employed an alternative scaling analysis based on the real-space off-diagonal Green's function at the maximal separation along the $x$ direction (see Fig.~\ref{fig:model_and_phase_diagram}(c)), $G_{AB}(r=L/2)$, as a function of $L$. To mitigate finite-size effects, we selected system sizes $L \in \{8, 14, 20\}$ that are not multiples of 3. As shown in Fig.~\ref{fig:fermion_anomalous_dimension}, this approach, which we benchmarked against the known scaling dimension of free fermions~\cite{suppl}, yields anomalous dimensions near the critical point that are in agreement with the value obtained from momentum space. Furthermore, slightly away from the critical point into the strong coupling regime ($J > J_c$), the scaling of $G_{AB}(r=L/2)$ deviates significantly from a power law and exhibits exponential decay, providing independent and robust confirmation of the fermion gap opening in the SMG phase.

The small value of $\eta_\psi \approx 0.07$ presents a significant theoretical puzzle. It stands in sharp contrast to the relatively large anomalous dimensions typically associated with strong-coupling fixed points in fermionic deconfined quantum critical point (fDQCP) theories. Moreover, our result differs quantitatively from the VMC estimate $\eta_\psi \approx 0.62$ reported by Hou and You~\cite{Hou2023} and from the large-$N$ prediction $\eta_\psi \approx 0.595$~\cite{Kaul2008}. These discrepancies suggest that the critical point observed here may belong to a distinct universality class not captured by the fDQCP framework with an $\mathrm{U}(1)$ gauge field, potentially necessitating a theoretical description involving non-Abelian gauge fields matching the $\mathrm{SU}(2)^3/\mathbb{Z}_2$ structure or indicating a weakly coupled nature despite the strong interactions required to open the gap.

{\it Discussion and Conclusion\/}\,---\,
In summary, our large-scale DQMC simulations provide robust evidence for symmetric mass generation in a bilayer honeycomb lattice model with $\mathrm{SU}(2)^3/\mathbb{Z}_2$ symmetry (Lie algebra $3\,\mathfrak{su}(2)$). We have demonstrated a direct, continuous transition from a Dirac semimetal to a symmetric insulator, characterized by the simultaneous opening of fermionic and bosonic gaps without the development of any local order parameter.

The critical exponents extracted from our study challenge current effective field theory descriptions. While the transition shares qualitative features with the proposed fDQCP scenario, the quantitative mismatch---our $\eta_\psi = 0.071(1)$ versus the VMC value $\sim 0.62$~\cite{Hou2023} and the large-$N$ prediction $\sim 0.595$~\cite{Kaul2008}---points toward a more complex low-energy effective theory. It is instructive to contrast our findings with a closely related bilayer honeycomb model possessing $\mathrm{Spin}(5)\times\mathrm{U}(1)/\mathbb{Z}_2$ symmetry (see the SM~\cite{suppl} for a detailed analysis). In that system, we find no direct SMG transition; instead, an intermediate phase characterized by inter-layer excitonic order with $\mathrm{U}(1)$ symmetry breaking emerges between the semimetal and the trivial insulator (the first transition belongs to the Gross-Neveu-XY universality class and the second to the (2+1)D XY class). This comparison highlights that the pure non-Abelian symmetry of the $\mathrm{SU}(2)^3/\mathbb{Z}_2$ model is essential for prohibiting bilinear condensates and enforcing the direct SMG transition: the absence of any $\mathrm{U}(1)$ factor prevents the formation of excitonic order that would otherwise intervene.

Our work establishes a concrete and numerically exact lattice realization of SMG in (2+1)D, offering a platform to test and refine field-theoretical descriptions of non-Landau phase transitions. The unexpectedly small $\eta_\psi$ raises the question of whether the critical theory involves a non-Abelian gauge field with gauge group matching the $\mathrm{SU}(2)^3/\mathbb{Z}_2$ global symmetry, rather than the $\mathrm{U}(1)$ gauge field assumed in the standard fDQCP framework. Resolving this requires new analytical approaches, such as non-Abelian gauge theory calculations at finite $N$, as well as further numerical studies on models with varying symmetry groups to map out how the hierarchy of global symmetries governs the SMG mechanism.

\begin{acknowledgments}
CHH and XYX are supported by the National Natural Science Foundation of China (Grants No. 12447103, No. 12274289), the National Key R\&D Program of China (Grants No. 2022YFA1402702, No. 2021YFA1401400), the Innovation Program for Quantum Science and Technology (under Grant No. 2021ZD0301902), Yangyang Development Fund, and Shanghai Jiao Tong University 2030 Initiative. YZY is supported by the National Science Foundation Grant No. DMR-2238360.
The computations in this paper were run on the Siyuan-1 and $\pi$ 2.0 clusters supported by the Center for High Performance Computing at Shanghai Jiao Tong University.
\end{acknowledgments}

\textit{Note added.} We are aware a work on related topics
that have appeared on arXiv after our posting~\cite{Li2026};
their results have finite overlap with ours and seem consistent.

\bibliography{main.bib}

\begin{thebibliography}{56}%
\makeatletter
\providecommand \@ifxundefined [1]{%
 \@ifx{#1\undefined}
}%
\providecommand \@ifnum [1]{%
 \ifnum #1\expandafter \@firstoftwo
 \else \expandafter \@secondoftwo
 \fi
}%
\providecommand \@ifx [1]{%
 \ifx #1\expandafter \@firstoftwo
 \else \expandafter \@secondoftwo
 \fi
}%
\providecommand \natexlab [1]{#1}%
\providecommand \enquote  [1]{``#1''}%
\providecommand \bibnamefont  [1]{#1}%
\providecommand \bibfnamefont [1]{#1}%
\providecommand \citenamefont [1]{#1}%
\providecommand \href@noop [0]{\@secondoftwo}%
\providecommand \href [0]{\begingroup \@sanitize@url \@href}%
\providecommand \@href[1]{\@@startlink{#1}\@@href}%
\providecommand \@@href[1]{\endgroup#1\@@endlink}%
\providecommand \@sanitize@url [0]{\catcode `\\12\catcode `\$12\catcode
  `\&12\catcode `\#12\catcode `\^12\catcode `\_12\catcode `\%12\relax}%
\providecommand \@@startlink[1]{}%
\providecommand \@@endlink[0]{}%
\providecommand \url  [0]{\begingroup\@sanitize@url \@url }%
\providecommand \@url [1]{\endgroup\@href {#1}{\urlprefix }}%
\providecommand \urlprefix  [0]{URL }%
\providecommand \Eprint [0]{\href }%
\providecommand \doibase [0]{https://doi.org/}%
\providecommand \selectlanguage [0]{\@gobble}%
\providecommand \bibinfo  [0]{\@secondoftwo}%
\providecommand \bibfield  [0]{\@secondoftwo}%
\providecommand \translation [1]{[#1]}%
\providecommand \BibitemOpen [0]{}%
\providecommand \bibitemStop [0]{}%
\providecommand \bibitemNoStop [0]{.\EOS\space}%
\providecommand \EOS [0]{\spacefactor3000\relax}%
\providecommand \BibitemShut  [1]{\csname bibitem#1\endcsname}%
\let\auto@bib@innerbib\@empty
\bibitem [{\citenamefont {Anderson}(1963)}]{Anderson1963}%
  \BibitemOpen
  \bibfield  {author} {\bibinfo {author} {\bibfnamefont {P.~W.}\ \bibnamefont
  {Anderson}},\ }\bibfield  {title} {\bibinfo {title} {Plasmons, gauge
  invariance, and mass},\ }\href {https://doi.org/10.1103/PhysRev.130.439}
  {\bibfield  {journal} {\bibinfo  {journal} {Phys. Rev.}\ }\textbf {\bibinfo
  {volume} {130}},\ \bibinfo {pages} {439} (\bibinfo {year}
  {1963})}\BibitemShut {NoStop}%
\bibitem [{\citenamefont {Englert}\ and\ \citenamefont
  {Brout}(1964)}]{Englert1964}%
  \BibitemOpen
  \bibfield  {author} {\bibinfo {author} {\bibfnamefont {F.}~\bibnamefont
  {Englert}}\ and\ \bibinfo {author} {\bibfnamefont {R.}~\bibnamefont
  {Brout}},\ }\bibfield  {title} {\bibinfo {title} {Broken symmetry and the
  mass of gauge vector mesons},\ }\href
  {https://doi.org/10.1103/PhysRevLett.13.321} {\bibfield  {journal} {\bibinfo
  {journal} {Phys. Rev. Lett.}\ }\textbf {\bibinfo {volume} {13}},\ \bibinfo
  {pages} {321} (\bibinfo {year} {1964})}\BibitemShut {NoStop}%
\bibitem [{\citenamefont {Higgs}(1964)}]{Higgs1964}%
  \BibitemOpen
  \bibfield  {author} {\bibinfo {author} {\bibfnamefont {P.~W.}\ \bibnamefont
  {Higgs}},\ }\bibfield  {title} {\bibinfo {title} {Broken symmetries and the
  masses of gauge bosons},\ }\href {https://doi.org/10.1103/PhysRevLett.13.508}
  {\bibfield  {journal} {\bibinfo  {journal} {Phys. Rev. Lett.}\ }\textbf
  {\bibinfo {volume} {13}},\ \bibinfo {pages} {508} (\bibinfo {year}
  {1964})}\BibitemShut {NoStop}%
\bibitem [{\citenamefont {Guralnik}\ \emph {et~al.}(1964)\citenamefont
  {Guralnik}, \citenamefont {Hagen},\ and\ \citenamefont
  {Kibble}}]{Guralnik1964}%
  \BibitemOpen
  \bibfield  {author} {\bibinfo {author} {\bibfnamefont {G.~S.}\ \bibnamefont
  {Guralnik}}, \bibinfo {author} {\bibfnamefont {C.~R.}\ \bibnamefont
  {Hagen}},\ and\ \bibinfo {author} {\bibfnamefont {T.~W.~B.}\ \bibnamefont
  {Kibble}},\ }\bibfield  {title} {\bibinfo {title} {Global conservation laws
  and massless particles},\ }\href {https://doi.org/10.1103/PhysRevLett.13.585}
  {\bibfield  {journal} {\bibinfo  {journal} {Phys. Rev. Lett.}\ }\textbf
  {\bibinfo {volume} {13}},\ \bibinfo {pages} {585} (\bibinfo {year}
  {1964})}\BibitemShut {NoStop}%
\bibitem [{\citenamefont {Wang}\ and\ \citenamefont
  {You}(2022)}]{wang2022symmetric}%
  \BibitemOpen
  \bibfield  {author} {\bibinfo {author} {\bibfnamefont {J.}~\bibnamefont
  {Wang}}\ and\ \bibinfo {author} {\bibfnamefont {Y.-Z.}\ \bibnamefont {You}},\
  }\bibfield  {title} {\bibinfo {title} {Symmetric mass generation},\
  }\href@noop {} {\bibfield  {journal} {\bibinfo  {journal} {Symmetry}\
  }\textbf {\bibinfo {volume} {14}},\ \bibinfo {pages} {1475} (\bibinfo {year}
  {2022})}\BibitemShut {NoStop}%
\bibitem [{\citenamefont {You}\ \emph {et~al.}(2018{\natexlab{a}})\citenamefont
  {You}, \citenamefont {He}, \citenamefont {Xu},\ and\ \citenamefont
  {Vishwanath}}]{You2018}%
  \BibitemOpen
  \bibfield  {author} {\bibinfo {author} {\bibfnamefont {Y.-Z.}\ \bibnamefont
  {You}}, \bibinfo {author} {\bibfnamefont {Y.-C.}\ \bibnamefont {He}},
  \bibinfo {author} {\bibfnamefont {C.}~\bibnamefont {Xu}},\ and\ \bibinfo
  {author} {\bibfnamefont {A.}~\bibnamefont {Vishwanath}},\ }\bibfield  {title}
  {\bibinfo {title} {Symmetric fermion mass generation as deconfined quantum
  criticality},\ }\href {https://doi.org/10.1103/PhysRevX.8.011026} {\bibfield
  {journal} {\bibinfo  {journal} {Phys. Rev. X}\ }\textbf {\bibinfo {volume}
  {8}},\ \bibinfo {pages} {011026} (\bibinfo {year}
  {2018}{\natexlab{a}})}\BibitemShut {NoStop}%
\bibitem [{\citenamefont {Tong}(2022)}]{tong2022comments}%
  \BibitemOpen
  \bibfield  {author} {\bibinfo {author} {\bibfnamefont {D.}~\bibnamefont
  {Tong}},\ }\bibfield  {title} {\bibinfo {title} {Comments on symmetric mass
  generation in 2d and 4d},\ }\href@noop {} {\bibfield  {journal} {\bibinfo
  {journal} {Journal of High Energy Physics}\ }\textbf {\bibinfo {volume}
  {2022}},\ \bibinfo {pages} {1} (\bibinfo {year} {2022})}\BibitemShut
  {NoStop}%
\bibitem [{\citenamefont {Zeng}\ \emph {et~al.}(2022)\citenamefont {Zeng},
  \citenamefont {Zhu}, \citenamefont {Wang},\ and\ \citenamefont
  {You}}]{Zeng2022}%
  \BibitemOpen
  \bibfield  {author} {\bibinfo {author} {\bibfnamefont {M.}~\bibnamefont
  {Zeng}}, \bibinfo {author} {\bibfnamefont {Z.}~\bibnamefont {Zhu}}, \bibinfo
  {author} {\bibfnamefont {J.}~\bibnamefont {Wang}},\ and\ \bibinfo {author}
  {\bibfnamefont {Y.-Z.}\ \bibnamefont {You}},\ }\bibfield  {title} {\bibinfo
  {title} {Symmetric mass generation in the $1+1$ dimensional chiral fermion
  3-4-5-0 model},\ }\href {https://doi.org/10.1103/PhysRevLett.128.185301}
  {\bibfield  {journal} {\bibinfo  {journal} {Phys. Rev. Lett.}\ }\textbf
  {\bibinfo {volume} {128}},\ \bibinfo {pages} {185301} (\bibinfo {year}
  {2022})}\BibitemShut {NoStop}%
\bibitem [{\citenamefont {Lu}\ \emph {et~al.}(2023)\citenamefont {Lu},
  \citenamefont {Zeng}, \citenamefont {Wang},\ and\ \citenamefont
  {You}}]{Lu2023}%
  \BibitemOpen
  \bibfield  {author} {\bibinfo {author} {\bibfnamefont {D.-C.}\ \bibnamefont
  {Lu}}, \bibinfo {author} {\bibfnamefont {M.}~\bibnamefont {Zeng}}, \bibinfo
  {author} {\bibfnamefont {J.}~\bibnamefont {Wang}},\ and\ \bibinfo {author}
  {\bibfnamefont {Y.-Z.}\ \bibnamefont {You}},\ }\bibfield  {title} {\bibinfo
  {title} {Fermi surface symmetric mass generation},\ }\href
  {https://doi.org/10.1103/PhysRevB.107.195133} {\bibfield  {journal} {\bibinfo
   {journal} {Phys. Rev. B}\ }\textbf {\bibinfo {volume} {107}},\ \bibinfo
  {pages} {195133} (\bibinfo {year} {2023})}\BibitemShut {NoStop}%
\bibitem [{\citenamefont {Xu}\ and\ \citenamefont {Xu}(2021)}]{xu2021green}%
  \BibitemOpen
  \bibfield  {author} {\bibinfo {author} {\bibfnamefont {Y.}~\bibnamefont
  {Xu}}\ and\ \bibinfo {author} {\bibfnamefont {C.}~\bibnamefont {Xu}},\
  }\bibfield  {title} {\bibinfo {title} {Green's function zero and symmetric
  mass generation},\ }\href@noop {} {\bibfield  {journal} {\bibinfo  {journal}
  {arXiv preprint arXiv:2103.15865}\ } (\bibinfo {year} {2021})}\BibitemShut
  {NoStop}%
\bibitem [{\citenamefont {You}\ \emph {et~al.}(2018{\natexlab{b}})\citenamefont
  {You}, \citenamefont {He}, \citenamefont {Vishwanath},\ and\ \citenamefont
  {Xu}}]{You2018from}%
  \BibitemOpen
  \bibfield  {author} {\bibinfo {author} {\bibfnamefont {Y.-Z.}\ \bibnamefont
  {You}}, \bibinfo {author} {\bibfnamefont {Y.-C.}\ \bibnamefont {He}},
  \bibinfo {author} {\bibfnamefont {A.}~\bibnamefont {Vishwanath}},\ and\
  \bibinfo {author} {\bibfnamefont {C.}~\bibnamefont {Xu}},\ }\bibfield
  {title} {\bibinfo {title} {From bosonic topological transition to symmetric
  fermion mass generation},\ }\href
  {https://doi.org/10.1103/PhysRevB.97.125112} {\bibfield  {journal} {\bibinfo
  {journal} {Phys. Rev. B}\ }\textbf {\bibinfo {volume} {97}},\ \bibinfo
  {pages} {125112} (\bibinfo {year} {2018}{\natexlab{b}})}\BibitemShut
  {NoStop}%
\bibitem [{\citenamefont {Eichten}\ and\ \citenamefont
  {Preskill}(1986)}]{Eichten1986}%
  \BibitemOpen
  \bibfield  {author} {\bibinfo {author} {\bibfnamefont {E.}~\bibnamefont
  {Eichten}}\ and\ \bibinfo {author} {\bibfnamefont {J.}~\bibnamefont
  {Preskill}},\ }\bibfield  {title} {\bibinfo {title} {Chiral gauge theories on
  the lattice},\ }\href {https://doi.org/10.1016/0550-3213(86)90207-5}
  {\bibfield  {journal} {\bibinfo  {journal} {Nucl. Phys. B}\ }\textbf
  {\bibinfo {volume} {268}},\ \bibinfo {pages} {179} (\bibinfo {year}
  {1986})}\BibitemShut {NoStop}%
\bibitem [{\citenamefont {Lee}\ \emph {et~al.}(1990{\natexlab{a}})\citenamefont
  {Lee}, \citenamefont {Shigemitsu},\ and\ \citenamefont {Shrock}}]{Lee1990}%
  \BibitemOpen
  \bibfield  {author} {\bibinfo {author} {\bibfnamefont {I.-H.}\ \bibnamefont
  {Lee}}, \bibinfo {author} {\bibfnamefont {J.}~\bibnamefont {Shigemitsu}},\
  and\ \bibinfo {author} {\bibfnamefont {R.~E.}\ \bibnamefont {Shrock}},\
  }\bibfield  {title} {\bibinfo {title} {Study of different lattice
  formulations of a {Yukawa} model with a real scalar field},\ }\href
  {https://doi.org/10.1016/0550-3213(90)90664-Y} {\bibfield  {journal}
  {\bibinfo  {journal} {Nucl. Phys. B}\ }\textbf {\bibinfo {volume} {334}},\
  \bibinfo {pages} {265} (\bibinfo {year} {1990}{\natexlab{a}})}\BibitemShut
  {NoStop}%
\bibitem [{\citenamefont {Lee}\ \emph {et~al.}(1990{\natexlab{b}})\citenamefont
  {Lee}, \citenamefont {Shigemitsu},\ and\ \citenamefont
  {Shrock}}]{lee1990lattice}%
  \BibitemOpen
  \bibfield  {author} {\bibinfo {author} {\bibfnamefont {I.-H.}\ \bibnamefont
  {Lee}}, \bibinfo {author} {\bibfnamefont {J.}~\bibnamefont {Shigemitsu}},\
  and\ \bibinfo {author} {\bibfnamefont {R.~E.}\ \bibnamefont {Shrock}},\
  }\bibfield  {title} {\bibinfo {title} {Lattice study of a {Yukawa} theory
  with a real scalar field},\ }\href@noop {} {\bibfield  {journal} {\bibinfo
  {journal} {Nuclear Physics B}\ }\textbf {\bibinfo {volume} {330}},\ \bibinfo
  {pages} {225} (\bibinfo {year} {1990}{\natexlab{b}})}\BibitemShut {NoStop}%
\bibitem [{\citenamefont {Bock}\ \emph {et~al.}(1990)\citenamefont {Bock},
  \citenamefont {De}, \citenamefont {Jansen}, \citenamefont {Jers{\'a}k},
  \citenamefont {Neuhaus},\ and\ \citenamefont {Smit}}]{bock1990phase}%
  \BibitemOpen
  \bibfield  {author} {\bibinfo {author} {\bibfnamefont {W.}~\bibnamefont
  {Bock}}, \bibinfo {author} {\bibfnamefont {A.~K.}\ \bibnamefont {De}},
  \bibinfo {author} {\bibfnamefont {K.}~\bibnamefont {Jansen}}, \bibinfo
  {author} {\bibfnamefont {J.}~\bibnamefont {Jers{\'a}k}}, \bibinfo {author}
  {\bibfnamefont {T.}~\bibnamefont {Neuhaus}},\ and\ \bibinfo {author}
  {\bibfnamefont {J.}~\bibnamefont {Smit}},\ }\bibfield  {title} {\bibinfo
  {title} {Phase diagram of a lattice su(2) $\otimes$ su(2) scalar-fermion
  model with naive and wilson fermions},\ }\href
  {https://doi.org/10.1016/0550-3213(90)90689-B} {\bibfield  {journal}
  {\bibinfo  {journal} {Nucl. Phys. B}\ }\textbf {\bibinfo {volume} {344}},\
  \bibinfo {pages} {207} (\bibinfo {year} {1990})}\BibitemShut {NoStop}%
\bibitem [{\citenamefont {Bock}\ and\ \citenamefont {De}(1990)}]{Bock1990}%
  \BibitemOpen
  \bibfield  {author} {\bibinfo {author} {\bibfnamefont {W.}~\bibnamefont
  {Bock}}\ and\ \bibinfo {author} {\bibfnamefont {A.~K.}\ \bibnamefont {De}},\
  }\bibfield  {title} {\bibinfo {title} {Unquenched investigation of fermion
  masses in a chiral fermion theory on the lattice},\ }\href
  {https://doi.org/10.1016/0370-2693(90)90135-S} {\bibfield  {journal}
  {\bibinfo  {journal} {Phys. Lett. B}\ }\textbf {\bibinfo {volume} {245}},\
  \bibinfo {pages} {207} (\bibinfo {year} {1990})}\BibitemShut {NoStop}%
\bibitem [{Aba()}]{Abada1991}%
  \BibitemOpen
  \href@noop {} {\ }\BibitemShut {NoStop}%
\bibitem [{\citenamefont {Hasenfratz}\ \emph {et~al.}(1991)\citenamefont
  {Hasenfratz}, \citenamefont {Hasenfratz}, \citenamefont {Jansen},
  \citenamefont {Kuti},\ and\ \citenamefont {Shen}}]{Hasenfratz1991}%
  \BibitemOpen
  \bibfield  {author} {\bibinfo {author} {\bibfnamefont {A.}~\bibnamefont
  {Hasenfratz}}, \bibinfo {author} {\bibfnamefont {P.}~\bibnamefont
  {Hasenfratz}}, \bibinfo {author} {\bibfnamefont {K.}~\bibnamefont {Jansen}},
  \bibinfo {author} {\bibfnamefont {J.}~\bibnamefont {Kuti}},\ and\ \bibinfo
  {author} {\bibfnamefont {Y.}~\bibnamefont {Shen}},\ }\bibfield  {title}
  {\bibinfo {title} {The equivalence of the top quark condensate and the
  elementary {Higgs} field},\ }\href
  {https://doi.org/10.1016/0550-3213(91)90607-Y} {\bibfield  {journal}
  {\bibinfo  {journal} {Nucl. Phys. B}\ }\textbf {\bibinfo {volume} {365}},\
  \bibinfo {pages} {79} (\bibinfo {year} {1991})}\BibitemShut {NoStop}%
\bibitem [{\citenamefont {Banks}\ and\ \citenamefont
  {Dabholkar}(1992)}]{Banks1992}%
  \BibitemOpen
  \bibfield  {author} {\bibinfo {author} {\bibfnamefont {T.}~\bibnamefont
  {Banks}}\ and\ \bibinfo {author} {\bibfnamefont {A.}~\bibnamefont
  {Dabholkar}},\ }\bibfield  {title} {\bibinfo {title} {Decoupling a fermion
  whose mass comes from a {Yukawa} coupling: Nonperturbative considerations},\
  }\href {https://doi.org/10.1103/PhysRevD.46.4016} {\bibfield  {journal}
  {\bibinfo  {journal} {Phys. Rev. D}\ }\textbf {\bibinfo {volume} {46}},\
  \bibinfo {pages} {4016} (\bibinfo {year} {1992})}\BibitemShut {NoStop}%
\bibitem [{\citenamefont {Golterman}\ \emph {et~al.}(1993)\citenamefont
  {Golterman}, \citenamefont {Petcher},\ and\ \citenamefont
  {Rivas}}]{Golterman1993}%
  \BibitemOpen
  \bibfield  {author} {\bibinfo {author} {\bibfnamefont {M.~F.~L.}\
  \bibnamefont {Golterman}}, \bibinfo {author} {\bibfnamefont {D.~N.}\
  \bibnamefont {Petcher}},\ and\ \bibinfo {author} {\bibfnamefont
  {E.}~\bibnamefont {Rivas}},\ }\bibfield  {title} {\bibinfo {title} {Absence
  of chiral fermions in the {Eichten-Preskill} model},\ }\href
  {https://doi.org/10.1016/0550-3213(93)90049-U} {\bibfield  {journal}
  {\bibinfo  {journal} {Nucl. Phys. B}\ }\textbf {\bibinfo {volume} {395}},\
  \bibinfo {pages} {596} (\bibinfo {year} {1993})}\BibitemShut {NoStop}%
\bibitem [{\citenamefont {You}\ and\ \citenamefont {Xu}(2014)}]{You2014}%
  \BibitemOpen
  \bibfield  {author} {\bibinfo {author} {\bibfnamefont {Y.-Z.}\ \bibnamefont
  {You}}\ and\ \bibinfo {author} {\bibfnamefont {C.}~\bibnamefont {Xu}},\
  }\bibfield  {title} {\bibinfo {title} {Symmetry-protected topological states
  of interacting fermions and bosons},\ }\href
  {https://doi.org/10.1103/PhysRevB.90.245120} {\bibfield  {journal} {\bibinfo
  {journal} {Phys. Rev. B}\ }\textbf {\bibinfo {volume} {90}},\ \bibinfo
  {pages} {245120} (\bibinfo {year} {2014})}\BibitemShut {NoStop}%
\bibitem [{\citenamefont {Fidkowski}\ and\ \citenamefont
  {Kitaev}(2010)}]{Fidkowski2010}%
  \BibitemOpen
  \bibfield  {author} {\bibinfo {author} {\bibfnamefont {L.}~\bibnamefont
  {Fidkowski}}\ and\ \bibinfo {author} {\bibfnamefont {A.}~\bibnamefont
  {Kitaev}},\ }\bibfield  {title} {\bibinfo {title} {Effects of interactions on
  the topological classification of free fermion systems},\ }\href
  {https://doi.org/10.1103/PhysRevB.81.134509} {\bibfield  {journal} {\bibinfo
  {journal} {Phys. Rev. B}\ }\textbf {\bibinfo {volume} {81}},\ \bibinfo
  {pages} {134509} (\bibinfo {year} {2010})}\BibitemShut {NoStop}%
\bibitem [{\citenamefont {Fidkowski}\ and\ \citenamefont
  {Kitaev}(2011)}]{Fidkowski2011}%
  \BibitemOpen
  \bibfield  {author} {\bibinfo {author} {\bibfnamefont {L.}~\bibnamefont
  {Fidkowski}}\ and\ \bibinfo {author} {\bibfnamefont {A.}~\bibnamefont
  {Kitaev}},\ }\bibfield  {title} {\bibinfo {title} {Topological phases of
  fermions in one dimension},\ }\href
  {https://doi.org/10.1103/PhysRevB.83.075103} {\bibfield  {journal} {\bibinfo
  {journal} {Phys. Rev. B}\ }\textbf {\bibinfo {volume} {83}},\ \bibinfo
  {pages} {075103} (\bibinfo {year} {2011})}\BibitemShut {NoStop}%
\bibitem [{\citenamefont {Wang}\ and\ \citenamefont {Wen}(2013)}]{Wang2013}%
  \BibitemOpen
  \bibfield  {author} {\bibinfo {author} {\bibfnamefont {J.}~\bibnamefont
  {Wang}}\ and\ \bibinfo {author} {\bibfnamefont {X.-G.}\ \bibnamefont {Wen}},\
  }\bibfield  {title} {\bibinfo {title} {Non-perturbative regularization of 1+
  1d anomaly-free chiral fermions and bosons: On the equivalence of anomaly
  matching conditions and boundary gapping rules},\ }\href@noop {} {\bibfield
  {journal} {\bibinfo  {journal} {arXiv preprint arXiv:1307.7480}\ } (\bibinfo
  {year} {2013})}\BibitemShut {NoStop}%
\bibitem [{\citenamefont {Slagle}\ \emph {et~al.}(2015)\citenamefont {Slagle},
  \citenamefont {You},\ and\ \citenamefont {Xu}}]{Slagle2015}%
  \BibitemOpen
  \bibfield  {author} {\bibinfo {author} {\bibfnamefont {K.}~\bibnamefont
  {Slagle}}, \bibinfo {author} {\bibfnamefont {Y.-Z.}\ \bibnamefont {You}},\
  and\ \bibinfo {author} {\bibfnamefont {C.}~\bibnamefont {Xu}},\ }\bibfield
  {title} {\bibinfo {title} {Exotic quantum phase transitions of strongly
  interacting topological insulators},\ }\href
  {https://doi.org/10.1103/PhysRevB.91.115121} {\bibfield  {journal} {\bibinfo
  {journal} {Phys. Rev. B}\ }\textbf {\bibinfo {volume} {91}},\ \bibinfo
  {pages} {115121} (\bibinfo {year} {2015})}\BibitemShut {NoStop}%
\bibitem [{\citenamefont {He}\ \emph {et~al.}(2016)\citenamefont {He},
  \citenamefont {Wu}, \citenamefont {You}, \citenamefont {Xu}, \citenamefont
  {Meng},\ and\ \citenamefont {Lu}}]{He2016}%
  \BibitemOpen
  \bibfield  {author} {\bibinfo {author} {\bibfnamefont {Y.-Y.}\ \bibnamefont
  {He}}, \bibinfo {author} {\bibfnamefont {H.-Q.}\ \bibnamefont {Wu}}, \bibinfo
  {author} {\bibfnamefont {Y.-Z.}\ \bibnamefont {You}}, \bibinfo {author}
  {\bibfnamefont {C.}~\bibnamefont {Xu}}, \bibinfo {author} {\bibfnamefont
  {Z.~Y.}\ \bibnamefont {Meng}},\ and\ \bibinfo {author} {\bibfnamefont
  {Z.-Y.}\ \bibnamefont {Lu}},\ }\bibfield  {title} {\bibinfo {title} {Quantum
  critical point of dirac fermion mass generation without spontaneous symmetry
  breaking},\ }\href {https://doi.org/10.1103/PhysRevB.94.241111} {\bibfield
  {journal} {\bibinfo  {journal} {Phys. Rev. B}\ }\textbf {\bibinfo {volume}
  {94}},\ \bibinfo {pages} {241111} (\bibinfo {year} {2016})}\BibitemShut
  {NoStop}%
\bibitem [{\citenamefont {Hou}\ and\ \citenamefont {You}(2023)}]{Hou2023}%
  \BibitemOpen
  \bibfield  {author} {\bibinfo {author} {\bibfnamefont {W.}~\bibnamefont
  {Hou}}\ and\ \bibinfo {author} {\bibfnamefont {Y.-Z.}\ \bibnamefont {You}},\
  }\bibfield  {title} {\bibinfo {title} {Variational monte carlo study of
  symmetric mass generation in a bilayer honeycomb lattice model},\ }\href
  {https://doi.org/10.1103/PhysRevB.108.125130} {\bibfield  {journal} {\bibinfo
   {journal} {Phys. Rev. B}\ }\textbf {\bibinfo {volume} {108}},\ \bibinfo
  {pages} {125130} (\bibinfo {year} {2023})}\BibitemShut {NoStop}%
\bibitem [{\citenamefont {Liu}\ \emph {et~al.}(2024)\citenamefont {Liu},
  \citenamefont {Da~Liao}, \citenamefont {Pan}, \citenamefont {Song},
  \citenamefont {Zhao}, \citenamefont {Jiang}, \citenamefont {Jian},
  \citenamefont {You}, \citenamefont {Assaad}, \citenamefont {Meng},\ and\
  \citenamefont {Xu}}]{Liu2024}%
  \BibitemOpen
  \bibfield  {author} {\bibinfo {author} {\bibfnamefont {Z.~H.}\ \bibnamefont
  {Liu}}, \bibinfo {author} {\bibfnamefont {Y.}~\bibnamefont {Da~Liao}},
  \bibinfo {author} {\bibfnamefont {G.}~\bibnamefont {Pan}}, \bibinfo {author}
  {\bibfnamefont {M.}~\bibnamefont {Song}}, \bibinfo {author} {\bibfnamefont
  {J.}~\bibnamefont {Zhao}}, \bibinfo {author} {\bibfnamefont {W.}~\bibnamefont
  {Jiang}}, \bibinfo {author} {\bibfnamefont {C.-M.}\ \bibnamefont {Jian}},
  \bibinfo {author} {\bibfnamefont {Y.-Z.}\ \bibnamefont {You}}, \bibinfo
  {author} {\bibfnamefont {F.~F.}\ \bibnamefont {Assaad}}, \bibinfo {author}
  {\bibfnamefont {Z.~Y.}\ \bibnamefont {Meng}},\ and\ \bibinfo {author}
  {\bibfnamefont {C.}~\bibnamefont {Xu}},\ }\bibfield  {title} {\bibinfo
  {title} {Disorder operator and r\'enyi entanglement entropy of symmetric mass
  generation},\ }\href {https://doi.org/10.1103/PhysRevLett.132.156503}
  {\bibfield  {journal} {\bibinfo  {journal} {Phys. Rev. Lett.}\ }\textbf
  {\bibinfo {volume} {132}},\ \bibinfo {pages} {156503} (\bibinfo {year}
  {2024})}\BibitemShut {NoStop}%
\bibitem [{\citenamefont {Catterall}(2016)}]{Catterall2016}%
  \BibitemOpen
  \bibfield  {author} {\bibinfo {author} {\bibfnamefont {S.}~\bibnamefont
  {Catterall}},\ }\bibfield  {title} {\bibinfo {title} {Fermion mass without
  symmetry breaking},\ }\href@noop {} {\bibfield  {journal} {\bibinfo
  {journal} {Journal of High Energy Physics}\ }\textbf {\bibinfo {volume}
  {2016}},\ \bibinfo {pages} {1} (\bibinfo {year} {2016})}\BibitemShut
  {NoStop}%
\bibitem [{\citenamefont {Ayyar}\ and\ \citenamefont
  {Chandrasekharan}(2015)}]{Ayyar2015}%
  \BibitemOpen
  \bibfield  {author} {\bibinfo {author} {\bibfnamefont {V.}~\bibnamefont
  {Ayyar}}\ and\ \bibinfo {author} {\bibfnamefont {S.}~\bibnamefont
  {Chandrasekharan}},\ }\bibfield  {title} {\bibinfo {title} {Massive fermions
  without fermion bilinear condensates},\ }\href
  {https://doi.org/10.1103/PhysRevD.91.065035} {\bibfield  {journal} {\bibinfo
  {journal} {Phys. Rev. D}\ }\textbf {\bibinfo {volume} {91}},\ \bibinfo
  {pages} {065035} (\bibinfo {year} {2015})}\BibitemShut {NoStop}%
\bibitem [{\citenamefont {Ayyar}\ and\ \citenamefont
  {Chandrasekharan}(2016)}]{Ayyar2016}%
  \BibitemOpen
  \bibfield  {author} {\bibinfo {author} {\bibfnamefont {V.}~\bibnamefont
  {Ayyar}}\ and\ \bibinfo {author} {\bibfnamefont {S.}~\bibnamefont
  {Chandrasekharan}},\ }\bibfield  {title} {\bibinfo {title} {Origin of fermion
  masses without spontaneous symmetry breaking},\ }\href
  {https://doi.org/10.1103/PhysRevD.93.081701} {\bibfield  {journal} {\bibinfo
  {journal} {Phys. Rev. D}\ }\textbf {\bibinfo {volume} {93}},\ \bibinfo
  {pages} {081701} (\bibinfo {year} {2016})}\BibitemShut {NoStop}%
\bibitem [{\citenamefont {Maiti}\ \emph {et~al.}(2026)\citenamefont {Maiti},
  \citenamefont {Banerjee}, \citenamefont {Chandrasekharan},\ and\
  \citenamefont {Marinkovic}}]{maiti2026phase}%
  \BibitemOpen
  \bibfield  {author} {\bibinfo {author} {\bibfnamefont {S.}~\bibnamefont
  {Maiti}}, \bibinfo {author} {\bibfnamefont {D.}~\bibnamefont {Banerjee}},
  \bibinfo {author} {\bibfnamefont {S.}~\bibnamefont {Chandrasekharan}},\ and\
  \bibinfo {author} {\bibfnamefont {M.~K.}\ \bibnamefont {Marinkovic}},\
  }\bibfield  {title} {\bibinfo {title} {Phase diagram of a lattice fermion
  model with symmetric mass generation},\ }\href@noop {} {\bibfield  {journal}
  {\bibinfo  {journal} {arXiv preprint arXiv:2602.18360}\ } (\bibinfo {year}
  {2026})}\BibitemShut {NoStop}%
\bibitem [{\citenamefont {Razamat}\ and\ \citenamefont
  {Tong}(2021)}]{Razamat2021}%
  \BibitemOpen
  \bibfield  {author} {\bibinfo {author} {\bibfnamefont {S.~S.}\ \bibnamefont
  {Razamat}}\ and\ \bibinfo {author} {\bibfnamefont {D.}~\bibnamefont {Tong}},\
  }\bibfield  {title} {\bibinfo {title} {Gapped chiral fermions},\ }\href
  {https://doi.org/10.1103/PhysRevX.11.011063} {\bibfield  {journal} {\bibinfo
  {journal} {Phys. Rev. X}\ }\textbf {\bibinfo {volume} {11}},\ \bibinfo
  {pages} {011063} (\bibinfo {year} {2021})}\BibitemShut {NoStop}%
\bibitem [{\citenamefont {Blankenbecler}\ \emph {et~al.}(1981)\citenamefont
  {Blankenbecler}, \citenamefont {Scalapino},\ and\ \citenamefont
  {Sugar}}]{blankenbeclerMonteCarloCalculations1981}%
  \BibitemOpen
  \bibfield  {author} {\bibinfo {author} {\bibfnamefont {R.}~\bibnamefont
  {Blankenbecler}}, \bibinfo {author} {\bibfnamefont {D.~J.}\ \bibnamefont
  {Scalapino}},\ and\ \bibinfo {author} {\bibfnamefont {R.~L.}\ \bibnamefont
  {Sugar}},\ }\bibfield  {title} {\bibinfo {title} {{Monte Carlo calculations
  of coupled boson-fermion systems. I}},\ }\href
  {https://doi.org/10.1103/PhysRevD.24.2278} {\bibfield  {journal} {\bibinfo
  {journal} {Phys. Rev. D}\ }\textbf {\bibinfo {volume} {24}},\ \bibinfo
  {pages} {2278} (\bibinfo {year} {1981})}\BibitemShut {NoStop}%
\bibitem [{\citenamefont {Scalapino}\ and\ \citenamefont
  {Sugar}(1981)}]{scalapinoMonteCarloCalculations1981}%
  \BibitemOpen
  \bibfield  {author} {\bibinfo {author} {\bibfnamefont {D.~J.}\ \bibnamefont
  {Scalapino}}\ and\ \bibinfo {author} {\bibfnamefont {R.~L.}\ \bibnamefont
  {Sugar}},\ }\bibfield  {title} {\bibinfo {title} {{Monte Carlo calculations
  of coupled boson-fermion systems. II}},\ }\href
  {https://doi.org/10.1103/PhysRevB.24.4295} {\bibfield  {journal} {\bibinfo
  {journal} {Phys. Rev. B}\ }\textbf {\bibinfo {volume} {24}},\ \bibinfo
  {pages} {4295} (\bibinfo {year} {1981})}\BibitemShut {NoStop}%
\bibitem [{\citenamefont {Assaad}\ and\ \citenamefont
  {Evertz}(2008)}]{assaadWorldlineDeterminantalQuantum2008}%
  \BibitemOpen
  \bibfield  {author} {\bibinfo {author} {\bibfnamefont {F.}~\bibnamefont
  {Assaad}}\ and\ \bibinfo {author} {\bibfnamefont {H.}~\bibnamefont
  {Evertz}},\ }\bibinfo {title} {World-line and determinantal quantum monte
  carlo methods for spins, phonons and electrons},\ in\ \href
  {https://doi.org/10.1007/978-3-540-74686-7_10} {\emph {\bibinfo {booktitle}
  {Computational Many-Particle Physics}}},\ \bibinfo {editor} {edited by\
  \bibinfo {editor} {\bibfnamefont {H.}~\bibnamefont {Fehske}}, \bibinfo
  {editor} {\bibfnamefont {R.}~\bibnamefont {Schneider}},\ and\ \bibinfo
  {editor} {\bibfnamefont {A.}~\bibnamefont {Wei{\ss}e}}}\ (\bibinfo
  {publisher} {Springer Berlin Heidelberg},\ \bibinfo {address} {Berlin,
  Heidelberg},\ \bibinfo {year} {2008})\ pp.\ \bibinfo {pages}
  {277--356}\BibitemShut {NoStop}%
\bibitem [{\citenamefont {Sugiyama}\ and\ \citenamefont
  {Koonin}(1986)}]{Ann.Phys.1986Sugiyama}%
  \BibitemOpen
  \bibfield  {author} {\bibinfo {author} {\bibfnamefont {G.}~\bibnamefont
  {Sugiyama}}\ and\ \bibinfo {author} {\bibfnamefont {S.~E.}\ \bibnamefont
  {Koonin}},\ }\bibfield  {title} {\bibinfo {title} {Auxiliary field
  {{Monte-Carlo}} for quantum many-body ground states},\ }\href
  {https://doi.org/10.1016/0003-4916(86)90107-7} {\bibfield  {journal}
  {\bibinfo  {journal} {Annals of Physics}\ }\textbf {\bibinfo {volume}
  {168}},\ \bibinfo {pages} {1} (\bibinfo {year} {1986})}\BibitemShut {NoStop}%
\bibitem [{\citenamefont {Sorella}\ \emph {et~al.}(1989)\citenamefont
  {Sorella}, \citenamefont {Baroni}, \citenamefont {Car},\ and\ \citenamefont
  {Parrinello}}]{Europhys.Lett.1989Sorella}%
  \BibitemOpen
  \bibfield  {author} {\bibinfo {author} {\bibfnamefont {S.}~\bibnamefont
  {Sorella}}, \bibinfo {author} {\bibfnamefont {S.}~\bibnamefont {Baroni}},
  \bibinfo {author} {\bibfnamefont {R.}~\bibnamefont {Car}},\ and\ \bibinfo
  {author} {\bibfnamefont {M.}~\bibnamefont {Parrinello}},\ }\bibfield  {title}
  {\bibinfo {title} {A novel technique for the simulation of interacting
  fermion systems},\ }\href {https://doi.org/10.1209/0295-5075/8/7/014}
  {\bibfield  {journal} {\bibinfo  {journal} {Europhysics Letters (EPL)}\
  }\textbf {\bibinfo {volume} {8}},\ \bibinfo {pages} {663} (\bibinfo {year}
  {1989})}\BibitemShut {NoStop}%
\bibitem [{\citenamefont {White}\ \emph {et~al.}(1989)\citenamefont {White},
  \citenamefont {Scalapino}, \citenamefont {Sugar}, \citenamefont {Loh},
  \citenamefont {Gubernatis},\ and\ \citenamefont
  {Scalettar}}]{Phys.Rev.B1989White}%
  \BibitemOpen
  \bibfield  {author} {\bibinfo {author} {\bibfnamefont {S.~R.}\ \bibnamefont
  {White}}, \bibinfo {author} {\bibfnamefont {D.~J.}\ \bibnamefont
  {Scalapino}}, \bibinfo {author} {\bibfnamefont {R.~L.}\ \bibnamefont
  {Sugar}}, \bibinfo {author} {\bibfnamefont {E.~Y.}\ \bibnamefont {Loh}},
  \bibinfo {author} {\bibfnamefont {J.~E.}\ \bibnamefont {Gubernatis}},\ and\
  \bibinfo {author} {\bibfnamefont {R.~T.}\ \bibnamefont {Scalettar}},\
  }\bibfield  {title} {\bibinfo {title} {Numerical study of the two-dimensional
  {{Hubbard}} model},\ }\href {https://doi.org/10.1103/PhysRevB.40.506}
  {\bibfield  {journal} {\bibinfo  {journal} {Physical Review B}\ }\textbf
  {\bibinfo {volume} {40}},\ \bibinfo {pages} {506} (\bibinfo {year}
  {1989})}\BibitemShut {NoStop}%
\bibitem [{\citenamefont {Wang}\ \emph {et~al.}(2026)\citenamefont {Wang},
  \citenamefont {Sun}, \citenamefont {He},\ and\ \citenamefont
  {Xu}}]{wang_resolving_2026}%
  \BibitemOpen
  \bibfield  {author} {\bibinfo {author} {\bibfnamefont {F.-H.}\ \bibnamefont
  {Wang}}, \bibinfo {author} {\bibfnamefont {F.}~\bibnamefont {Sun}}, \bibinfo
  {author} {\bibfnamefont {C.}~\bibnamefont {He}},\ and\ \bibinfo {author}
  {\bibfnamefont {X.~Y.}\ \bibnamefont {Xu}},\ }\href
  {https://doi.org/10.48550/arXiv.2602.03656} {\bibinfo {title} {{Resolving
  Quantum Criticality in the Honeycomb Hubbard Model}}} (\bibinfo {year}
  {2026}),\ \Eprint {https://arxiv.org/abs/2602.03656} {arXiv:2602.03656
  [cond-mat]} \BibitemShut {NoStop}%
\bibitem [{sup()}]{suppl}%
  \BibitemOpen
  \href@noop {} {\bibinfo  {journal} {See {\textrm{Supplemental Material}}(SM)
  for more details.}\ }\BibitemShut {NoStop}%
\bibitem [{\citenamefont {He}\ \emph {et~al.}(2026)\citenamefont {He},
  \citenamefont {You},\ and\ \citenamefont {Xu}}]{He2026b}%
  \BibitemOpen
\bibfield  {journal} {  }\bibfield  {author} {\bibinfo {author} {\bibfnamefont
  {C.-H.}\ \bibnamefont {He}}, \bibinfo {author} {\bibfnamefont {Y.-Z.}\
  \bibnamefont {You}},\ and\ \bibinfo {author} {\bibfnamefont {X.~Y.}\
  \bibnamefont {Xu}},\ }\bibfield  {title} {\bibinfo {title} {Continuous
  symmetry analysis and systematic identification of candidate order parameters
  for interacting fermion models},\ }\href@noop {} {\bibfield  {journal}
  {\bibinfo  {journal} {arXiv preprint arXiv:2603.18285}\ } (\bibinfo {year}
  {2026})}\BibitemShut {NoStop}%
\bibitem [{\citenamefont {Harada}(2011)}]{harada_bayesian_2011}%
  \BibitemOpen
  \bibfield  {author} {\bibinfo {author} {\bibfnamefont {K.}~\bibnamefont
  {Harada}},\ }\bibfield  {title} {\bibinfo {title} {Bayesian inference in the
  scaling analysis of critical phenomena},\ }\href
  {https://doi.org/10.1103/PhysRevE.84.056704} {\bibfield  {journal} {\bibinfo
  {journal} {Physical Review E}\ }\textbf {\bibinfo {volume} {84}},\ \bibinfo
  {pages} {56704} (\bibinfo {year} {2011})}\BibitemShut {NoStop}%
\bibitem [{\citenamefont {Harada}(2015)}]{harada_kernel_2015}%
  \BibitemOpen
  \bibfield  {author} {\bibinfo {author} {\bibfnamefont {K.}~\bibnamefont
  {Harada}},\ }\bibfield  {title} {\bibinfo {title} {Kernel method for
  corrections to scaling},\ }\href {https://doi.org/10.1103/PhysRevE.92.012106}
  {\bibfield  {journal} {\bibinfo  {journal} {Physical Review E}\ }\textbf
  {\bibinfo {volume} {92}},\ \bibinfo {pages} {012106} (\bibinfo {year}
  {2015})}\BibitemShut {NoStop}%
\bibitem [{\citenamefont {Kaul}\ and\ \citenamefont
  {Sachdev}(2008)}]{Kaul2008}%
  \BibitemOpen
  \bibfield  {author} {\bibinfo {author} {\bibfnamefont {R.~K.}\ \bibnamefont
  {Kaul}}\ and\ \bibinfo {author} {\bibfnamefont {S.}~\bibnamefont {Sachdev}},\
  }\bibfield  {title} {\bibinfo {title} {Quantum criticality of u(1) gauge
  theories with fermionic and bosonic matter in two spatial dimensions},\
  }\href {https://doi.org/10.1103/PhysRevB.77.155105} {\bibfield  {journal}
  {\bibinfo  {journal} {Phys. Rev. B}\ }\textbf {\bibinfo {volume} {77}},\
  \bibinfo {pages} {155105} (\bibinfo {year} {2008})}\BibitemShut {NoStop}%
\bibitem [{\citenamefont {Li}\ \emph {et~al.}(2026)\citenamefont {Li},
  \citenamefont {Yu}, \citenamefont {Li},\ and\ \citenamefont {Yin}}]{Li2026}%
  \BibitemOpen
  \bibfield  {author} {\bibinfo {author} {\bibfnamefont {Z.-X.}\ \bibnamefont
  {Li}}, \bibinfo {author} {\bibfnamefont {Y.-K.}\ \bibnamefont {Yu}}, \bibinfo
  {author} {\bibfnamefont {Z.-X.}\ \bibnamefont {Li}},\ and\ \bibinfo {author}
  {\bibfnamefont {S.}~\bibnamefont {Yin}},\ }\bibfield  {title} {\bibinfo
  {title} {Symmetric mass generation transition and its nonequilibrium critical
  dynamics in a bilayer honeycomb lattice model},\ }\href@noop {} {\bibfield
  {journal} {\bibinfo  {journal} {arXiv preprint arXiv:2603.22736}\ } (\bibinfo
  {year} {2026})}\BibitemShut {NoStop}%
\bibitem [{\citenamefont {Chang}\ \emph {et~al.}(2023)\citenamefont {Chang},
  \citenamefont {Guo}, \citenamefont {You},\ and\ \citenamefont
  {Li}}]{chang_fermi_2023}%
  \BibitemOpen
  \bibfield  {author} {\bibinfo {author} {\bibfnamefont {W.-X.}\ \bibnamefont
  {Chang}}, \bibinfo {author} {\bibfnamefont {S.}~\bibnamefont {Guo}}, \bibinfo
  {author} {\bibfnamefont {Y.-Z.}\ \bibnamefont {You}},\ and\ \bibinfo {author}
  {\bibfnamefont {Z.-X.}\ \bibnamefont {Li}},\ }\href
  {https://doi.org/10.48550/arXiv.2311.09970} {\bibinfo {title} {Fermi surface
  symmetric mass generation: A quantum monte-carlo study}} (\bibinfo {year}
  {2023}),\ \Eprint {https://arxiv.org/abs/2311.09970} {arXiv:2311.09970
  [cond-mat]} \BibitemShut {NoStop}%
\bibitem [{\citenamefont {Assaad}(2002)}]{assaad_depleted_2002}%
  \BibitemOpen
  \bibfield  {author} {\bibinfo {author} {\bibfnamefont {F.~F.}\ \bibnamefont
  {Assaad}},\ }\bibfield  {title} {\bibinfo {title} {Depleted kondo lattices:
  Quantum monte carlo and mean-field calculations},\ }\href
  {https://doi.org/10.1103/PhysRevB.65.115104} {\bibfield  {journal} {\bibinfo
  {journal} {Physical Review B}\ }\textbf {\bibinfo {volume} {65}},\ \bibinfo
  {pages} {115104} (\bibinfo {year} {2002})}\BibitemShut {NoStop}%
\bibitem [{\citenamefont {Li}\ \emph {et~al.}(2017)\citenamefont {Li},
  \citenamefont {Jiang}, \citenamefont {Jian},\ and\ \citenamefont
  {Yao}}]{li_fermioninduced_2017}%
  \BibitemOpen
  \bibfield  {author} {\bibinfo {author} {\bibfnamefont {Z.-X.}\ \bibnamefont
  {Li}}, \bibinfo {author} {\bibfnamefont {Y.-F.}\ \bibnamefont {Jiang}},
  \bibinfo {author} {\bibfnamefont {S.-K.}\ \bibnamefont {Jian}},\ and\
  \bibinfo {author} {\bibfnamefont {H.}~\bibnamefont {Yao}},\ }\bibfield
  {title} {\bibinfo {title} {Fermion-induced quantum critical points},\ }\href
  {https://doi.org/10.1038/s41467-017-00167-6} {\bibfield  {journal} {\bibinfo
  {journal} {Nature Communications}\ }\textbf {\bibinfo {volume} {8}},\
  \bibinfo {pages} {314} (\bibinfo {year} {2017})}\BibitemShut {NoStop}%
\bibitem [{\citenamefont {Da~Liao}\ \emph {et~al.}(2019)\citenamefont
  {Da~Liao}, \citenamefont {Meng},\ and\ \citenamefont
  {Xu}}]{daliao_valence_2019}%
  \BibitemOpen
  \bibfield  {author} {\bibinfo {author} {\bibfnamefont {Y.}~\bibnamefont
  {Da~Liao}}, \bibinfo {author} {\bibfnamefont {Z.~Y.}\ \bibnamefont {Meng}},\
  and\ \bibinfo {author} {\bibfnamefont {X.~Y.}\ \bibnamefont {Xu}},\
  }\bibfield  {title} {\bibinfo {title} {Valence bond orders at charge
  neutrality in a possible two-orbital extended hubbard model for twisted
  bilayer graphene},\ }\href {https://doi.org/10.1103/PhysRevLett.123.157601}
  {\bibfield  {journal} {\bibinfo  {journal} {Physical Review Letters}\
  }\textbf {\bibinfo {volume} {123}},\ \bibinfo {pages} {157601} (\bibinfo
  {year} {2019})}\BibitemShut {NoStop}%
\bibitem [{\citenamefont {Hawashin}\ \emph {et~al.}(2025)\citenamefont
  {Hawashin}, \citenamefont {Scherer},\ and\ \citenamefont
  {Janssen}}]{hawashin_grossneveuxy_2025}%
  \BibitemOpen
  \bibfield  {author} {\bibinfo {author} {\bibfnamefont {B.}~\bibnamefont
  {Hawashin}}, \bibinfo {author} {\bibfnamefont {M.~M.}\ \bibnamefont
  {Scherer}},\ and\ \bibinfo {author} {\bibfnamefont {L.}~\bibnamefont
  {Janssen}},\ }\bibfield  {title} {\bibinfo {title} {Gross-neveu-xy quantum
  criticality in moir\textbackslash 'e dirac materials},\ }\href
  {https://doi.org/10.1103/PhysRevB.111.205129} {\bibfield  {journal} {\bibinfo
   {journal} {Physical Review B}\ }\textbf {\bibinfo {volume} {111}},\ \bibinfo
  {pages} {205129} (\bibinfo {year} {2025})}\BibitemShut {NoStop}%
\bibitem [{\citenamefont {Zerf}\ \emph {et~al.}(2017)\citenamefont {Zerf},
  \citenamefont {Mihaila}, \citenamefont {Marquard}, \citenamefont {Herbut},\
  and\ \citenamefont {Scherer}}]{zerf_fourloop_2017}%
  \BibitemOpen
  \bibfield  {author} {\bibinfo {author} {\bibfnamefont {N.}~\bibnamefont
  {Zerf}}, \bibinfo {author} {\bibfnamefont {L.~N.}\ \bibnamefont {Mihaila}},
  \bibinfo {author} {\bibfnamefont {P.}~\bibnamefont {Marquard}}, \bibinfo
  {author} {\bibfnamefont {I.~F.}\ \bibnamefont {Herbut}},\ and\ \bibinfo
  {author} {\bibfnamefont {M.~M.}\ \bibnamefont {Scherer}},\ }\bibfield
  {title} {\bibinfo {title} {Four-loop critical exponents for the
  gross-neveu-yukawa models},\ }\href
  {https://doi.org/10.1103/PhysRevD.96.096010} {\bibfield  {journal} {\bibinfo
  {journal} {Physical Review D}\ }\textbf {\bibinfo {volume} {96}},\ \bibinfo
  {pages} {096010} (\bibinfo {year} {2017})}\BibitemShut {NoStop}%
\bibitem [{\citenamefont {Gracey}(2021)}]{gracey_critical_2021}%
  \BibitemOpen
  \bibfield  {author} {\bibinfo {author} {\bibfnamefont {J.~A.}\ \bibnamefont
  {Gracey}},\ }\bibfield  {title} {\bibinfo {title} {Critical exponent
  {$\ensuremath{\eta}$} at {$O(1/{N}^{3})$} in the chiral xy model using the
  large {$N$} conformal bootstrap},\ }\href
  {https://doi.org/10.1103/PhysRevD.103.065018} {\bibfield  {journal} {\bibinfo
   {journal} {Physical Review D}\ }\textbf {\bibinfo {volume} {103}},\ \bibinfo
  {pages} {065018} (\bibinfo {year} {2021})}\BibitemShut {NoStop}%
\bibitem [{\citenamefont {Classen}\ \emph {et~al.}(2017)\citenamefont
  {Classen}, \citenamefont {Herbut},\ and\ \citenamefont
  {Scherer}}]{classen_fluctuationinduced_2017}%
  \BibitemOpen
  \bibfield  {author} {\bibinfo {author} {\bibfnamefont {L.}~\bibnamefont
  {Classen}}, \bibinfo {author} {\bibfnamefont {I.~F.}\ \bibnamefont
  {Herbut}},\ and\ \bibinfo {author} {\bibfnamefont {M.~M.}\ \bibnamefont
  {Scherer}},\ }\bibfield  {title} {\bibinfo {title} {Fluctuation-induced
  continuous transition and quantum criticality in dirac semimetals},\ }\href
  {https://doi.org/10.1103/PhysRevB.96.115132} {\bibfield  {journal} {\bibinfo
  {journal} {Physical Review B}\ }\textbf {\bibinfo {volume} {96}},\ \bibinfo
  {pages} {115132} (\bibinfo {year} {2017})}\BibitemShut {NoStop}%
\bibitem [{\citenamefont {Hasenbusch}(2025)}]{hasenbusch_eliminating_2025}%
  \BibitemOpen
  \bibfield  {author} {\bibinfo {author} {\bibfnamefont {M.}~\bibnamefont
  {Hasenbusch}},\ }\bibfield  {title} {\bibinfo {title} {Eliminating leading
  and subleading corrections to scaling in the three-dimensional xy
  universality class},\ }\href {https://doi.org/10.1103/1d4g-483z} {\bibfield
  {journal} {\bibinfo  {journal} {Physical Review B}\ }\textbf {\bibinfo
  {volume} {112}},\ \bibinfo {pages} {184512} (\bibinfo {year}
  {2025})}\BibitemShut {NoStop}%
\bibitem [{\citenamefont {Chester}\ \emph {et~al.}(2020)\citenamefont
  {Chester}, \citenamefont {Landry}, \citenamefont {Liu}, \citenamefont
  {Poland}, \citenamefont {{Simmons-Duffin}}, \citenamefont {Su},\ and\
  \citenamefont {Vichi}}]{chester_carving_2020}%
  \BibitemOpen
  \bibfield  {author} {\bibinfo {author} {\bibfnamefont {S.~M.}\ \bibnamefont
  {Chester}}, \bibinfo {author} {\bibfnamefont {W.}~\bibnamefont {Landry}},
  \bibinfo {author} {\bibfnamefont {J.}~\bibnamefont {Liu}}, \bibinfo {author}
  {\bibfnamefont {D.}~\bibnamefont {Poland}}, \bibinfo {author} {\bibfnamefont
  {D.}~\bibnamefont {{Simmons-Duffin}}}, \bibinfo {author} {\bibfnamefont
  {N.}~\bibnamefont {Su}},\ and\ \bibinfo {author} {\bibfnamefont
  {A.}~\bibnamefont {Vichi}},\ }\bibfield  {title} {\bibinfo {title} {Carving
  out ope space and precise o(2) model critical exponents},\ }\href
  {https://doi.org/10.1007/JHEP06(2020)142} {\bibfield  {journal} {\bibinfo
  {journal} {Journal of High Energy Physics}\ }\textbf {\bibinfo {volume}
  {2020}},\ \bibinfo {pages} {142} (\bibinfo {year} {2020})}\BibitemShut
  {NoStop}%
\end{thebibliography}%

\clearpage
\onecolumngrid

\setcounter{equation}{0}
\setcounter{figure}{0}
\setcounter{table}{0}
\setcounter{section}{0}
\renewcommand{\theequation}{S\arabic{equation}}
\renewcommand{\thefigure}{S\arabic{figure}}
\renewcommand{\thetable}{S\arabic{table}}
\renewcommand{\thesection}{S\arabic{section}}
\renewcommand{\bibnumfmt}[1]{[S#1]}
\renewcommand{\citenumfont}[1]{S#1}

\begin{center}
\textbf{\large Supplemental Material: Symmetric Mass Generation in a Bilayer Honeycomb Lattice with $\mathrm{SU}(2)\times\mathrm{SU}(2)\times\mathrm{SU}(2)/\mathbb{Z}_2$ Symmetry}
\end{center}

\section{Projection Quantum Monte Carlo Method}

To investigate the phase diagram of the bilayer honeycomb lattice model and search for evidence of symmetric mass generation (SMG), we employ the determinant quantum Monte Carlo (DQMC) method. DQMC is a powerful numerical technique for simulating interacting fermionic systems at both zero and finite temperature, allowing for unbiased calculations of thermodynamic quantities and correlation functions. We focus on ground-state properties and implement the projection version of DQMC, i.e., projection quantum Monte Carlo (PQMC). Observable expectation values are calculated as
\begin{equation}
\langle O \rangle= \frac{\langle \Psi_0 |O|\Psi_0\rangle}{\langle \Psi_0 |\Psi_0\rangle},
\end{equation}
where $|\Psi_0\rangle$ is obtained by projecting a trial wave function, $|\Psi_0\rangle = e^{-\Theta H}|\Psi_T\rangle$, and $|\Psi_T\rangle$ has a finite overlap with $|\Psi_0\rangle$. The parameter $\Theta$ is chosen large enough to project to the ground state. In PQMC, the Trotter decomposition and Hubbard-Stratonovich (HS) transformation are used, and the above observables can be expressed as a weighted sum over auxiliary fields. Markov chain Monte Carlo is then performed for importance sampling of the auxiliary fields.

The Trotter decomposition separates the non-commuting kinetic and interaction parts of the Hamiltonian as
\begin{equation}
    e^{-2\Theta H}=\left(e^{-\frac{\Delta\tau}{2}H_{0}}e^{-\Delta\tau H_{\text{int}}}e^{-\frac{\Delta\tau}{2}H_{0}}\right)^{L_{\tau}}+\mathcal{O}\left(\left(\Delta\tau\right)^{2}\right),
\end{equation}
where $L_{\tau}=2\Theta/\Delta\tau$ is the number of imaginary time slices. To perform the HS transformation, we first rewrite the interaction term as 
\begin{equation}
H_\text{int}=\frac{J}{4}\sum_{i}\sum_{\alpha=x,y,z}\left[\left(S_{i,1}^{\alpha}+S_{i,2}^{\alpha}\right)^{2}-\left(S_{i,1}^{\alpha}-S_{i,2}^{\alpha}\right)^{2}\right].
\end{equation}
Again, using the Trotter decomposition, we separate the above interaction term,
\begin{equation}
    e^{-\Delta\tau H_{\text{int}}}=\prod_{i}\prod_{\alpha=x,y,z}\exp\left(-\Delta\tau\frac{J}{4}\left(S_{i,1}^{\alpha}+S_{i,2}^{\alpha}\right)^{2}\right)\exp\left(\Delta\tau\frac{J}{4}\left(S_{i,1}^{\alpha}-S_{i,2}^{\alpha}\right)^{2}\right)+\mathcal{O}\left(\left(\Delta\tau\right)^{2}\right).
\end{equation}
We then use the following HS transformation:
\begin{equation}
e^{\lambda A^{2}}=\frac{1}{4}\sum_{s=\pm1,\pm2}\gamma(s)e^{\sqrt{\lambda}\eta(s)A}+\mathcal{O}(\lambda^{4}),
\end{equation}
where $A=S_{i,1}^{\alpha}\pm S_{i,2}^{\alpha}$ and $\lambda=\mp\Delta\tau J/4$. The coefficients are $\gamma(\pm1)=1+\frac{\sqrt{6}}{3}$, $\gamma(\pm2)=1-\frac{\sqrt{6}}{3}$, $\eta(\pm1)=\pm\sqrt{2\left(3-\sqrt{6}\right)}$, and $\eta(\pm2)=\pm\sqrt{2\left(3+\sqrt{6}\right)}$. Chang et al.~\cite{chang_fermi_2023} showed that this HS transformation leads to an antiunitary symmetry of the spacetime fermion-bilinear Hamiltonian, which guarantees the absence of the sign problem for finite-temperature DQMC. For PQMC, the trial wave function is typically required to respect the same antiunitary symmetry to remain sign-problem-free. At half-filling, the absence of the sign problem can be understood more directly, which also guides the choice of $|\Psi_T\rangle$. The model has a particle-hole symmetry at half-filling, so the chemical potential is set to $\mu=0$. Under the new basis $\left\{ \tilde{c}\right\} $ obtained by performing a partial particle-hole transformation on the layer-2 fermions, 
\begin{gather}
    c_{i,1,\sigma}=\tilde{c}_{i,1,\sigma},\\c_{i,2,\sigma}=\left(-1\right)^{i}\tilde{c}_{i,2,\sigma}^{\dagger},
\end{gather}
where $(-1)^i$ represents $\pm 1$ on sublattice A/B. In this new basis, the fermion-bilinear operators decouple into a block-diagonal form with respect to the layer index, and the two blocks are complex conjugates of each other. If we choose the trial wave function $|\Psi_T\rangle$ with the same property (which is straightforward), it suffices to show that the PQMC simulation is free of the sign problem.

We choose the trial wave function $|\Psi_T\rangle$ (for one block; the other block is obtained by complex conjugation) as the ground state of the noninteracting Hamiltonian $H_0'$, with a tiny spin flip term introduced to lift degeneracy at Dirac points,
\begin{gather}
    H_{0}^{\prime}=-t\sum_{\left\langle j,k\right\rangle }\left(c_{j,\uparrow}^{\dagger},c_{j,\downarrow}^{\dagger}\right)e^{i\phi_{jk}\sigma^{x}}\begin{pmatrix}c_{k,\uparrow}\\
c_{k,\downarrow}
\end{pmatrix}+\mathrm{h.c.},\\\phi_{jk}=\frac{2\pi}{\Phi_{0}}\int_{j}^{k}\vec{A}\cdot d\vec{r},\\\vec{A}=\left(\frac{\mathscr{A}_{a}\vec{a}}{L_{a}}+\frac{\mathscr{A}_{b}\vec{b}}{L_{b}}\right)\Phi_{0}.
\end{gather}
Here, $\Phi_{0}=h/e$ is the magnetic flux quantum, and $\vec{a}$ and $\vec{b}$ are the two basis vectors of the unit cell. $L_{a}$ and $L_{b}$ denote the lengths of the system along the $\vec{a}$ and $\vec{b}$ directions, respectively. The dimensionless parameters $\mathscr{A}_{a}$ and $\mathscr{A}_{b}$ represent the magnetic fluxes (in units of $\Phi_0$) threading the cyclic loops along the $\vec{a}$ and $\vec{b}$ directions, which are set to $\mathscr{A}_{a}=10^{-4}$ and $\mathscr{A}_{b}=0$ in our calculations. We set the projection parameter $2\Theta = L+30$ and $\Delta\tau=0.1$ to obtain sufficiently accurate results throughout the calculation, where $L$ is the linear system size.

\section{Possible Symmetry-Breaking Orders}

As discussed in the main text, the model has a continuous symmetry $\mathrm{SU}(2)^3/\mathbb{Z}_2$. To diagnose possible symmetry-breaking phases, we enumerate fermion-bilinear order parameters defined within a unit cell. The internal degrees of freedom can be organized as
\begin{equation}
    \underset{\text{sublattice}}{\begin{bmatrix}A\\
B
\end{bmatrix}}\otimes\underset{\text{Majorana}}{\begin{bmatrix}\Re c\\
\Im c
\end{bmatrix}}\otimes\underset{\text{layer}}{\begin{bmatrix}1\\
2
\end{bmatrix}}\otimes\underset{\text{spin}}{\begin{bmatrix}\uparrow\\
\downarrow
\end{bmatrix}},
\end{equation}
which gives $2^{4}=16$ real (Majorana) components and hence $\binom{16}{2}=120$ independent Majorana bilinears. The Hamiltonian also has two additional $\mathbb{Z}_{2}$ discrete symmetries associated with the sublattice and layer degrees of freedom. From the perspective of group representation theory, the order parameter can be viewed as a non-trivial irreducible representation of the symmetry group acting on the space of degrees of freedom under consideration. Correspondingly, the problem of determining the order parameter can be transformed into the decomposition of the Lie group representation into irreducible representations. The hopping term
\begin{equation}
    \sum_{i,l,\sigma}\left(c_{i,A,l,\sigma}^{\dagger}c_{i,B,l,\sigma}+\mathrm{h.c.}\right),
\end{equation}
where $i$ labels the unit cell (suppressed below), $A/B$ denote the two sublattices, $l=1,2$ is the layer index, and $\sigma=\uparrow,\downarrow$ denotes the spin, transforms as a scalar under the full symmetry group and thus is not a symmetry-breaking order parameter. The remaining $119$ nontrivial bilinears can be grouped into $19$ multiplets (of different dimensions), which we list below:
\begin{equation}\label{eq:A-1}
    \tilde{A}_{1}=\begin{pmatrix}i\left(c_{A,2,\uparrow}^{\dag}c_{A,2,\downarrow}^{\dag}+c_{B,2,\uparrow}^{\dag}c_{B,2,\downarrow}^{\dag}+c_{A,2,\uparrow}c_{A,2,\downarrow}+c_{B,2,\uparrow}c_{B,2,\downarrow}\right)\\
c_{A,2,\uparrow}^{\dag}c_{A,2,\downarrow}^{\dag}+c_{B,2,\uparrow}^{\dag}c_{B,2,\downarrow}^{\dag}-c_{A,2,\uparrow}c_{A,2,\downarrow}-c_{B,2,\uparrow}c_{B,2,\downarrow}\\
\left(c_{A,2,\uparrow}^{\dag}c_{A,2,\uparrow}+c_{A,2,\downarrow}^{\dag}c_{A,2,\downarrow}\right)-\left(c_{B,2,\uparrow}^{\dag}c_{B,2,\uparrow}+c_{B,2,\downarrow}^{\dag}c_{B,2,\downarrow}\right)\\
c_{A,1,\uparrow}^{\dag}c_{A,1,\downarrow}^{\dag}+c_{B,1,\uparrow}^{\dag}c_{B,1,\downarrow}^{\dag}-c_{A,1,\uparrow}c_{A,1,\downarrow}-c_{B,1,\uparrow}c_{B,1,\downarrow}\\
i\left(c_{A,1,\uparrow}^{\dag}c_{A,1,\downarrow}^{\dag}+c_{B,1,\uparrow}^{\dag}c_{B,1,\downarrow}^{\dag}+c_{A,1,\uparrow}c_{A,1,\downarrow}+c_{B,1,\uparrow}c_{B,1,\downarrow}\right)\\
\left(c_{A,1,\uparrow}^{\dag}c_{A,1,\uparrow}+c_{A,1,\downarrow}^{\dag}c_{A,1,\downarrow}\right)-\left(c_{B,1,\uparrow}^{\dag}c_{B,1,\uparrow}+c_{B,1,\downarrow}^{\dag}c_{B,1,\downarrow}\right)
\end{pmatrix},
\end{equation}
\begin{equation}\label{eq:A-2}
    \tilde{A}_{2}=\begin{pmatrix}c_{A,2,\uparrow}^{\dag}c_{A,2,\downarrow}^{\dag}-c_{B,2,\uparrow}^{\dag}c_{B,2,\downarrow}^{\dag}-c_{A,2,\uparrow}c_{A,2,\downarrow}+c_{B,2,\uparrow}c_{B,2,\downarrow}\\
i\left(c_{A,2,\uparrow}^{\dag}c_{A,2,\downarrow}^{\dag}-c_{B,2,\uparrow}^{\dag}c_{B,2,\downarrow}^{\dag}+c_{A,2,\uparrow}c_{A,2,\downarrow}-c_{B,2,\uparrow}c_{B,2,\downarrow}\right)\\
c_{A,2,\uparrow}^{\dag}c_{A,2,\uparrow}+c_{A,2,\downarrow}^{\dag}c_{A,2,\downarrow}+c_{B,2,\uparrow}^{\dag}c_{B,2,\uparrow}+c_{B,2,\downarrow}^{\dag}c_{B,2,\downarrow}-2\\
c_{A,1,\uparrow}^{\dag}c_{A,1,\downarrow}^{\dag}-c_{B,1,\uparrow}^{\dag}c_{B,1,\downarrow}^{\dag}-c_{A,1,\uparrow}c_{A,1,\downarrow}+c_{B,1,\uparrow}c_{B,1,\downarrow}\\
i\left(c_{A,1,\uparrow}^{\dag}c_{A,1,\downarrow}^{\dag}-c_{B,1,\uparrow}^{\dag}c_{B,1,\downarrow}^{\dag}+c_{A,1,\uparrow}c_{A,1,\downarrow}-c_{B,1,\uparrow}c_{B,1,\downarrow}\right)\\
c_{A,1,\uparrow}^{\dag}c_{A,1,\uparrow}+c_{A,1,\downarrow}^{\dag}c_{A,1,\downarrow}+c_{B,1,\uparrow}^{\dag}c_{B,1,\uparrow}+c_{B,1,\downarrow}^{\dag}c_{B,1,\downarrow}-2
\end{pmatrix},
\end{equation}
\begin{equation} \label{A-3}
    \tilde{A}_{3}=\begin{pmatrix}c_{A,2,\uparrow}^{\dag}c_{B,2,\downarrow}^{\dag}-c_{A,2,\downarrow}^{\dag}c_{B,2,\uparrow}^{\dag}-c_{A,2,\uparrow}c_{B,2,\downarrow}+c_{A,2,\downarrow}c_{B,2,\uparrow}\\
i\left(c_{A,2,\uparrow}^{\dag}c_{B,2,\downarrow}^{\dag}-c_{A,2,\downarrow}^{\dag}c_{B,2,\uparrow}^{\dag}+c_{A,2,\uparrow}c_{B,2,\downarrow}-c_{A,2,\downarrow}c_{B,2,\uparrow}\right)\\
i\left(c_{A,2,\uparrow}^{\dag}c_{B,2,\uparrow}+c_{A,2,\downarrow}^{\dag}c_{B,2,\downarrow}-c_{B,2,\uparrow}^{\dag}c_{A,2,\uparrow}-c_{B,2,\downarrow}^{\dag}c_{A,2,\downarrow}\right)\\
c_{A,1,\uparrow}^{\dag}c_{B,1,\downarrow}^{\dag}-c_{A,1,\downarrow}^{\dag}c_{B,1,\uparrow}^{\dag}-c_{A,1,\uparrow}c_{B,1,\downarrow}+c_{A,1,\downarrow}c_{B,1,\uparrow}\\
i\left(c_{A,1,\uparrow}^{\dag}c_{B,1,\downarrow}^{\dag}-c_{A,1,\downarrow}^{\dag}c_{B,1,\uparrow}^{\dag}+c_{A,1,\uparrow}c_{B,1,\downarrow}-c_{A,1,\downarrow}c_{B,1,\uparrow}\right)\\
i\left(c_{A,1,\uparrow}^{\dag}c_{B,1,\uparrow}+c_{A,1,\downarrow}^{\dag}c_{B,1,\downarrow}-c_{B,1,\uparrow}^{\dag}c_{A,1,\uparrow}-c_{B,1,\downarrow}^{\dag}c_{A,1,\downarrow}\right)
\end{pmatrix},
\end{equation}
\begin{equation}
    \tilde{A}_{4}=\begin{pmatrix}i\begin{bmatrix}\left(c_{A,1,\uparrow}^{\dag}c_{A,1,\downarrow}-c_{B,1,\uparrow}^{\dag}c_{B,1,\downarrow}\right)-\left(c_{A,2,\uparrow}^{\dag}c_{A,2,\downarrow}-c_{B,2,\uparrow}^{\dag}c_{B,2,\downarrow}\right)\\
-\left(c_{A,1,\downarrow}^{\dag}c_{A,1,\uparrow}-c_{B,1,\downarrow}^{\dag}c_{B,1,\uparrow}\right)+\left(c_{A,2,\downarrow}^{\dag}c_{A,2,\uparrow}-c_{B,2,\downarrow}^{\dag}c_{B,2,\uparrow}\right)
\end{bmatrix}\\
\begin{bmatrix}\left(c_{A,1,\uparrow}^{\dag}c_{A,1,\downarrow}-c_{B,1,\uparrow}^{\dag}c_{B,1,\downarrow}\right)-\left(c_{A,2,\uparrow}^{\dag}c_{A,2,\downarrow}-c_{B,2,\uparrow}^{\dag}c_{B,2,\downarrow}\right)\\
+\left(c_{A,1,\downarrow}^{\dag}c_{A,1,\uparrow}-c_{B,1,\downarrow}^{\dag}c_{B,1,\uparrow}\right)-\left(c_{A,2,\downarrow}^{\dag}c_{A,2,\uparrow}-c_{B,2,\downarrow}^{\dag}c_{B,2,\uparrow}\right)
\end{bmatrix}\\
\begin{bmatrix}\left(c_{A,1,\uparrow}^{\dag}c_{A,1,\uparrow}-c_{A,1,\downarrow}^{\dag}c_{A,1,\downarrow}\right)-\left(c_{B,1,\uparrow}^{\dag}c_{B,1,\uparrow}-c_{B,1,\downarrow}^{\dag}c_{B,1,\downarrow}\right)\\
-\left[\left(c_{A,2,\uparrow}^{\dag}c_{A,2,\uparrow}-c_{A,2,\downarrow}^{\dag}c_{A,2,\downarrow}\right)+\left(c_{B,2,\uparrow}^{\dag}c_{B,2,\uparrow}-c_{B,2,\downarrow}^{\dag}c_{B,2,\downarrow}\right)\right]
\end{bmatrix}
\end{pmatrix},
\end{equation}
\begin{equation}\label{eq:A-5}
    \tilde{A}_{5}=\begin{pmatrix}i\begin{bmatrix}\left(c_{A,1,\uparrow}^{\dag}c_{A,1,\downarrow}-c_{B,1,\uparrow}^{\dag}c_{B,1,\downarrow}\right)+\left(c_{A,2,\uparrow}^{\dag}c_{A,2,\downarrow}-c_{B,2,\uparrow}^{\dag}c_{B,2,\downarrow}\right)\\
-\left(c_{A,1,\downarrow}^{\dag}c_{A,1,\uparrow}-c_{B,1,\downarrow}^{\dag}c_{B,1,\uparrow}\right)-\left(c_{A,2,\downarrow}^{\dag}c_{A,2,\uparrow}-c_{B,2,\downarrow}^{\dag}c_{B,2,\uparrow}\right)
\end{bmatrix}\\
\begin{bmatrix}\left(c_{A,1,\uparrow}^{\dag}c_{A,1,\downarrow}-c_{B,1,\uparrow}^{\dag}c_{B,1,\downarrow}\right)+\left(c_{A,2,\uparrow}^{\dag}c_{A,2,\downarrow}-c_{B,2,\uparrow}^{\dag}c_{B,2,\downarrow}\right)\\
+\left(c_{A,1,\downarrow}^{\dag}c_{A,1,\uparrow}-c_{B,1,\downarrow}^{\dag}c_{B,1,\uparrow}\right)+\left(c_{A,2,\downarrow}^{\dag}c_{A,2,\uparrow}-c_{B,2,\downarrow}^{\dag}c_{B,2,\uparrow}\right)
\end{bmatrix}\\
\begin{bmatrix}\left(c_{A,1,\uparrow}^{\dag}c_{A,1,\uparrow}-c_{A,1,\downarrow}^{\dag}c_{A,1,\downarrow}\right)-\left(c_{B,1,\uparrow}^{\dag}c_{B,1,\uparrow}-c_{B,1,\downarrow}^{\dag}c_{B,1,\downarrow}\right)\\
+\left(c_{A,2,\uparrow}^{\dag}c_{A,2,\uparrow}-c_{A,2,\downarrow}^{\dag}c_{A,2,\downarrow}\right)-\left(c_{B,2,\uparrow}^{\dag}c_{B,2,\uparrow}-c_{B,2,\downarrow}^{\dag}c_{B,2,\downarrow}\right)
\end{bmatrix}
\end{pmatrix},
\end{equation}
\begin{equation}
    \tilde{A}_{6}=\begin{pmatrix}i\begin{bmatrix}\left(c_{A,1,\uparrow}^{\dag}c_{A,1,\downarrow}+c_{B,1,\uparrow}^{\dag}c_{B,1,\downarrow}\right)-\left(c_{A,2,\uparrow}^{\dag}c_{A,2,\downarrow}+c_{B,2,\uparrow}^{\dag}c_{B,2,\downarrow}\right)\\
-\left(c_{A,1,\downarrow}^{\dag}c_{A,1,\uparrow}+c_{B,1,\downarrow}^{\dag}c_{B,1,\uparrow}\right)+\left(c_{A,2,\downarrow}^{\dag}c_{A,2,\uparrow}+c_{B,2,\downarrow}^{\dag}c_{B,2,\uparrow}\right)
\end{bmatrix}\\
\begin{bmatrix}\left(c_{A,1,\uparrow}^{\dag}c_{A,1,\downarrow}+c_{B,1,\uparrow}^{\dag}c_{B,1,\downarrow}\right)-\left(c_{A,2,\uparrow}^{\dag}c_{A,2,\downarrow}+c_{B,2,\uparrow}^{\dag}c_{B,2,\downarrow}\right)\\
+\left(c_{A,1,\downarrow}^{\dag}c_{A,1,\uparrow}+c_{B,1,\downarrow}^{\dag}c_{B,1,\uparrow}\right)-\left(c_{A,2,\downarrow}^{\dag}c_{A,2,\uparrow}+c_{B,2,\downarrow}^{\dag}c_{B,2,\uparrow}\right)
\end{bmatrix}\\
\begin{bmatrix}\left(c_{A,1,\uparrow}^{\dag}c_{A,1,\uparrow}-c_{A,1,\downarrow}^{\dag}c_{A,1,\downarrow}\right)+\left(c_{B,1,\uparrow}^{\dag}c_{B,1,\uparrow}-c_{B,1,\downarrow}^{\dag}c_{B,1,\downarrow}\right)\\
-\left(c_{A,2,\uparrow}^{\dag}c_{A,2,\uparrow}-c_{A,2,\downarrow}^{\dag}c_{A,2,\downarrow}\right)-\left(c_{B,2,\uparrow}^{\dag}c_{B,2,\uparrow}-c_{B,2,\downarrow}^{\dag}c_{B,2,\downarrow}\right)
\end{bmatrix}
\end{pmatrix},
\end{equation}
\begin{equation}\label{eq:A-7}
    \tilde{A}_{7}=\begin{pmatrix}i\begin{bmatrix}\left(c_{A,1,\uparrow}^{\dag}c_{A,1,\downarrow}+c_{B,1,\uparrow}^{\dag}c_{B,1,\downarrow}\right)+\left(c_{A,2,\uparrow}^{\dag}c_{A,2,\downarrow}+c_{B,2,\uparrow}^{\dag}c_{B,2,\downarrow}\right)\\
-\left(c_{A,1,\downarrow}^{\dag}c_{A,1,\uparrow}+c_{B,1,\downarrow}^{\dag}c_{B,1,\uparrow}\right)-\left(c_{A,2,\downarrow}^{\dag}c_{A,2,\uparrow}+c_{B,2,\downarrow}^{\dag}c_{B,2,\uparrow}\right)
\end{bmatrix}\\
\begin{bmatrix}\left(c_{A,1,\uparrow}^{\dag}c_{A,1,\downarrow}+c_{B,1,\uparrow}^{\dag}c_{B,1,\downarrow}\right)+\left(c_{A,2,\uparrow}^{\dag}c_{A,2,\downarrow}+c_{B,2,\uparrow}^{\dag}c_{B,2,\downarrow}\right)\\
+\left(c_{A,1,\downarrow}^{\dag}c_{A,1,\uparrow}+c_{B,1,\downarrow}^{\dag}c_{B,1,\uparrow}\right)+\left(c_{A,2,\downarrow}^{\dag}c_{A,2,\uparrow}+c_{B,2,\downarrow}^{\dag}c_{B,2,\uparrow}\right)
\end{bmatrix}\\
\begin{bmatrix}\left(c_{A,1,\uparrow}^{\dag}c_{A,1,\uparrow}-c_{A,1,\downarrow}^{\dag}c_{A,1,\downarrow}\right)+\left(c_{B,1,\uparrow}^{\dag}c_{B,1,\uparrow}-c_{B,1,\downarrow}^{\dag}c_{B,1,\downarrow}\right)\\
+\left(c_{A,2,\uparrow}^{\dag}c_{A,2,\uparrow}-c_{A,2,\downarrow}^{\dag}c_{A,2,\downarrow}\right)+\left(c_{B,2,\uparrow}^{\dag}c_{B,2,\uparrow}-c_{B,2,\downarrow}^{\dag}c_{B,2,\downarrow}\right)
\end{bmatrix}
\end{pmatrix},
\end{equation}
\begin{equation}
    \tilde{A}_{8}=\begin{pmatrix}i\begin{bmatrix}\left(c_{A,1,\uparrow}^{\dag}c_{B,1,\downarrow}-c_{B,1,\uparrow}^{\dag}c_{A,1,\downarrow}\right)-\left(c_{A,2,\uparrow}^{\dag}c_{B,2,\downarrow}-c_{B,2,\uparrow}^{\dag}c_{A,2,\downarrow}\right)\\
+\left(c_{A,1,\downarrow}^{\dag}c_{B,1,\uparrow}-c_{B,1,\downarrow}^{\dag}c_{A,1,\uparrow}\right)-\left(c_{A,2,\downarrow}^{\dag}c_{B,2,\uparrow}-c_{B,2,\downarrow}^{\dag}c_{A,2,\uparrow}\right)
\end{bmatrix}\\
\begin{bmatrix}\left(c_{A,1,\uparrow}^{\dag}c_{B,1,\downarrow}-c_{B,1,\uparrow}^{\dag}c_{A,1,\downarrow}\right)-\left(c_{A,2,\uparrow}^{\dag}c_{B,2,\downarrow}-c_{B,2,\uparrow}^{\dag}c_{A,2,\downarrow}\right)\\
-\left(c_{A,1,\downarrow}^{\dag}c_{B,1,\uparrow}-c_{B,1,\downarrow}^{\dag}c_{A,1,\uparrow}\right)+\left(c_{A,2,\downarrow}^{\dag}c_{B,2,\uparrow}-c_{B,2,\downarrow}^{\dag}c_{A,2,\uparrow}\right)
\end{bmatrix}\\
i\begin{bmatrix}\left(c_{A,1,\uparrow}^{\dag}c_{B,1,\uparrow}-c_{A,1,\downarrow}^{\dag}c_{B,1,\downarrow}-c_{B,1,\uparrow}^{\dag}c_{A,1,\uparrow}+c_{B,1,\downarrow}^{\dag}c_{A,1,\downarrow}\right)\\
-\left(c_{A,2,\uparrow}^{\dag}c_{B,2,\uparrow}-c_{A,2,\downarrow}^{\dag}c_{B,2,\downarrow}-c_{B,2,\uparrow}^{\dag}c_{A,2,\uparrow}+c_{B,2,\downarrow}^{\dag}c_{A,2,\downarrow}\right)
\end{bmatrix}
\end{pmatrix},
\end{equation}
\begin{equation}
    \tilde{A}_{9}=\begin{pmatrix}i\begin{bmatrix}\left(c_{A,1,\uparrow}^{\dag}c_{B,1,\downarrow}-c_{B,1,\uparrow}^{\dag}c_{A,1,\downarrow}\right)+\left(c_{A,2,\uparrow}^{\dag}c_{B,2,\downarrow}-c_{B,2,\uparrow}^{\dag}c_{A,2,\downarrow}\right)\\
+\left(c_{A,1,\downarrow}^{\dag}c_{B,1,\uparrow}-c_{B,1,\downarrow}^{\dag}c_{A,1,\uparrow}\right)+\left(c_{A,2,\downarrow}^{\dag}c_{B,2,\uparrow}-c_{B,2,\downarrow}^{\dag}c_{A,2,\uparrow}\right)
\end{bmatrix}\\
\begin{bmatrix}\left(c_{A,1,\uparrow}^{\dag}c_{B,1,\downarrow}-c_{B,1,\uparrow}^{\dag}c_{A,1,\downarrow}\right)+\left(c_{A,2,\uparrow}^{\dag}c_{B,2,\downarrow}-c_{B,2,\uparrow}^{\dag}c_{A,2,\downarrow}\right)\\
-\left(c_{A,1,\downarrow}^{\dag}c_{B,1,\uparrow}-c_{B,1,\downarrow}^{\dag}c_{A,1,\uparrow}\right)-\left(c_{A,2,\downarrow}^{\dag}c_{B,2,\uparrow}-c_{B,2,\downarrow}^{\dag}c_{A,2,\uparrow}\right)
\end{bmatrix}\\
i\begin{bmatrix}\left(c_{A,1,\uparrow}^{\dag}c_{B,1,\uparrow}-c_{A,1,\downarrow}^{\dag}c_{B,1,\downarrow}-c_{B,1,\uparrow}^{\dag}c_{A,1,\uparrow}+c_{B,1,\downarrow}^{\dag}c_{A,1,\downarrow}\right)\\
+\left(c_{A,2,\uparrow}^{\dag}c_{B,2,\uparrow}-c_{A,2,\downarrow}^{\dag}c_{B,2,\downarrow}-c_{B,2,\uparrow}^{\dag}c_{A,2,\uparrow}+c_{B,2,\downarrow}^{\dag}c_{A,2,\downarrow}\right)
\end{bmatrix}
\end{pmatrix},
\end{equation}
\begin{equation}\label{eq:A-10}
    \tilde{A}_{10}=\begin{pmatrix}i\begin{bmatrix}\left(c_{A,1,\uparrow}^{\dag}c_{A,2,\uparrow}+c_{A,1,\downarrow}^{\dag}c_{A,2,\downarrow}\right)-\left(c_{B,1,\uparrow}^{\dag}c_{B,2,\uparrow}+c_{B,1,\downarrow}^{\dag}c_{B,2,\downarrow}\right)\\
-\left(c_{A,2,\uparrow}^{\dag}c_{A,1,\uparrow}+c_{A,2,\downarrow}^{\dag}c_{A,1,\downarrow}\right)+\left(c_{B,2,\uparrow}^{\dag}c_{B,1,\uparrow}+c_{B,2,\downarrow}^{\dag}c_{B,1,\downarrow}\right)
\end{bmatrix}\\
\begin{bmatrix}\left(c_{A,1,\uparrow}^{\dag}c_{A,2,\uparrow}+c_{A,1,\downarrow}^{\dag}c_{A,2,\downarrow}\right)-\left(c_{B,1,\uparrow}^{\dag}c_{B,2,\uparrow}+c_{B,1,\downarrow}^{\dag}c_{B,2,\downarrow}\right)\\
+\left(c_{A,2,\uparrow}^{\dag}c_{A,1,\uparrow}+c_{A,2,\downarrow}^{\dag}c_{A,1,\downarrow}\right)-\left(c_{B,2,\uparrow}^{\dag}c_{B,1,\uparrow}+c_{B,2,\downarrow}^{\dag}c_{B,1,\downarrow}\right)
\end{bmatrix}\\
i\begin{bmatrix}\left(c_{A,1,\uparrow}^{\dag}c_{A,2,\downarrow}^{\dagger}-c_{A,1,\downarrow}^{\dag}c_{A,2,\uparrow}^{\dag}\right)+\left(c_{B,1,\uparrow}^{\dag}c_{B,2,\downarrow}^{\dag}-c_{B,1,\downarrow}^{\dag}c_{B,2,\uparrow}^{\dag}\right)\\
+\left(c_{A,1,\uparrow}c_{A,2,\downarrow}-c_{A,1,\downarrow}c_{A,2,\uparrow}\right)+\left(c_{B,1,\uparrow}c_{B,2,\downarrow}-c_{B,1,\downarrow}c_{B,2,\uparrow}\right)
\end{bmatrix}\\
\begin{bmatrix}\left(c_{A,1,\uparrow}^{\dag}c_{A,2,\downarrow}^{\dagger}-c_{A,1,\downarrow}^{\dag}c_{A,2,\uparrow}^{\dag}\right)+\left(c_{B,1,\uparrow}^{\dag}c_{B,2,\downarrow}^{\dag}-c_{B,1,\downarrow}^{\dag}c_{B,2,\uparrow}^{\dag}\right)\\
-\left(c_{A,1,\uparrow}c_{A,2,\downarrow}-c_{A,1,\downarrow}c_{A,2,\uparrow}\right)-\left(c_{B,1,\uparrow}c_{B,2,\downarrow}-c_{B,1,\downarrow}c_{B,2,\uparrow}\right)
\end{bmatrix}
\end{pmatrix},
\end{equation}
\begin{equation}\label{eq:A-11}
    \tilde{A}_{11}=\begin{pmatrix}i\begin{bmatrix}\left(c_{A,1,\uparrow}^{\dag}c_{A,2,\uparrow}+c_{A,1,\downarrow}^{\dag}c_{A,2,\downarrow}\right)+\left(c_{B,1,\uparrow}^{\dag}c_{B,2,\uparrow}+c_{B,1,\downarrow}^{\dag}c_{B,2,\downarrow}\right)\\
-\left(c_{A,2,\uparrow}^{\dag}c_{A,1,\uparrow}+c_{A,2,\downarrow}^{\dag}c_{A,1,\downarrow}\right)-\left(c_{B,2,\uparrow}^{\dag}c_{B,1,\uparrow}+c_{B,2,\downarrow}^{\dag}c_{B,1,\downarrow}\right)
\end{bmatrix}\\
\begin{bmatrix}\left(c_{A,1,\uparrow}^{\dag}c_{A,2,\uparrow}+c_{A,1,\downarrow}^{\dag}c_{A,2,\downarrow}\right)+\left(c_{B,1,\uparrow}^{\dag}c_{B,2,\uparrow}+c_{B,1,\downarrow}^{\dag}c_{B,2,\downarrow}\right)\\
+\left(c_{A,2,\uparrow}^{\dag}c_{A,1,\uparrow}+c_{A,2,\downarrow}^{\dag}c_{A,1,\downarrow}\right)+\left(c_{B,2,\uparrow}^{\dag}c_{B,1,\uparrow}+c_{B,2,\downarrow}^{\dag}c_{B,1,\downarrow}\right)
\end{bmatrix}\\
i\begin{bmatrix}\left(c_{A,1,\uparrow}^{\dag}c_{A,2,\downarrow}^{\dagger}-c_{A,1,\downarrow}^{\dag}c_{A,2,\uparrow}^{\dag}\right)-\left(c_{B,1,\uparrow}^{\dag}c_{B,2,\downarrow}^{\dag}-c_{B,1,\downarrow}^{\dag}c_{B,2,\uparrow}^{\dag}\right)\\
+\left(c_{A,1,\uparrow}c_{A,2,\downarrow}-c_{A,1,\downarrow}c_{A,2,\uparrow}\right)-\left(c_{B,1,\uparrow}c_{B,2,\downarrow}-c_{B,1,\downarrow}c_{B,2,\uparrow}\right)
\end{bmatrix}\\
\begin{bmatrix}\left(c_{A,1,\uparrow}^{\dag}c_{A,2,\downarrow}^{\dagger}-c_{A,1,\downarrow}^{\dag}c_{A,2,\uparrow}^{\dag}\right)-\left(c_{B,1,\uparrow}^{\dag}c_{B,2,\downarrow}^{\dag}-c_{B,1,\downarrow}^{\dag}c_{B,2,\uparrow}^{\dag}\right)\\
-\left(c_{A,1,\uparrow}c_{A,2,\downarrow}-c_{A,1,\downarrow}c_{A,2,\uparrow}\right)+\left(c_{B,1,\uparrow}c_{B,2,\downarrow}-c_{B,1,\downarrow}c_{B,2,\uparrow}\right)
\end{bmatrix}
\end{pmatrix},
\end{equation}
\begin{equation}
    \tilde{A}_{12}=\begin{pmatrix}i\begin{bmatrix}\left(c_{A,1,\uparrow}^{\dag}c_{B,2,\uparrow}+c_{A,1,\downarrow}^{\dag}c_{B,2,\downarrow}\right)-\left(c_{B,1,\uparrow}^{\dag}c_{A,2,\uparrow}+c_{B,1,\downarrow}^{\dag}c_{A,2,\downarrow}\right)\\
-\left(c_{B,2,\uparrow}^{\dag}c_{A,1,\uparrow}+c_{B,2,\downarrow}^{\dag}c_{A,1,\downarrow}\right)+\left(c_{A,2,\uparrow}^{\dag}c_{B,1,\uparrow}+c_{A,2,\downarrow}^{\dag}c_{B,1,\downarrow}\right)
\end{bmatrix}\\
\begin{bmatrix}\left(c_{A,1,\uparrow}^{\dag}c_{B,2,\uparrow}+c_{A,1,\downarrow}^{\dag}c_{B,2,\downarrow}\right)-\left(c_{B,1,\uparrow}^{\dag}c_{A,2,\uparrow}+c_{B,1,\downarrow}^{\dag}c_{A,2,\downarrow}\right)\\
+\left(c_{B,2,\uparrow}^{\dag}c_{A,1,\uparrow}+c_{B,2,\downarrow}^{\dag}c_{A,1,\downarrow}\right)-\left(c_{A,2,\uparrow}^{\dag}c_{B,1,\uparrow}+c_{A,2,\downarrow}^{\dag}c_{B,1,\downarrow}\right)
\end{bmatrix}\\
i\begin{bmatrix}\left(c_{A,1,\uparrow}^{\dag}c_{B,2,\downarrow}^{\dag}-c_{A,1,\downarrow}^{\dag}c_{B,2,\uparrow}^{\dag}\right)+\left(c_{A,2,\uparrow}^{\dag}c_{B,1,\downarrow}^{\dag}-c_{A,2,\downarrow}^{\dag}c_{B,1,\uparrow}^{\dag}\right)\\
+\left(c_{A,1,\uparrow}c_{B,2,\downarrow}-c_{A,1,\downarrow}c_{B,2,\uparrow}\right)+\left(c_{A,2,\uparrow}c_{B,1,\downarrow}-c_{A,2,\downarrow}c_{B,1,\uparrow}\right)
\end{bmatrix}\\
\begin{bmatrix}\left(c_{A,1,\uparrow}^{\dag}c_{B,2,\downarrow}^{\dag}-c_{A,1,\downarrow}^{\dag}c_{B,2,\uparrow}^{\dag}\right)+\left(c_{A,2,\uparrow}^{\dag}c_{B,1,\downarrow}^{\dag}-c_{A,2,\downarrow}^{\dag}c_{B,1,\uparrow}^{\dag}\right)\\
-\left(c_{A,1,\uparrow}c_{B,2,\downarrow}-c_{A,1,\downarrow}c_{B,2,\uparrow}\right)-\left(c_{A,2,\uparrow}c_{B,1,\downarrow}-c_{A,2,\downarrow}c_{B,1,\uparrow}\right)
\end{bmatrix}
\end{pmatrix},
\end{equation}
\begin{equation}
    \tilde{A}_{13}=\begin{pmatrix}i\begin{bmatrix}\left(c_{A,1,\uparrow}^{\dag}c_{B,2,\uparrow}+c_{A,1,\downarrow}^{\dag}c_{B,2,\downarrow}\right)+\left(c_{B,1,\uparrow}^{\dag}c_{A,2,\uparrow}+c_{B,1,\downarrow}^{\dag}c_{A,2,\downarrow}\right)\\
-\left(c_{B,2,\uparrow}^{\dag}c_{A,1,\uparrow}+c_{B,2,\downarrow}^{\dag}c_{A,1,\downarrow}\right)-\left(c_{A,2,\uparrow}^{\dag}c_{B,1,\uparrow}+c_{A,2,\downarrow}^{\dag}c_{B,1,\downarrow}\right)
\end{bmatrix}\\
\begin{bmatrix}\left(c_{A,1,\uparrow}^{\dag}c_{B,2,\uparrow}+c_{A,1,\downarrow}^{\dag}c_{B,2,\downarrow}\right)+\left(c_{B,1,\uparrow}^{\dag}c_{A,2,\uparrow}+c_{B,1,\downarrow}^{\dag}c_{A,2,\downarrow}\right)\\
+\left(c_{B,2,\uparrow}^{\dag}c_{A,1,\uparrow}+c_{B,2,\downarrow}^{\dag}c_{A,1,\downarrow}\right)+\left(c_{A,2,\uparrow}^{\dag}c_{B,1,\uparrow}+c_{A,2,\downarrow}^{\dag}c_{B,1,\downarrow}\right)
\end{bmatrix}\\
i\begin{bmatrix}\left(c_{A,1,\uparrow}^{\dag}c_{B,2,\downarrow}^{\dag}-c_{A,1,\downarrow}^{\dag}c_{B,2,\uparrow}^{\dag}\right)-\left(c_{A,2,\uparrow}^{\dag}c_{B,1,\downarrow}^{\dag}-c_{A,2,\downarrow}^{\dag}c_{B,1,\uparrow}^{\dag}\right)\\
+\left(c_{A,1,\uparrow}c_{B,2,\downarrow}-c_{A,1,\downarrow}c_{B,2,\uparrow}\right)-\left(c_{A,2,\uparrow}c_{B,1,\downarrow}-c_{A,2,\downarrow}c_{B,1,\uparrow}\right)
\end{bmatrix}\\
\begin{bmatrix}\left(c_{A,1,\uparrow}^{\dag}c_{B,2,\downarrow}^{\dag}-c_{A,1,\downarrow}^{\dag}c_{B,2,\uparrow}^{\dag}\right)-\left(c_{A,2,\uparrow}^{\dag}c_{B,1,\downarrow}^{\dag}-c_{A,2,\downarrow}^{\dag}c_{B,1,\uparrow}^{\dag}\right)\\
-\left(c_{A,1,\uparrow}c_{B,2,\downarrow}-c_{A,1,\downarrow}c_{B,2,\uparrow}\right)+\left(c_{A,2,\uparrow}c_{B,1,\downarrow}-c_{A,2,\downarrow}c_{B,1,\uparrow}\right)
\end{bmatrix}
\end{pmatrix},
\end{equation}
\begin{equation}
    \tilde{A}_{14}=\begin{pmatrix}\frac{i}{2}\left(c_{A,1,\uparrow}^{\dag}c_{B,1,\downarrow}+c_{B,1,\uparrow}^{\dag}c_{A,1,\downarrow}-c_{A,1,\downarrow}^{\dag}c_{B,1,\uparrow}-c_{B,1,\downarrow}^{\dag}c_{A,1,\uparrow}\right)\\
\frac{1}{2}\left(c_{A,1,\uparrow}^{\dag}c_{B,1,\downarrow}+c_{B,1,\uparrow}^{\dag}c_{A,1,\downarrow}+c_{A,1,\downarrow}^{\dag}c_{B,1,\uparrow}+c_{B,1,\downarrow}^{\dag}c_{A,1,\uparrow}\right)\\
\frac{1}{2}\left(c_{A,1,\uparrow}^{\dag}c_{B,1,\uparrow}-c_{A,1,\downarrow}^{\dag}c_{B,1,\downarrow}+c_{B,1,\uparrow}^{\dag}c_{A,1,\uparrow}-c_{B,1,\downarrow}^{\dag}c_{A,1,\downarrow}\right)\\
i\left(c_{A,1,\uparrow}^{\dag}c_{B,1,\uparrow}^{\dag}+c_{A,1,\uparrow}c_{B,1,\uparrow}\right)\\
i\left(c_{A,1,\downarrow}^{\dag}c_{B,1,\downarrow}^{\dag}+c_{A,1,\downarrow}c_{B,1,\downarrow}\right)\\
c_{A,1,\uparrow}^{\dag}c_{B,1,\uparrow}^{\dag}-c_{A,1,\uparrow}c_{B,1,\uparrow}\\
c_{A,1,\downarrow}^{\dag}c_{B,1,\downarrow}^{\dag}-c_{A,1,\downarrow}c_{B,1,\downarrow}\\
\frac{i}{2}\left(c_{A,1,\uparrow}^{\dag}c_{B,1,\downarrow}^{\dag}+c_{A,1,\downarrow}^{\dag}c_{B,1,\uparrow}^{\dag}+c_{A,1,\uparrow}c_{B,1,\downarrow}+c_{A,1,\downarrow}c_{B,1,\uparrow}\right)\\
\frac{1}{2}\left(c_{A,1,\uparrow}^{\dag}c_{B,1,\downarrow}^{\dag}+c_{A,1,\downarrow}^{\dag}c_{B,1,\uparrow}^{\dag}-c_{A,1,\uparrow}c_{B,1,\downarrow}-c_{A,1,\downarrow}c_{B,1,\uparrow}\right)\\
\frac{i}{2}\left(c_{A,2,\uparrow}^{\dag}c_{B,2,\downarrow}+c_{B,2,\uparrow}^{\dag}c_{A,2,\downarrow}-c_{A,2,\downarrow}^{\dag}c_{B,2,\uparrow}-c_{B,2,\downarrow}^{\dag}c_{A,2,\uparrow}\right)\\
\frac{1}{2}\left(c_{A,2,\uparrow}^{\dag}c_{B,2,\downarrow}+c_{B,2,\uparrow}^{\dag}c_{A,2,\downarrow}+c_{A,2,\downarrow}^{\dag}c_{B,2,\uparrow}+c_{B,2,\downarrow}^{\dag}c_{A,2,\uparrow}\right)\\
\frac{1}{2}\left(c_{A,2,\uparrow}^{\dag}c_{B,2,\uparrow}-c_{A,2,\downarrow}^{\dag}c_{B,2,\downarrow}+c_{B,2,\uparrow}^{\dag}c_{A,2,\uparrow}-c_{B,2,\downarrow}^{\dag}c_{A,2,\downarrow}\right)\\
i\left(c_{A,2,\uparrow}^{\dag}c_{B,2,\uparrow}^{\dag}+c_{A,2,\uparrow}c_{B,2,\uparrow}\right)\\
i\left(c_{A,2,\downarrow}^{\dag}c_{B,2,\downarrow}^{\dag}+c_{A,2,\downarrow}c_{B,2,\downarrow}\right)\\
c_{A,2,\uparrow}^{\dag}c_{B,2,\uparrow}^{\dag}-c_{A,2,\uparrow}c_{B,2,\uparrow}\\
c_{A,2,\downarrow}^{\dag}c_{B,2,\downarrow}^{\dag}-c_{A,2,\downarrow}c_{B,2,\downarrow}\\
\frac{i}{2}\left(c_{A,2,\uparrow}^{\dag}c_{B,2,\downarrow}^{\dag}+c_{A,2,\downarrow}^{\dag}c_{B,2,\uparrow}^{\dag}+c_{A,2,\uparrow}c_{B,2,\downarrow}+c_{A,2,\downarrow}c_{B,2,\uparrow}\right)\\
\frac{1}{2}\left(c_{A,2,\uparrow}^{\dag}c_{B,2,\downarrow}^{\dag}+c_{A,2,\downarrow}^{\dag}c_{B,2,\uparrow}^{\dag}-c_{A,2,\uparrow}c_{B,2,\downarrow}-c_{A,2,\downarrow}c_{B,2,\uparrow}\right)
\end{pmatrix},
\end{equation}
\begin{equation}
    \tilde{A}_{15}=\begin{pmatrix}i\left(c_{A,1,\uparrow}^{\dag}c_{A,2,\downarrow}-c_{B,1,\uparrow}^{\dag}c_{B,2,\downarrow}-c_{A,2,\downarrow}^{\dag}c_{A,1,\uparrow}+c_{B,2,\downarrow}^{\dag}c_{B,1,\uparrow}\right)\\
i\left(c_{A,1,\downarrow}^{\dag}c_{A,2,\uparrow}-c_{B,1,\downarrow}^{\dag}c_{B,2,\uparrow}-c_{A,2,\uparrow}^{\dag}c_{A,1,\downarrow}+c_{B,2,\uparrow}^{\dag}c_{B,1,\downarrow}\right)\\
c_{A,1,\downarrow}^{\dag}c_{A,2,\uparrow}-c_{B,1,\downarrow}^{\dag}c_{B,2,\uparrow}+c_{A,2,\uparrow}^{\dag}c_{A,1,\downarrow}-c_{B,2,\uparrow}^{\dag}c_{B,1,\downarrow}\\
c_{A,1,\uparrow}^{\dag}c_{A,2,\downarrow}-c_{B,1,\uparrow}^{\dag}c_{B,2,\downarrow}+c_{A,2,\downarrow}^{\dag}c_{A,1,\uparrow}-c_{B,2,\downarrow}^{\dag}c_{B,1,\uparrow}\\
\frac{i}{2}\begin{bmatrix}\left(c_{A,1,\uparrow}^{\dag}c_{A,2,\uparrow}-c_{A,1,\downarrow}^{\dag}c_{A,2,\downarrow}\right)-\left(c_{B,1,\uparrow}^{\dag}c_{B,2,\uparrow}-c_{B,1,\downarrow}^{\dag}c_{B,2,\downarrow}\right)\\
-\left(c_{A,2,\uparrow}^{\dag}c_{A,1,\uparrow}-c_{A,2,\downarrow}^{\dag}c_{A,1,\downarrow}\right)+\left(c_{B,2,\uparrow}^{\dag}c_{B,1,\uparrow}-c_{B,2,\downarrow}^{\dag}c_{B,1,\downarrow}\right)
\end{bmatrix}\\
\frac{1}{2}\begin{bmatrix}\left(c_{A,1,\uparrow}^{\dag}c_{A,2,\uparrow}-c_{A,1,\downarrow}^{\dag}c_{A,2,\downarrow}\right)-\left(c_{B,1,\uparrow}^{\dag}c_{B,2,\uparrow}-c_{B,1,\downarrow}^{\dag}c_{B,2,\downarrow}\right)\\
+\left(c_{A,2,\uparrow}^{\dag}c_{A,1,\uparrow}-c_{A,2,\downarrow}^{\dag}c_{A,1,\downarrow}\right)-\left(c_{B,2,\uparrow}^{\dag}c_{B,1,\uparrow}-c_{B,2,\downarrow}^{\dag}c_{B,1,\downarrow}\right)
\end{bmatrix}\\
i\left[\left(c_{A,1,\uparrow}^{\dag}c_{A,2,\uparrow}^{\dag}+c_{B,1,\uparrow}^{\dag}c_{B,2,\uparrow}^{\dag}\right)+\left(c_{A,1,\uparrow}c_{A,2,\uparrow}+c_{B,1,\uparrow}c_{B,2,\uparrow}\right)\right]\\
i\left[\left(c_{A,1,\downarrow}^{\dag}c_{A,2,\downarrow}^{\dag}+c_{B,1,\downarrow}^{\dag}c_{B,2,\downarrow}^{\dag}\right)+\left(c_{A,1,\downarrow}c_{A,2,\downarrow}+c_{B,1,\downarrow}c_{B,2,\downarrow}\right)\right]\\
\left(c_{A,1,\uparrow}^{\dag}c_{A,2,\uparrow}^{\dag}+c_{B,1,\uparrow}^{\dag}c_{B,2,\uparrow}^{\dag}\right)-\left(c_{A,1,\uparrow}c_{A,2,\uparrow}+c_{B,1,\uparrow}c_{B,2,\uparrow}\right)\\
\left(c_{A,1,\downarrow}^{\dag}c_{A,2,\downarrow}^{\dag}+c_{B,1,\downarrow}^{\dag}c_{B,2,\downarrow}^{\dag}\right)-\left(c_{A,1,\downarrow}c_{A,2,\downarrow}+c_{B,1,\downarrow}c_{B,2,\downarrow}\right)\\
\frac{i}{2}\begin{bmatrix}\left(c_{A,1,\uparrow}^{\dag}c_{A,2,\downarrow}^{\dag}+c_{A,1,\downarrow}^{\dag}c_{A,2,\uparrow}^{\dag}\right)+\left(c_{B,1,\uparrow}^{\dag}c_{B,2,\downarrow}^{\dag}+c_{B,1,\downarrow}^{\dag}c_{B,2,\uparrow}^{\dag}\right)\\
+\left(c_{A,1,\uparrow}c_{A,2,\downarrow}+c_{A,1,\downarrow}c_{A,2,\uparrow}\right)+\left(c_{B,1,\uparrow}c_{B,2,\downarrow}+c_{B,1,\downarrow}c_{B,2,\uparrow}\right)
\end{bmatrix}\\
\frac{1}{2}\begin{bmatrix}\left(c_{A,1,\uparrow}^{\dag}c_{A,2,\downarrow}^{\dag}+c_{A,1,\downarrow}^{\dag}c_{A,2,\uparrow}^{\dag}\right)+\left(c_{B,1,\uparrow}^{\dag}c_{B,2,\downarrow}^{\dag}+c_{B,1,\downarrow}^{\dag}c_{B,2,\uparrow}^{\dag}\right)\\
-\left(c_{A,1,\uparrow}c_{A,2,\downarrow}+c_{A,1,\downarrow}c_{A,2,\uparrow}\right)-\left(c_{B,1,\uparrow}c_{B,2,\downarrow}+c_{B,1,\downarrow}c_{B,2,\uparrow}\right)
\end{bmatrix}
\end{pmatrix},
\end{equation}
\begin{equation}
    \tilde{A}_{16}=\begin{pmatrix}i\left(c_{A,1,\uparrow}^{\dag}c_{A,2,\downarrow}+c_{B,1,\uparrow}^{\dag}c_{B,2,\downarrow}-c_{A,2,\downarrow}^{\dag}c_{A,1,\uparrow}-c_{B,2,\downarrow}^{\dag}c_{B,1,\uparrow}\right)\\
i\left(c_{A,1,\downarrow}^{\dag}c_{A,2,\uparrow}+c_{B,1,\downarrow}^{\dag}c_{B,2,\uparrow}-c_{A,2,\uparrow}^{\dag}c_{A,1,\downarrow}-c_{B,2,\uparrow}^{\dag}c_{B,1,\downarrow}\right)\\
c_{A,1,\downarrow}^{\dag}c_{A,2,\uparrow}+c_{B,1,\downarrow}^{\dag}c_{B,2,\uparrow}+c_{A,2,\uparrow}^{\dag}c_{A,1,\downarrow}+c_{B,2,\uparrow}^{\dag}c_{B,1,\downarrow}\\
c_{A,1,\uparrow}^{\dag}c_{A,2,\downarrow}+c_{B,1,\uparrow}^{\dag}c_{B,2,\downarrow}+c_{A,2,\downarrow}^{\dag}c_{A,1,\uparrow}+c_{B,2,\downarrow}^{\dag}c_{B,1,\uparrow}\\
\frac{i}{2}\begin{bmatrix}\left(c_{A,1,\uparrow}^{\dag}c_{A,2,\uparrow}-c_{A,1,\downarrow}^{\dag}c_{A,2,\downarrow}\right)+\left(c_{B,1,\uparrow}^{\dag}c_{B,2,\uparrow}-c_{B,1,\downarrow}^{\dag}c_{B,2,\downarrow}\right)\\
-\left(c_{A,2,\uparrow}^{\dag}c_{A,1,\uparrow}-c_{A,2,\downarrow}^{\dag}c_{A,1,\downarrow}\right)-\left(c_{B,2,\uparrow}^{\dag}c_{B,1,\uparrow}-c_{B,2,\downarrow}^{\dag}c_{B,1,\downarrow}\right)
\end{bmatrix}\\
\frac{1}{2}\begin{bmatrix}\left(c_{A,1,\uparrow}^{\dag}c_{A,2,\uparrow}-c_{A,1,\downarrow}^{\dag}c_{A,2,\downarrow}\right)+\left(c_{B,1,\uparrow}^{\dag}c_{B,2,\uparrow}-c_{B,1,\downarrow}^{\dag}c_{B,2,\downarrow}\right)\\
+\left(c_{A,2,\uparrow}^{\dag}c_{A,1,\uparrow}-c_{A,2,\downarrow}^{\dag}c_{A,1,\downarrow}\right)+\left(c_{B,2,\uparrow}^{\dag}c_{B,1,\uparrow}-c_{B,2,\downarrow}^{\dag}c_{B,1,\downarrow}\right)
\end{bmatrix}\\
i\left[\left(c_{A,1,\uparrow}^{\dag}c_{A,2,\uparrow}^{\dag}-c_{B,1,\uparrow}^{\dag}c_{B,2,\uparrow}^{\dag}\right)+\left(c_{A,1,\uparrow}c_{A,2,\uparrow}-c_{B,1,\uparrow}c_{B,2,\uparrow}\right)\right]\\
i\left[\left(c_{A,1,\downarrow}^{\dag}c_{A,2,\downarrow}^{\dag}-c_{B,1,\downarrow}^{\dag}c_{B,2,\downarrow}^{\dag}\right)+\left(c_{A,1,\downarrow}c_{A,2,\downarrow}-c_{B,1,\downarrow}c_{B,2,\downarrow}\right)\right]\\
\left(c_{A,1,\uparrow}^{\dag}c_{A,2,\uparrow}^{\dag}-c_{B,1,\uparrow}^{\dag}c_{B,2,\uparrow}^{\dag}\right)-\left(c_{A,1,\uparrow}c_{A,2,\uparrow}-c_{B,1,\uparrow}c_{B,2,\uparrow}\right)\\
\left(c_{A,1,\downarrow}^{\dag}c_{A,2,\downarrow}^{\dag}-c_{B,1,\downarrow}^{\dag}c_{B,2,\downarrow}^{\dag}\right)-\left(c_{A,1,\downarrow}c_{A,2,\downarrow}-c_{B,1,\downarrow}c_{B,2,\downarrow}\right)\\
\frac{i}{2}\begin{bmatrix}\left(c_{A,1,\uparrow}^{\dag}c_{A,2,\downarrow}^{\dag}+c_{A,1,\downarrow}^{\dag}c_{A,2,\uparrow}^{\dag}\right)-\left(c_{B,1,\uparrow}^{\dag}c_{B,2,\downarrow}^{\dag}+c_{B,1,\downarrow}^{\dag}c_{B,2,\uparrow}^{\dag}\right)\\
+\left(c_{A,1,\uparrow}c_{A,2,\downarrow}+c_{A,1,\downarrow}c_{A,2,\uparrow}\right)-\left(c_{B,1,\uparrow}c_{B,2,\downarrow}+c_{B,1,\downarrow}c_{B,2,\uparrow}\right)
\end{bmatrix}\\
\frac{1}{2}\begin{bmatrix}\left(c_{A,1,\uparrow}^{\dag}c_{A,2,\downarrow}^{\dag}+c_{A,1,\downarrow}^{\dag}c_{A,2,\uparrow}^{\dag}\right)-\left(c_{B,1,\uparrow}^{\dag}c_{B,2,\downarrow}^{\dag}+c_{B,1,\downarrow}^{\dag}c_{B,2,\uparrow}^{\dag}\right)\\
-\left(c_{A,1,\uparrow}c_{A,2,\downarrow}+c_{A,1,\downarrow}c_{A,2,\uparrow}\right)+\left(c_{B,1,\uparrow}c_{B,2,\downarrow}+c_{B,1,\downarrow}c_{B,2,\uparrow}\right)
\end{bmatrix}
\end{pmatrix},
\end{equation}
\begin{equation}
    \tilde{A}_{17}=\begin{pmatrix}i\left[\left(c_{A,1,\uparrow}^{\dag}c_{B,2,\downarrow}-c_{B,1,\uparrow}^{\dag}c_{A,2,\downarrow}\right)-\left(c_{B,2,\downarrow}^{\dag}c_{A,1,\uparrow}-c_{A,2,\downarrow}^{\dag}c_{B,1,\uparrow}\right)\right]\\
i\left[\left(c_{A,1,\downarrow}^{\dag}c_{B,2,\uparrow}-c_{B,1,\downarrow}^{\dag}c_{A,2,\uparrow}\right)-\left(c_{B,2,\uparrow}^{\dag}c_{A,1,\downarrow}-c_{A,2,\uparrow}^{\dag}c_{B,1,\downarrow}\right)\right]\\
\left(c_{A,1,\downarrow}^{\dag}c_{B,2,\uparrow}-c_{B,1,\downarrow}^{\dag}c_{A,2,\uparrow}\right)+\left(c_{B,2,\uparrow}^{\dag}c_{A,1,\downarrow}-c_{A,2,\uparrow}^{\dag}c_{B,1,\downarrow}\right)\\
\left(c_{A,1,\uparrow}^{\dag}c_{B,2,\downarrow}-c_{B,1,\uparrow}^{\dag}c_{A,2,\downarrow}\right)+\left(c_{B,2,\downarrow}^{\dag}c_{A,1,\uparrow}-c_{A,2,\downarrow}^{\dag}c_{B,1,\uparrow}\right)\\
\frac{1}{2}\begin{bmatrix}\left(c_{A,1,\uparrow}^{\dag}c_{B,2,\uparrow}-c_{A,1,\downarrow}^{\dag}c_{B,2,\downarrow}\right)-\left(c_{B,1,\uparrow}^{\dag}c_{A,2,\uparrow}-c_{B,1,\downarrow}^{\dag}c_{A,2,\downarrow}\right)\\
+\left(c_{B,2,\uparrow}^{\dag}c_{A,1,\uparrow}-c_{B,2,\downarrow}^{\dag}c_{A,1,\downarrow}\right)-\left(c_{A,2,\uparrow}^{\dag}c_{B,1,\uparrow}-c_{A,2,\downarrow}^{\dag}c_{B,1,\downarrow}\right)
\end{bmatrix}\\
\frac{i}{2}\begin{bmatrix}\left(c_{A,1,\uparrow}^{\dag}c_{B,2,\uparrow}-c_{A,1,\downarrow}^{\dag}c_{B,2,\downarrow}\right)-\left(c_{B,1,\uparrow}^{\dag}c_{A,2,\uparrow}-c_{B,1,\downarrow}^{\dag}c_{A,2,\downarrow}\right)\\
-\left(c_{B,2,\uparrow}^{\dag}c_{A,1,\uparrow}-c_{B,2,\downarrow}^{\dag}c_{A,1,\downarrow}\right)+\left(c_{A,2,\uparrow}^{\dag}c_{B,1,\uparrow}-c_{A,2,\downarrow}^{\dag}c_{B,1,\downarrow}\right)
\end{bmatrix}\\
i\left[\left(c_{A,1,\uparrow}^{\dag}c_{B,2,\uparrow}^{\dag}-c_{A,2,\uparrow}^{\dag}c_{B,1,\uparrow}^{\dag}\right)+\left(c_{A,1,\uparrow}c_{B,2,\uparrow}-c_{A,2,\uparrow}c_{B,1,\uparrow}\right)\right]\\
i\left[\left(c_{A,1,\downarrow}^{\dag}c_{B,2,\downarrow}^{\dag}-c_{A,2,\downarrow}^{\dag}c_{B,1,\downarrow}^{\dag}\right)+\left(c_{A,1,\downarrow}c_{B,2,\downarrow}-c_{A,2,\downarrow}c_{B,1,\downarrow}\right)\right]\\
\left(c_{A,1,\uparrow}^{\dag}c_{B,2,\uparrow}^{\dag}-c_{A,2,\uparrow}^{\dag}c_{B,1,\uparrow}^{\dag}\right)-\left(c_{A,1,\uparrow}c_{B,2,\uparrow}-c_{A,2,\uparrow}c_{B,1,\uparrow}\right)\\
\left(c_{A,1,\downarrow}^{\dag}c_{B,2,\downarrow}^{\dag}-c_{A,2,\downarrow}^{\dag}c_{B,1,\downarrow}^{\dag}\right)-\left(c_{A,1,\downarrow}c_{B,2,\downarrow}-c_{A,2,\downarrow}c_{B,1,\downarrow}\right)\\
\frac{i}{2}\begin{bmatrix}\left(c_{A,1,\uparrow}^{\dag}c_{B,2,\downarrow}^{\dag}+c_{A,1,\downarrow}^{\dag}c_{B,2,\uparrow}^{\dag}\right)-\left(c_{A,2,\uparrow}^{\dag}c_{B,1,\downarrow}^{\dag}+c_{A,2,\downarrow}^{\dag}c_{B,1,\uparrow}^{\dag}\right)\\
+\left(c_{A,1,\uparrow}c_{B,2,\downarrow}+c_{A,1,\downarrow}c_{B,2,\uparrow}\right)-\left(c_{A,2,\uparrow}c_{B,1,\downarrow}+c_{A,2,\downarrow}c_{B,1,\uparrow}\right)
\end{bmatrix}\\
\frac{1}{2}\begin{bmatrix}\left(c_{A,1,\uparrow}^{\dag}c_{B,2,\downarrow}^{\dag}+c_{A,1,\downarrow}^{\dag}c_{B,2,\uparrow}^{\dag}\right)-\left(c_{A,2,\uparrow}^{\dag}c_{B,1,\downarrow}^{\dag}+c_{A,2,\downarrow}^{\dag}c_{B,1,\uparrow}^{\dag}\right)\\
-\left(c_{A,1,\uparrow}c_{B,2,\downarrow}+c_{A,1,\downarrow}c_{B,2,\uparrow}\right)+\left(c_{A,2,\uparrow}c_{B,1,\downarrow}+c_{A,2,\downarrow}c_{B,1,\uparrow}\right)
\end{bmatrix}
\end{pmatrix},
\end{equation}
\begin{equation}
    \tilde{A}_{18}=\begin{pmatrix}i\left[\left(c_{A,1,\uparrow}^{\dag}c_{B,2,\downarrow}+c_{B,1,\uparrow}^{\dag}c_{A,2,\downarrow}\right)-\left(c_{B,2,\downarrow}^{\dag}c_{A,1,\uparrow}+c_{A,2,\downarrow}^{\dag}c_{B,1,\uparrow}\right)\right]\\
i\left[\left(c_{A,1,\downarrow}^{\dag}c_{B,2,\uparrow}+c_{B,1,\downarrow}^{\dag}c_{A,2,\uparrow}\right)-\left(c_{B,2,\uparrow}^{\dag}c_{A,1,\downarrow}+c_{A,2,\uparrow}^{\dag}c_{B,1,\downarrow}\right)\right]\\
\left(c_{A,1,\downarrow}^{\dag}c_{B,2,\uparrow}+c_{B,1,\downarrow}^{\dag}c_{A,2,\uparrow}\right)+\left(c_{B,2,\uparrow}^{\dag}c_{A,1,\downarrow}+c_{A,2,\uparrow}^{\dag}c_{B,1,\downarrow}\right)\\
\left(c_{A,1,\uparrow}^{\dag}c_{B,2,\downarrow}+c_{B,1,\uparrow}^{\dag}c_{A,2,\downarrow}\right)+\left(c_{B,2,\downarrow}^{\dag}c_{A,1,\uparrow}+c_{A,2,\downarrow}^{\dag}c_{B,1,\uparrow}\right)\\
\frac{1}{2}\begin{bmatrix}\left(c_{A,1,\uparrow}^{\dag}c_{B,2,\uparrow}-c_{A,1,\downarrow}^{\dag}c_{B,2,\downarrow}\right)+\left(c_{B,1,\uparrow}^{\dag}c_{A,2,\uparrow}-c_{B,1,\downarrow}^{\dag}c_{A,2,\downarrow}\right)\\
+\left(c_{B,2,\uparrow}^{\dag}c_{A,1,\uparrow}-c_{B,2,\downarrow}^{\dag}c_{A,1,\downarrow}\right)+\left(c_{A,2,\uparrow}^{\dag}c_{B,1,\uparrow}-c_{A,2,\downarrow}^{\dag}c_{B,1,\downarrow}\right)
\end{bmatrix}\\
\frac{i}{2}\begin{bmatrix}\left(c_{A,1,\uparrow}^{\dag}c_{B,2,\uparrow}-c_{A,1,\downarrow}^{\dag}c_{B,2,\downarrow}\right)+\left(c_{B,1,\uparrow}^{\dag}c_{A,2,\uparrow}-c_{B,1,\downarrow}^{\dag}c_{A,2,\downarrow}\right)\\
-\left(c_{B,2,\uparrow}^{\dag}c_{A,1,\uparrow}-c_{B,2,\downarrow}^{\dag}c_{A,1,\downarrow}\right)-\left(c_{A,2,\uparrow}^{\dag}c_{B,1,\uparrow}-c_{A,2,\downarrow}^{\dag}c_{B,1,\downarrow}\right)
\end{bmatrix}\\
i\left(c_{A,1,\uparrow}^{\dag}c_{B,2,\uparrow}^{\dag}+c_{A,2,\uparrow}^{\dag}c_{B,1,\uparrow}^{\dag}+c_{A,1,\uparrow}c_{B,2,\uparrow}+c_{A,2,\uparrow}c_{B,1,\uparrow}\right)\\
i\left(c_{A,1,\downarrow}^{\dag}c_{B,2,\downarrow}^{\dag}+c_{A,2,\downarrow}^{\dag}c_{B,1,\downarrow}^{\dag}+c_{A,1,\downarrow}c_{B,2,\downarrow}+c_{A,2,\downarrow}c_{B,1,\downarrow}\right)\\
c_{A,1,\uparrow}^{\dag}c_{B,2,\uparrow}^{\dag}+c_{A,2,\uparrow}^{\dag}c_{B,1,\uparrow}^{\dag}-c_{A,1,\uparrow}c_{B,2,\uparrow}-c_{A,2,\uparrow}c_{B,1,\uparrow}\\
c_{A,1,\downarrow}^{\dag}c_{B,2,\downarrow}^{\dag}+c_{A,2,\downarrow}^{\dag}c_{B,1,\downarrow}^{\dag}-c_{A,1,\downarrow}c_{B,2,\downarrow}-c_{A,2,\downarrow}c_{B,1,\downarrow}\\
\frac{i}{2}\begin{bmatrix}\left(c_{A,1,\uparrow}^{\dag}c_{B,2,\downarrow}^{\dag}+c_{A,1,\downarrow}^{\dag}c_{B,2,\uparrow}^{\dag}\right)+\left(c_{A,2,\uparrow}^{\dag}c_{B,1,\downarrow}^{\dag}+c_{A,2,\downarrow}^{\dag}c_{B,1,\uparrow}^{\dag}\right)\\
+\left(c_{A,1,\uparrow}c_{B,2,\downarrow}+c_{A,1,\downarrow}c_{B,2,\uparrow}\right)+\left(c_{A,2,\uparrow}c_{B,1,\downarrow}+c_{A,2,\downarrow}c_{B,1,\uparrow}\right)
\end{bmatrix}\\
\frac{1}{2}\begin{bmatrix}\left(c_{A,1,\uparrow}^{\dag}c_{B,2,\downarrow}^{\dag}+c_{A,1,\downarrow}^{\dag}c_{B,2,\uparrow}^{\dag}\right)+\left(c_{A,2,\uparrow}^{\dag}c_{B,1,\downarrow}^{\dag}+c_{A,2,\downarrow}^{\dag}c_{B,1,\uparrow}^{\dag}\right)\\
-\left(c_{A,1,\uparrow}c_{B,2,\downarrow}+c_{A,1,\downarrow}c_{B,2,\uparrow}\right)-\left(c_{A,2,\uparrow}c_{B,1,\downarrow}+c_{A,2,\downarrow}c_{B,1,\uparrow}\right)
\end{bmatrix}
\end{pmatrix},
\end{equation}
\begin{equation}\label{eq:A-19}
    \begin{aligned}
        \tilde{A}_{19}=&\left(c_{A,1,\uparrow}^{\dag}c_{B,1,\uparrow}+c_{A,1,\downarrow}^{\dag}c_{B,1,\downarrow}+c_{B,1,\uparrow}^{\dag}c_{A,1,\uparrow}+c_{B,1,\downarrow}^{\dag}c_{A,1,\downarrow}\right)\\&-\left(c_{A,2,\uparrow}^{\dag}c_{B,2,\uparrow}+c_{A,2,\downarrow}^{\dag}c_{B,2,\downarrow}+c_{B,2,\uparrow}^{\dag}c_{A,2,\uparrow}+c_{B,2,\downarrow}^{\dag}c_{A,2,\downarrow}\right).
    \end{aligned}
\end{equation}
The charge density wave (CDW),  spin density wave (SDW), and ferromagnetic (FM) orders discussed in the main text correspond to specific components of $\tilde{A}_1$, $\tilde{A}_5$, and $\tilde{A}_7$ (see Eqs.~\ref{eq:A-1}, \ref{eq:A-5}, and \ref{eq:A-7}). The $\phi$ order mentioned in the main text corresponds to $\tilde{A}_{10}$ (Eq.~\ref{eq:A-10}). 

\begin{figure}[!htp]
    \centering
    \includegraphics[width=1\textwidth]{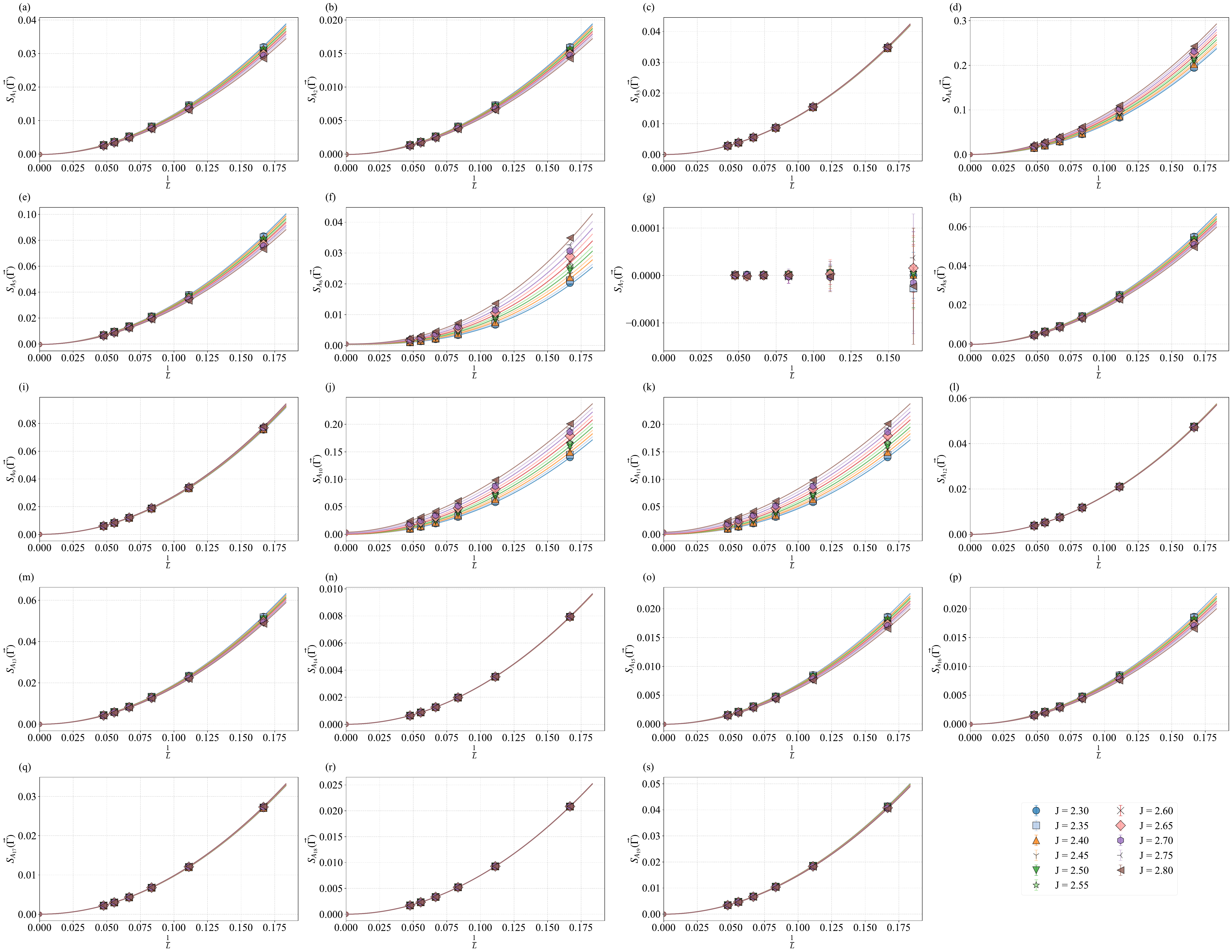}
    \caption{Panels (a)-(s) show that for each order parameter defined in Eqs.~\ref{eq:A-1}--\ref{eq:A-19}, the corresponding structure factor at $\vec{\Gamma}$ is extrapolated as a function of the inverse system size $1/L$ using the power-law ansatz $S_{A}\left(\vec{\Gamma}\right)=a\left(1/L\right)^{b}+c$. The extrapolated constants are consistent with zero, indicating that no corresponding long-range order emerges in the whole range we considered ( $J=2.3\text{--}2.8$ ).}
    \label{fig:smg-order-parameters}
\end{figure}

For an order-parameter multiplet $A$, we diagnose long-range order using its (normalized) structure factor $S_{A}\left(\vec{k}\right)$:
\begin{equation}
    S_{A}\left(\vec{k}\right)=\frac{1}{L^{4}}\sum_{j,l}\frac{1}{\dim A}\left\langle A\left(\vec{r}_{j}\right)\cdot A\left(\vec{r}_{l}\right)\right\rangle e^{\mathrm{i}\vec{k}\cdot\left(\vec{r}_{j}-\vec{r}_{l}\right)}.
\end{equation}
To determine whether the system possesses the corresponding long-range order, we extrapolate $S_{A}\left(\vec{\Gamma}\right)$ (with $\vec{\Gamma}=\left(0,0\right)$) to the thermodynamic limit $L\to\infty$. If the extrapolated value is consistent with zero within error bars, the system exhibits no $A$ long-range order; otherwise, it is considered ordered. Fig.~\ref{fig:smg-order-parameters} shows that all candidate orders vanish in the whole range we considered ( $J=2.3\text{--}2.8$ ).

\section{Continuous Phase Transition}

\begin{figure}[!htp]
    \centering
    \includegraphics[width=0.65\textwidth]{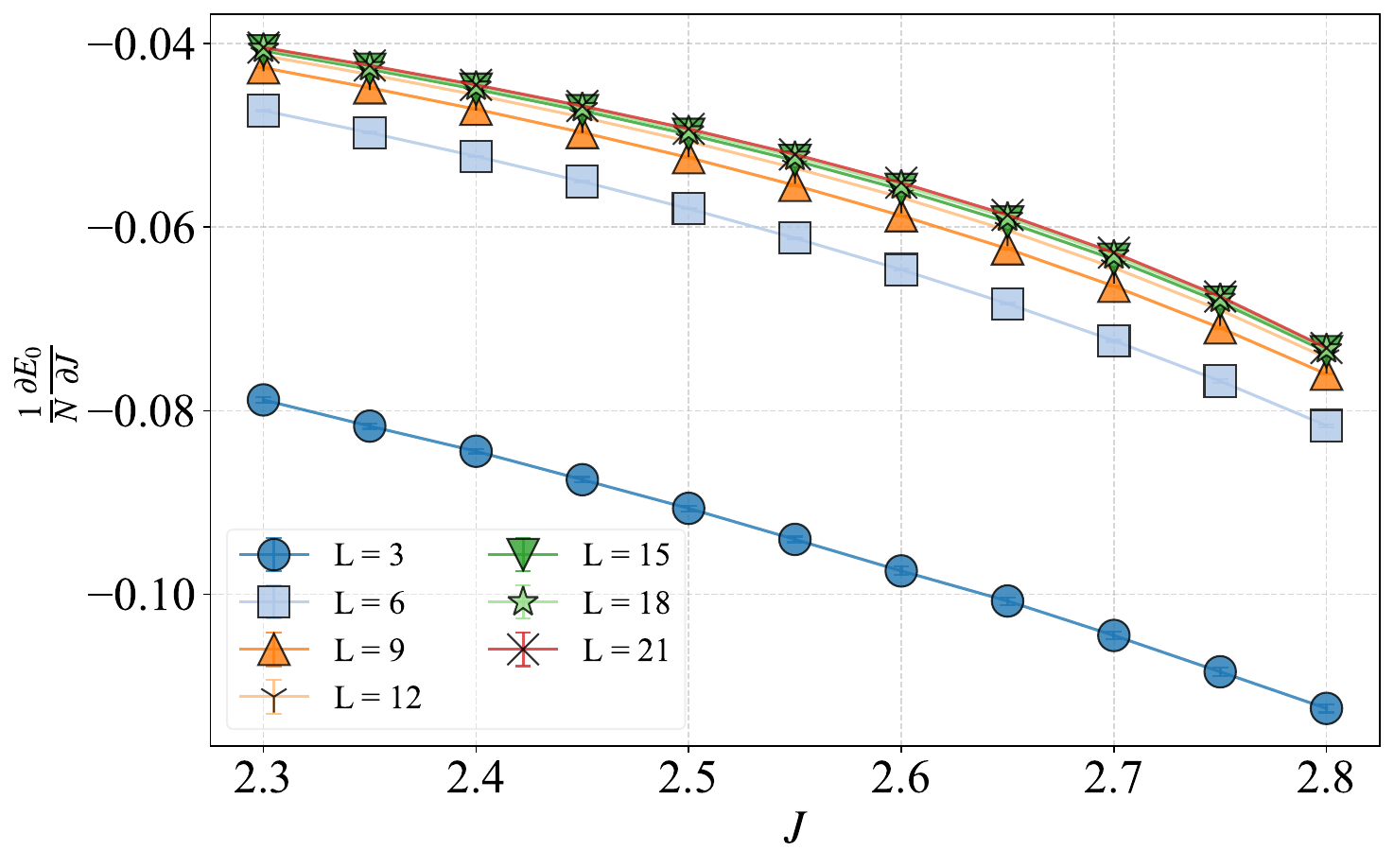}
    \caption{Derivative of the ground-state energy density, $\frac{1}{N}\frac{\partial E_{0}}{\partial J}$, as a function of interaction strength $J$ for different system sizes $L$. In the range $J=2.3\text{--}2.8$, the maximum difference between the two largest sizes ($L=21$ and $18$) is below $3\times10^{-4}$, indicating small finite-size effects. The smooth $J$ dependence without visible kinks or discontinuities supports a continuous SMG transition.}
    \label{fig:energy-derivative}
\end{figure}

The transition between the Dirac semimetal (DSM) and the SMG phase is consistent with a continuous transition. In finite-size scaling, a first-order quantum phase transition in $d=2$ with dynamical exponent $z=1$ is characterized by an effective exponent $\nu=1/(d+z)=1/3$, whereas our estimate $\nu=1.14(2)$ is much larger. Here, we provide additional evidence by examining the first derivative of the ground-state energy density $E_{0}/N$ with respect to $J$. By the Hellmann-Feynman theorem,
\begin{equation}
    \frac{1}{N}\frac{\partial E_{0}}{\partial J}=\frac{1}{N}\left\langle \frac{\partial H}{\partial J}\right\rangle =\frac{1}{N}\sum_{i}\left\langle \vec{S}_{i,1}\cdot\vec{S}_{i,2}\right\rangle,
\end{equation}
where an $L\times L$ system contains $L^{2}$ unit cells and $N=4L^{2}$ lattice sites. The index $i$ runs over the interlayer bonds (equivalently, the sites in a single layer), so the sum contains $N/2$ terms. In a first-order transition, this quantity would develop a discontinuity in the thermodynamic limit. Fig.~\ref{fig:energy-derivative} shows that the curves for different system sizes vary smoothly with $J$. In particular, for the two largest sizes ($L=21$ and $18$), the maximum difference between the two $\frac{1}{N}\frac{\partial E_{0}}{\partial J}$ curves is below $3\times10^{-4}$, indicating that finite-size effects are small and supporting a continuous phase transition.

\section{Imaginary-time Correlation Function}

In the main text, we extract the single-particle gap $\Delta_{sp}$ and the bosonic gap $\Delta_{b}$ from the long-imaginary-time behavior of the fermionic Green's function $G\left(\vec{K},\tau\right)$ and the bosonic correlation function $P\left(\vec{\Gamma},\tau\right)$, respectively. Here, we present representative data for these correlation functions.

\begin{figure}[!htp]
    \centering
    \includegraphics[width=1\textwidth]{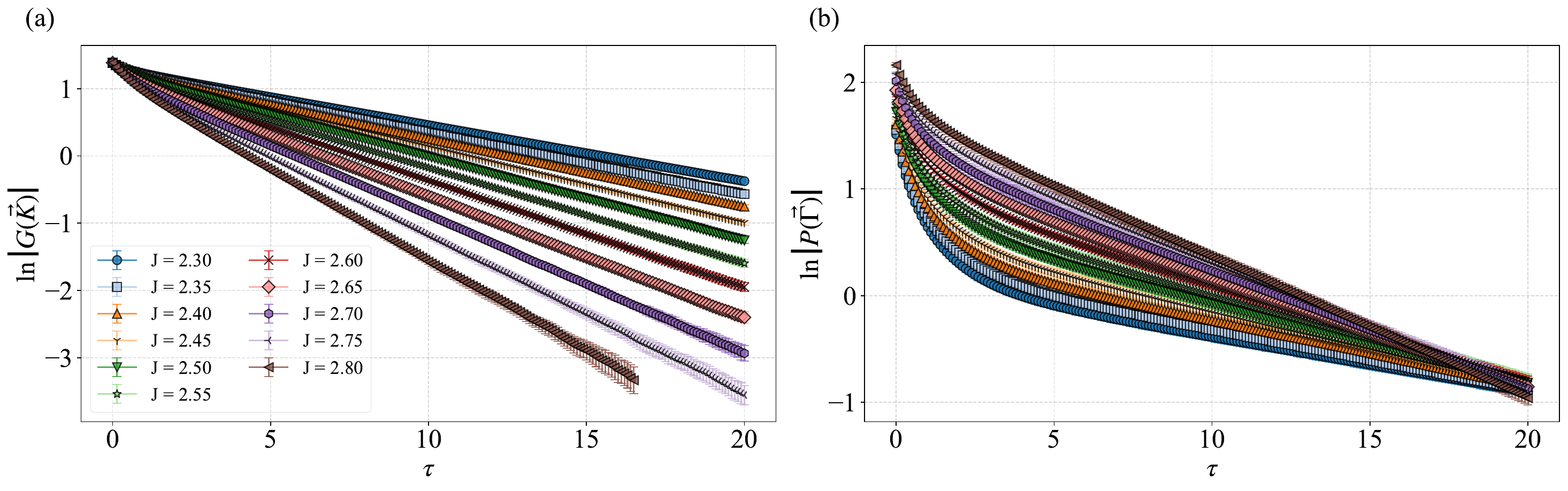}
    \caption{(a) $\ln \left|G\left(\vec{K},\tau\right)\right|$ and (b) $\ln \left| P\left(\vec{\Gamma},\tau\right) \right|$ as functions of imaginary time $\tau$ for a system size $L=12$ at various interaction strengths $J$. The linear regime in the semi-log plot sets in at a smaller $\tau$ and exhibits less noise for $G\left(\vec{K},\tau\right)$ than for $P\left(\vec{\Gamma},\tau\right)$, leading to a more accurate extraction of the single-particle gap.}
    \label{fig:imaginary_time_decay}
\end{figure}

To visualize the asymptotic exponential decay, we plot $\ln \left|G\left(\vec{K},\tau\right)\right|$ and $\ln \left| P\left(\vec{\Gamma},\tau\right) \right|$ as functions of $\tau$ in Fig.~\ref{fig:imaginary_time_decay}. At sufficiently large $\tau$, both curves show an approximately linear dependence on $\tau$, consistent with an exponential decay, from which the excitation gaps are extracted from the slopes. Compared with the fermionic Green's function, the bosonic correlator typically requires longer imaginary time to reach the asymptotic regime and exhibits larger statistical noise, especially for small $J$ and large $L$, leading to larger uncertainties in $\Delta_{b}$.

\section{Fermion Scaling Dimension}

\begin{figure}[!htp]
    \centering
    \includegraphics[width=1\textwidth]{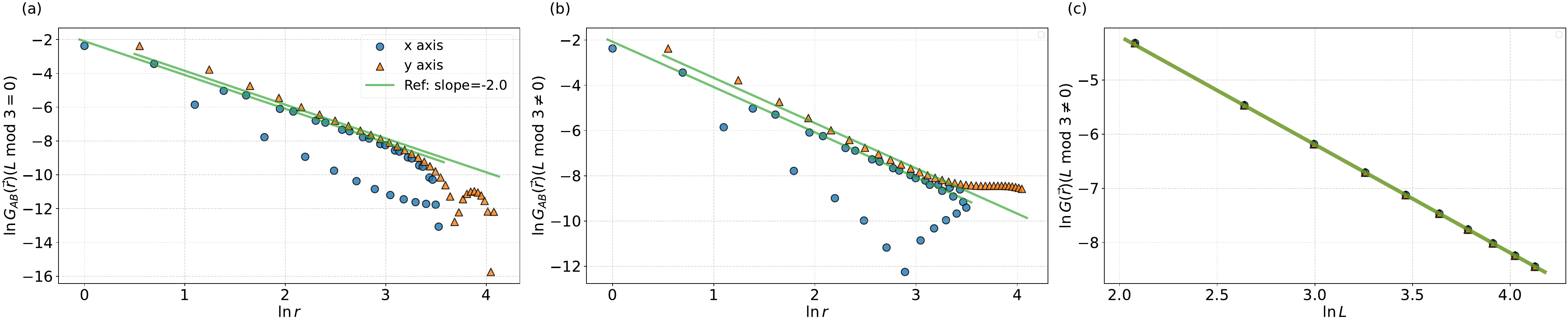}
    \caption{Log-log plot of the off-diagonal fermionic Green's function $G_{AB}\left(\vec{r}\right)$ for noninteracting fermions. (a) $G_{AB}\left(\vec{r}\right)$ along the $x$ and $y$ axes for $L=69$ (a multiple of three). The solid lines show the low-energy theoretical prediction. (b) Comparison of $G_{AB}\left(\vec{r}\right)$ for $L=68$ (not divisible by three), which exhibits smaller finite-size effects. (c) $G_{AB}\left(\left|\vec{r}_{\max}\right|\sim\frac{L}{2}\right)$ at the largest separation on the torus for system sizes $L=8+6n$ with $n=1,2,\ldots,10$. This estimator has the smallest finite-size effects and yields the Dirac-fermion scaling dimension $\Delta_{\psi}=1$.}
    \label{fig:fermion_scaling_dimension}
\end{figure}

The scaling dimension of the fermion field, denoted by $\Delta_{\psi}$, can be extracted from the equal-time Green's function in real space, $G\left(\vec{r}\right)=\left\langle c^{\dagger}\left(\vec{r}\right)c\left(0\right)\right\rangle$. At half filling on the honeycomb lattice, we focus on the off-diagonal Green's function between the two sublattices,
\begin{equation}
    G_{AB}\left(\vec{r}\right)=\left\langle c_{A}^{\dagger}\left(\vec{r}\right)c_{B}\left(0\right)\right\rangle \sim\frac{1}{r^{2\Delta_{\psi}}},
\end{equation}
in the thermodynamic limit. In the DSM phase, the fermion scaling dimension is $\Delta_{\psi}=1$. To benchmark the extraction procedure, we first consider noninteracting fermions. The conventions of the real-space and momentum-space coordinates are shown in Fig.~1(c) of the main text. Under the low-energy approximation, $G_{AB}\left(\vec{r}\right)$ takes the form
\begin{equation} \label{eq:G_AB}
    G_{AB}^{\left(0\right)}\left(\vec{r}\right)=\frac{S_{\text{cell}}}{4\pi}\left(e^{-i\vec{K}\cdot\vec{r}}\frac{i x+y}{r^{3}}+e^{-i\vec{K}^{\prime}\cdot\vec{r}}\frac{-i x+y}{r^{3}}\right),
\end{equation}
where $S_{\text{cell}}$ is the unit-cell area, and $\vec{K}=2\pi\left(\frac{2}{3},0\right)$ and $\vec{K}^{\prime}=-2\pi\left(\frac{2}{3},0\right)$ are the two Dirac points. We first consider system sizes that are multiples of three, for which the discretized momenta include the Dirac points. Along the two coordinate axes ($\vec{r}=(x,0)$ and $\vec{r}=(0,y)$), Eq.~\eqref{eq:G_AB} reduces to
\begin{gather}
    G_{AB}^{\left(0\right)}\left(x\right)=-\frac{\sqrt{3}}{2}\frac{S_{\text{cell}}}{4\pi}\frac{\sin\left(\frac{4\pi}{3}x\right)}{x^{2}},\\G_{AB}^{\left(0\right)}\left(y\right)=\frac{S_{\text{cell}}}{2\pi}\frac{1}{y^{2}}.
\end{gather}
This indicates that $G_{AB}$ decays as a power law along the $y$ axis and exhibits oscillations along the $x$ axis, consistent with the numerical results shown in Fig.~\ref{fig:fermion_scaling_dimension}(a). Moreover, we find that system sizes not divisible by three exhibit smaller finite-size effects, as shown in Fig.~\ref{fig:fermion_scaling_dimension}(b); we therefore use such system sizes in the following analysis. Since the asymptotic form in Eq.~\eqref{eq:G_AB} holds strictly in the thermodynamic limit, we further suppress finite-size effects by evaluating $G_{AB}$ at the largest separation $\vec{r}_{\max}$ on the torus (with $|\vec{r}_{\max}|\sim L/2$). The corresponding finite-size scaling reads
\begin{equation}
    G_{AB}\left(\left|\vec{r}_{\max}\right|\sim\frac{L}{2}\right)\sim\frac{1}{L^{2\Delta_{\psi}}}.
\end{equation}
Fig.~\ref{fig:fermion_scaling_dimension}(c) demonstrates that this procedure reproduces the Dirac-fermion value $\Delta_{\psi}=1$ for noninteracting fermions to high accuracy.

\section{Bilayer Honeycomb Lattice Model with \texorpdfstring{$\mathrm{Spin}(5) \times \mathrm{U}(1)/\mathbb{Z}_2$}{Spin(5) x U(1)/Z_2} Symmetry}

As discussed in the main text, a key finding of our work is that the pure non-Abelian symmetry $\mathrm{SU}(2)^3/\mathbb{Z}_2$ is essential for the direct SMG transition. To substantiate this claim, we study a related bilayer honeycomb lattice model with $\mathrm{Spin}(5) \times \mathrm{U}(1)/\mathbb{Z}_2$ symmetry, which also satisfies the anomaly-cancellation condition but contains a $\mathrm{U}(1)$ factor. As we show below, this model develops an intermediate symmetry-broken phase, demonstrating that the absence of a $\mathrm{U}(1)$ factor in the symmetry group is crucial for SMG. The Hamiltonian is defined as
\begin{equation}
    H = H_{0}+H_{\text{int}}^{\prime},
\end{equation}
where the kinetic part is the nearest-neighbor hopping on two decoupled honeycomb layers,
\begin{equation}
    H_{0}=-t\sum_{\langle i,j\rangle,\alpha,\sigma}\left(c_{i,\alpha,\sigma}^{\dagger}c_{j,\alpha,\sigma}+\mathrm{h.c.}\right),
\end{equation}
and the on-site interlayer interaction (including a density-density term) takes the form
\begin{equation}
    H_{\text{int}}^{\prime}=J\sum_{i}\left[\vec{S}_{i,1}\cdot\vec{S}_{i,2}+\frac{1}{4}\left(\rho_{i,1}\rho_{i,2}-1\right)\right].
\end{equation}
Here $n_{i,\alpha}=\sum_{\sigma}c_{i,\alpha,\sigma}^{\dagger}c_{i,\alpha,\sigma}$, $\rho_{i,\alpha}=n_{i,\alpha}-1$, and $\vec{S}_{i,\alpha} = \frac{1}{2} \sum_{\sigma, \sigma'} c^\dagger_{i,\alpha,\sigma} \vec{\sigma}_{\sigma\sigma'} c_{i,\alpha,\sigma'}$. The indices $i$, $\alpha$, and $\sigma$ are defined as in the main text.

We focus on half filling ($\mu=0$), where the model has particle-hole symmetry. In the free-fermion limit, the low-energy physics is governed by massless Dirac fermions, characterizing the system as a DSM phase. In the strong-coupling limit $J\rightarrow+\infty$, the interaction acts independently on each interlayer bond. At half filling, the two particles on each bond form a spin singlet $\frac{1}{\sqrt{2}}\left(\left|\uparrow\downarrow\right\rangle -\left|\downarrow\uparrow\right\rangle \right)$, and the ground state becomes a product state of these local singlets. This ground state is unique, gapped, symmetry-preserving, and topologically trivial, corresponding to the SMG phase. In the following, we investigate the phase diagram and critical behavior of the model.

\subsection{Sign-problem-free PQMC}

We employ the same PQMC framework as described above. For the interaction $H_{\text{int}}^{\prime}$, it is convenient to introduce the interlayer exciton operator
\begin{equation}
    D_{i}=\sum_{\sigma}c_{i,1,\sigma}^{\dagger}c_{i,2,\sigma},
\end{equation}
under which the interaction can be rewritten as
\begin{equation}
    H_{\text{int}}^{\prime}=-\frac{J}{4}\sum_{i}\left(D_{i}^{\dagger}D_{i}+D_{i}D_{i}^{\dagger}\right)=-\frac{J}{8}\sum_{i}\left[\left(D_{i}+D_{i}^{\dagger}\right)^{2}-\left(D_{i}-D_{i}^{\dagger}\right)^{2}\right].
\end{equation}
After the Trotter decomposition and HS decoupling, the fermion-bilinear Hamiltonian factorizes into two spin sectors. At half filling, we perform a partial particle-hole transformation on the spin $\downarrow$ fermions,
\begin{gather}
    c_{i,\alpha,\uparrow}=\tilde{c}_{i,\alpha,\uparrow},\\c_{i,\alpha,\downarrow}=\left(-1\right)^{i}\left(-1\right)^{\alpha-1}\tilde{c}_{i,\alpha,\downarrow}^{\dagger},
\end{gather}
with $(-1)^{i}=\pm 1$ on sublattice A/B and $(-1)^{\alpha-1}=\pm 1$ on layer $\alpha=1/2$, one finds that the two spin sectors are complex conjugates of each other. Therefore, if the trial wave function $|\Psi_{T}\rangle$ respects the same complex-conjugation structure, the PQMC simulation is free of the sign problem. In practice, we choose $|\Psi_{T}\rangle$ for one spin sector and take the other as its complex conjugate.

Assaad proposed that threading a magnetic flux perpendicular to a two-dimensional lattice can mitigate finite-size effects in DQMC calculations. He provided an intuitive explanation based on Landau levels: in the presence of a single magnetic flux quantum, the degeneracy of the Landau levels is unity. This implies that, with the exception of the $\varepsilon_n=0$ level, all other energy levels become non-degenerate. This distribution of non-degenerate levels effectively mimics the continuous energy spectrum of an infinite system, thereby significantly improving the convergence of thermodynamic quantities~\cite{assaad_depleted_2002}. In our PQMC calculations, we adopt this method, where the applied flux introduces Peierls phase factors into the hopping matrix elements of the system:
\begin{equation} \label{eq:h0-flux}
    H_{0}^{\left(\text{flux}\right)}=-t\sum_{\left\langle i,j\right\rangle ,\alpha}e^{i\frac{2\pi}{\Phi_{0}}\int_{i}^{j}\vec{A}\left(\vec{r}\right)\cdot d\vec{r}}c_{i,\alpha}^{\dagger}c_{j,\alpha}+\mathrm{h.c.}.
\end{equation}
For systems of various sizes, we consistently thread a single magnetic flux quantum $\Phi_{0}=h/e$ and select the vector potential $\vec{A}$ using the Landau gauge:
\begin{gather}
    B=\frac{\Phi_{0}}{L_{a}L_{b}\left|\vec{a}\times\vec{b}\right|}, \\ \vec{A}=\left(-By,0,0\right),
\end{gather}
where $\vec{a}$ and $\vec{b}$ are the primitive lattice vectors (see Fig.~1(c) in the main text), and $L_{a}$ and $L_{b}$ denote the linear dimensions of the system along the $\vec{a}$ and $\vec{b}$ directions, respectively.

In accordance with the principles for selecting $\left|\Psi_{T}\right\rangle$ outlined above, we restrict our consideration to the choice of $\left|\Psi_{T}\right\rangle$ for a single spin sector. To lift the degeneracy of $H_0$ and minimize the projection length required to evolve $\left|\Psi_{T}\right\rangle$ to the ground state, we employ distinct trial wave functions depending on the parameter regime:
\begin{enumerate}
    \item For the range $J=2.3 \text{--} 2.8$, the trial wave function is chosen as the ground state of Eq.~\eqref{eq:h0-flux} with an applied magnetic flux $\Phi=10^{-4}\Phi_{0}$. We set the projection length to $2\Theta=\max\{17,L+5\}$ and the Trotter time step to $\Delta\tau=0.1$.
    \item For the range $J=3.2 \text{--} 3.8$, we construct a Hamiltonian that introduces disorder into the hopping matrix elements:
    \begin{equation}
        H_{0}^{\prime\prime}=-\sum_{\left\langle i,j\right\rangle ,\alpha}\left(t+\delta\,r_{i,j}\right)c_{i,\alpha}^{\dagger}c_{j,\alpha} + \mathrm{h.c.}
    \end{equation}
    where $\delta$ denotes the disorder strength, set to $\delta =10^{-4}t$ in our calculations, and $r_{i,j}$ represents a random number uniformly distributed within the interval $(0,1)$. The ground state of $H_{0}^{\prime\prime}$ is adopted as the trial wave function. We choose a projection length of $2\Theta=L+10$ and maintain the Trotter time step at $\Delta\tau=0.1$.
\end{enumerate}

\subsection{Results and Phase Diagram}

Within the interaction ranges $J=2.3\text{--}2.8$ and $3.2\text{--}3.8$, we find an emerging interlayer exciton condensation (EC) order parameter defined as
\begin{equation}\label{eq:EC_order}
    \phi_{\mathrm{EC}}\left(\vec{r}_{j}\right)=\begin{pmatrix}
    \mathrm{i}\left[\left(-1\right)^{j}\sum_{\sigma}c_{j,1,\sigma}^{\dagger}c_{j,2,\sigma}-\mathrm{h.c.}\right]\\
    \left[\left(-1\right)^{j}\sum_{\sigma}c_{j,1,\sigma}^{\dagger}c_{j,2,\sigma}+\mathrm{h.c.}\right]
    \end{pmatrix}
\end{equation}
Here $(-1)^{j}=\pm 1$ on sublattice A/B. This indicates an intermediate symmetry-broken phase separating the DSM at weak coupling from the SMG phase at strong coupling. With increasing $J$, the system first undergoes a DSM-to-EC transition where the $\mathrm{U}(1)$ symmetry is spontaneously broken, and then a second transition where the EC order is suppressed and the $\mathrm{U}(1)$ symmetry is restored, entering the SMG phase.

\begin{figure}[!htp]
    \centering
    \includegraphics[width=0.65\textwidth]{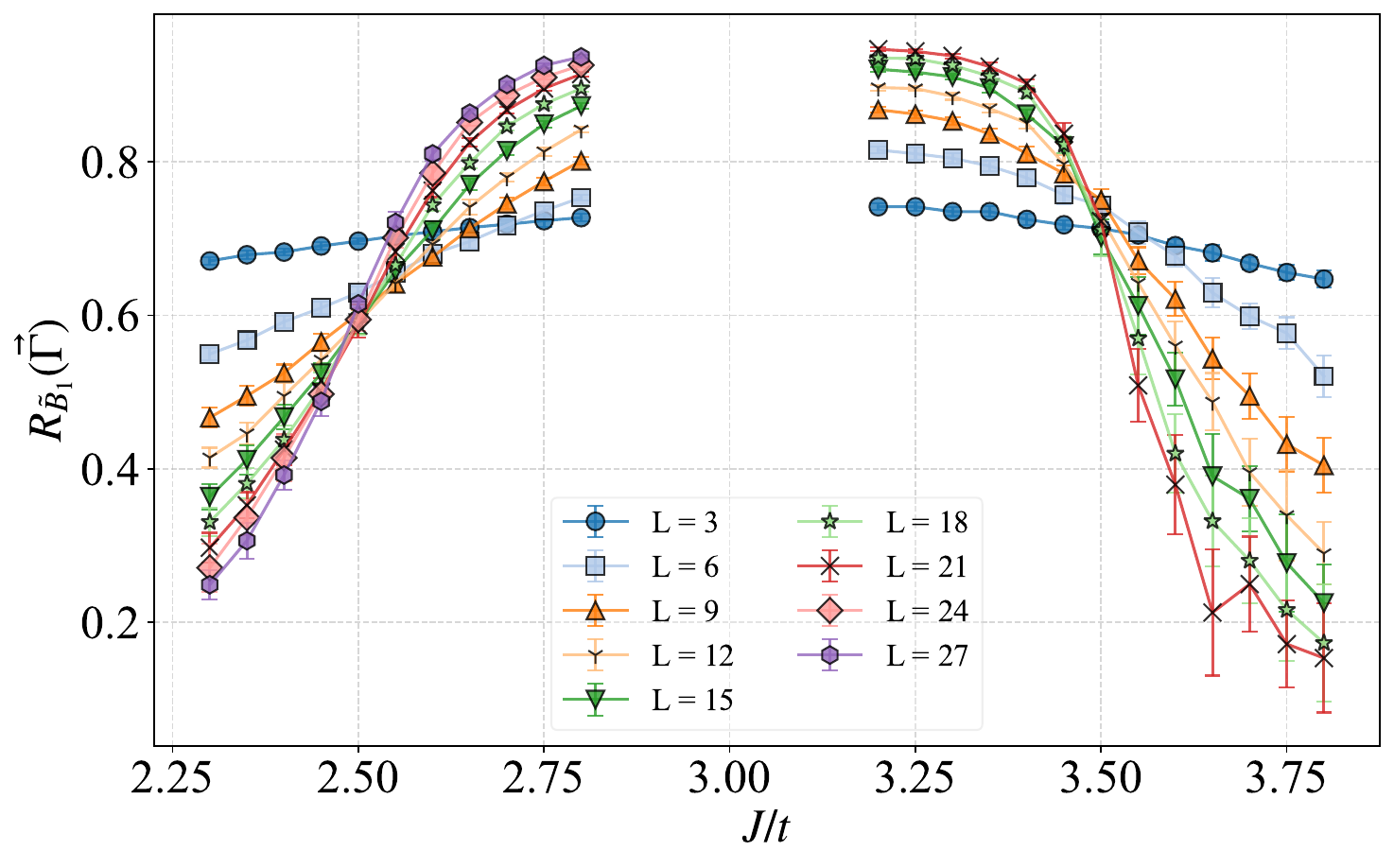}
    \caption{Correlation ratio $R_{\mathrm{EC}}(\vec{\Gamma})$ of the interlayer EC order at the $\vec{\Gamma}$ point as a function of $J$. The curves for different system sizes exhibit crossings in two regions, demarcating three distinct regimes: (1) in the leftmost region, $R_{\mathrm{EC}}(\vec{\Gamma},L)$ decreases with increasing $L$, indicating a DSM phase; (2) in the intermediate region, $R_{\mathrm{EC}}(\vec{\Gamma},L)$ increases with $L$, signifying the establishment of EC order (excitonic insulator phase); and (3) on the right, $R_{\mathrm{EC}}(\vec{\Gamma},L)$ decreases with $L$, corresponding to the SMG phase. The two crossing regions mark the DSM-to-EC and EC-to-SMG phase transitions, respectively.}
    \label{fig:correlation-ratio}
\end{figure}

We locate the two transitions using the correlation ratio associated with the EC order parameter in Eq.~\eqref{eq:EC_order}. We define the (normalized) structure factor
\begin{equation} \label{eq:structure-factor}
    S_{\mathrm{EC}}\left(\vec{k}\right)=\frac{1}{L^{4}}\sum_{j,l}\left\langle \phi_{\mathrm{EC}}\left(\vec{r}_{j}\right)\cdot\phi_{\mathrm{EC}}\left(\vec{r}_{l}\right)\right\rangle e^{\mathrm{i}\vec{k}\cdot\left(\vec{r}_{j}-\vec{r}_{l}\right)},
\end{equation}
and the correlation ratio at the $\vec{\Gamma}=\left(0,0\right)$ point,
\begin{equation} \label{eq:ssb-correlation-ratio}
    R_{\mathrm{EC}}\left(\vec{\Gamma}\right)=1-\frac{S_{\mathrm{EC}}\left(\vec{\Gamma}+\Delta\vec{k}\right)}{S_{\mathrm{EC}}\left(\vec{\Gamma}\right)},
\end{equation}
where $\Delta\vec{k}$ is the smallest nonzero momentum with $\left|\Delta\vec{k}\right|\sim 1/L$. In the disordered (ordered) phase, $R_{\mathrm{EC}}\left(\vec{\Gamma}\right)\to 0$ ($1$) as $L\to\infty$. As an RG invariant, it obeys the scaling form
\begin{equation} \label{eq:ssb-correlation-ratio-scaling}
    R_{\mathrm{EC}}\left(\vec{\Gamma}\right)\sim g_{R}\left(L^{\frac{1}{\nu}}\left(J-J_{c}\right)\right),
\end{equation}
such that curves for different $L$ cross near the critical point when finite-size corrections are small. Figure~\ref{fig:correlation-ratio} shows two crossing regions, indicating two phase transitions and three regimes: the DSM, an intermediate excitonic insulator phase with $\mathrm{U}(1)$ symmetry breaking, and the SMG phase.

\begin{figure}[!htp]
    \centering
    \includegraphics[width=1\textwidth]{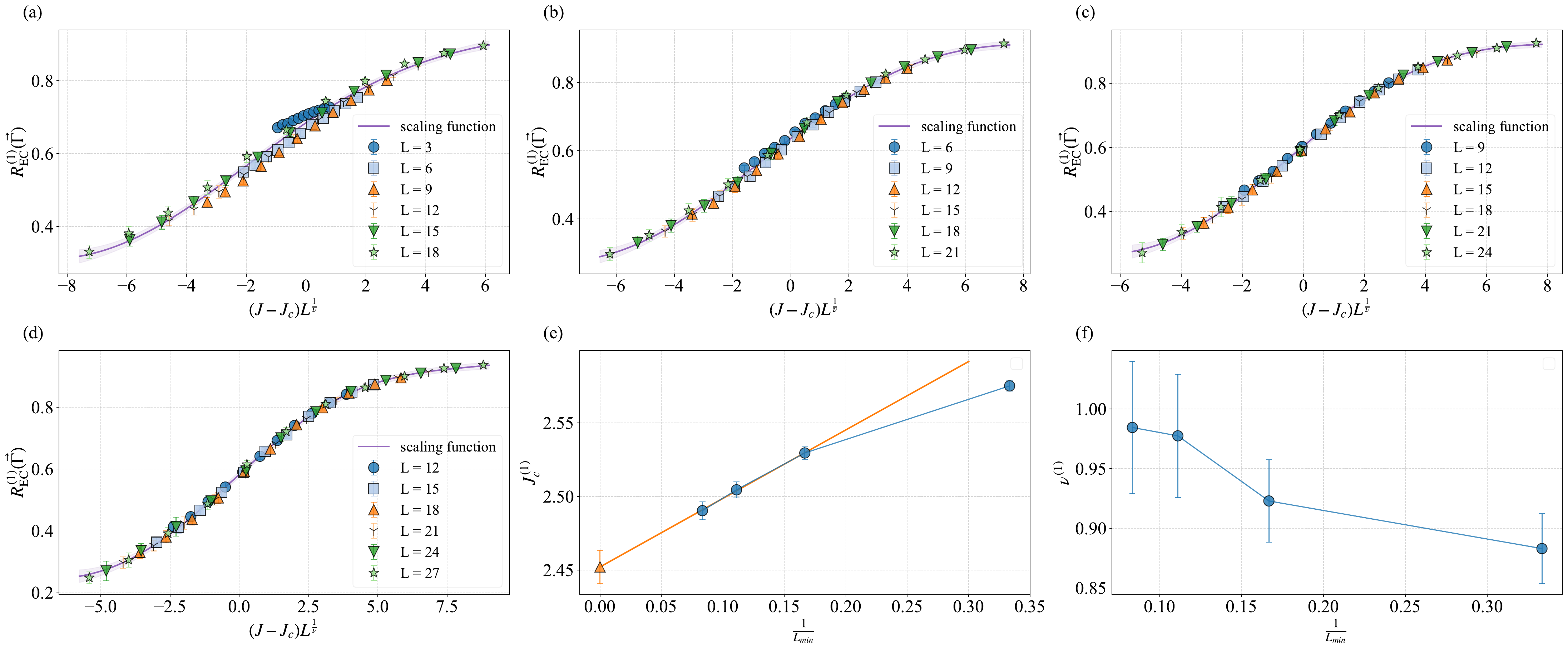}
    \caption{Finite-size scaling analysis for the first critical point. Panels (a)–(d) show the data collapse of the correlation ratio $R_{\mathrm{EC}}^{\left(1\right)}\left(\vec{\Gamma}\right)$ with varying minimum system sizes included in the fit ($L_{\min}$). A fitting window of six sizes with a step of 3 was used (e.g., for $L_{\min}=3$, $L=3, 6, \dots, 18$). (a) $L_{\min} = 3$, (b) $L_{\min} = 6$, (c) $L_{\min} = 9$, and (d) $L_{\min} = 12$. Panels (e) and (f) plot the fitted critical point $J_{c}^{\left(1\right)}$ and correlation length exponent $\nu^{\left(1\right)}$ as a function of the inverse minimum size $1/L_{\min}$. $J_{c}^{\left(1\right)}$ decreases with increasing $L_{\min}$ without convergence at $L_{\min}=12$; a linear extrapolation ($J_{c}=a/L_{\min}+b$) yields $J_{c}^{\left(1\right)}=2.45(2)$. The exponent $\nu^{\left(1\right)}$ exhibits signs of convergence at large $L_{\min}$, with the value at $L_{\min}=12$ determined to be $\nu^{\left(1\right)}=0.98(6)$.}
    \label{fig:ssb-correlation-ratio-fss}
\end{figure}

\begin{figure}[!htp]
    \centering
    \includegraphics[width=1\textwidth]{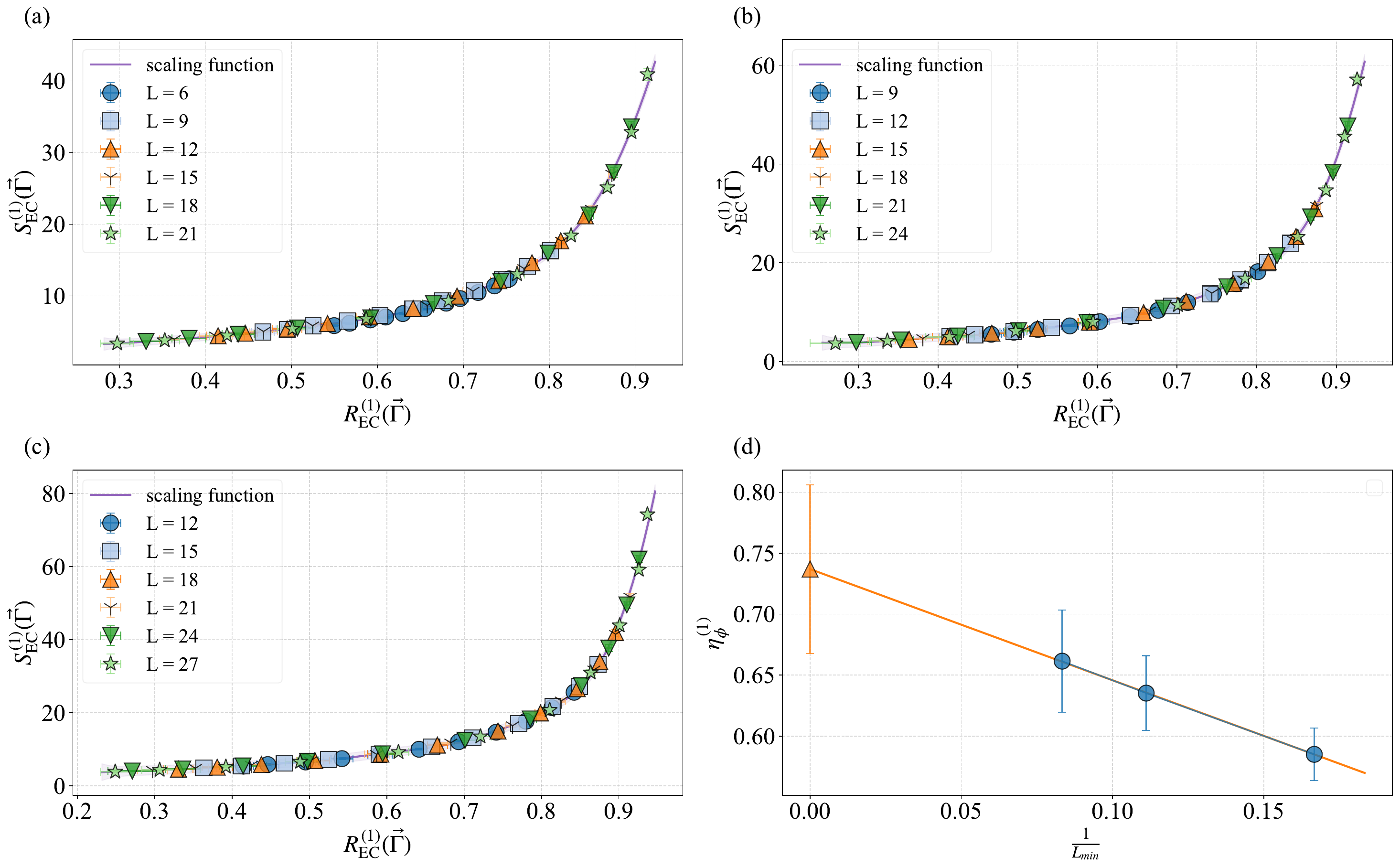}
    \caption{Finite-size scaling of the structure factor at the first phase transition point. (a)–(c) Data collapse of $S_{\mathrm{EC}}^{\left(1\right)}\left(\vec{\Gamma}\right)$ for various minimum system sizes $L_{\min}$, using a fitting window of six sizes. (a) $L_{\min} = 6$, (b) $L_{\min} = 9$, and (c) $L_{\min} = 12$. (d) Evolution of the anomalous dimension $\eta_{\phi}^{\left(1\right)}$ versus the inverse minimum size $1/L_{\min}$. The absence of convergence in $\eta_{\phi}^{\left(1\right)}$ as $1/L_{\min} \to 0$ necessitates a linear extrapolation, resulting in $\eta_{\phi}^{\left(1\right)}=0.74\left(7\right)$.}
    \label{fig:ssb-structure-factor-crossing-1-fss}
\end{figure}

To characterize the critical behavior, we perform finite-size scaling and data-collapse analyses proposed by Harada~\cite{harada_bayesian_2011,harada_kernel_2015}. The correlation length exponent can be obtained from Eq.~\eqref{eq:ssb-correlation-ratio-scaling}, as shown in Fig.~\ref{fig:ssb-correlation-ratio-fss}. The boson anomalous dimension $\eta_{\phi}$ can be extracted from the structure factor scaling,
\begin{equation} \label{eq:ssb-structure-factor-scaling}
    S_{\mathrm{EC}}\left(\vec{\Gamma}\right)\sim L^{-\left(d+z-2+\eta_{\phi}\right)}g_{S}\left(L^{\frac{1}{\nu}}\left(J-J_{c}\right)\right).
\end{equation}
At the critical point, an emergent Lorentz symmetry is expected, so we set $z=1$ and use $d=2$. Reparametrizing the scaling variable using the RG invariant in Eq.~\eqref{eq:ssb-correlation-ratio-scaling} yields
\begin{equation} \label{eq:ssb-structure-factor-scaling-z1}
    S_{\mathrm{EC}}\left(\vec{\Gamma}\right)\sim L^{-\left(1+\eta_{\phi}\right)}\tilde{g}_{S}\left(R_{\mathrm{EC}}\left(\vec{\Gamma}\right)\right).
\end{equation}

\begin{figure}[!htp]
    \centering
    \includegraphics[width=1\textwidth]{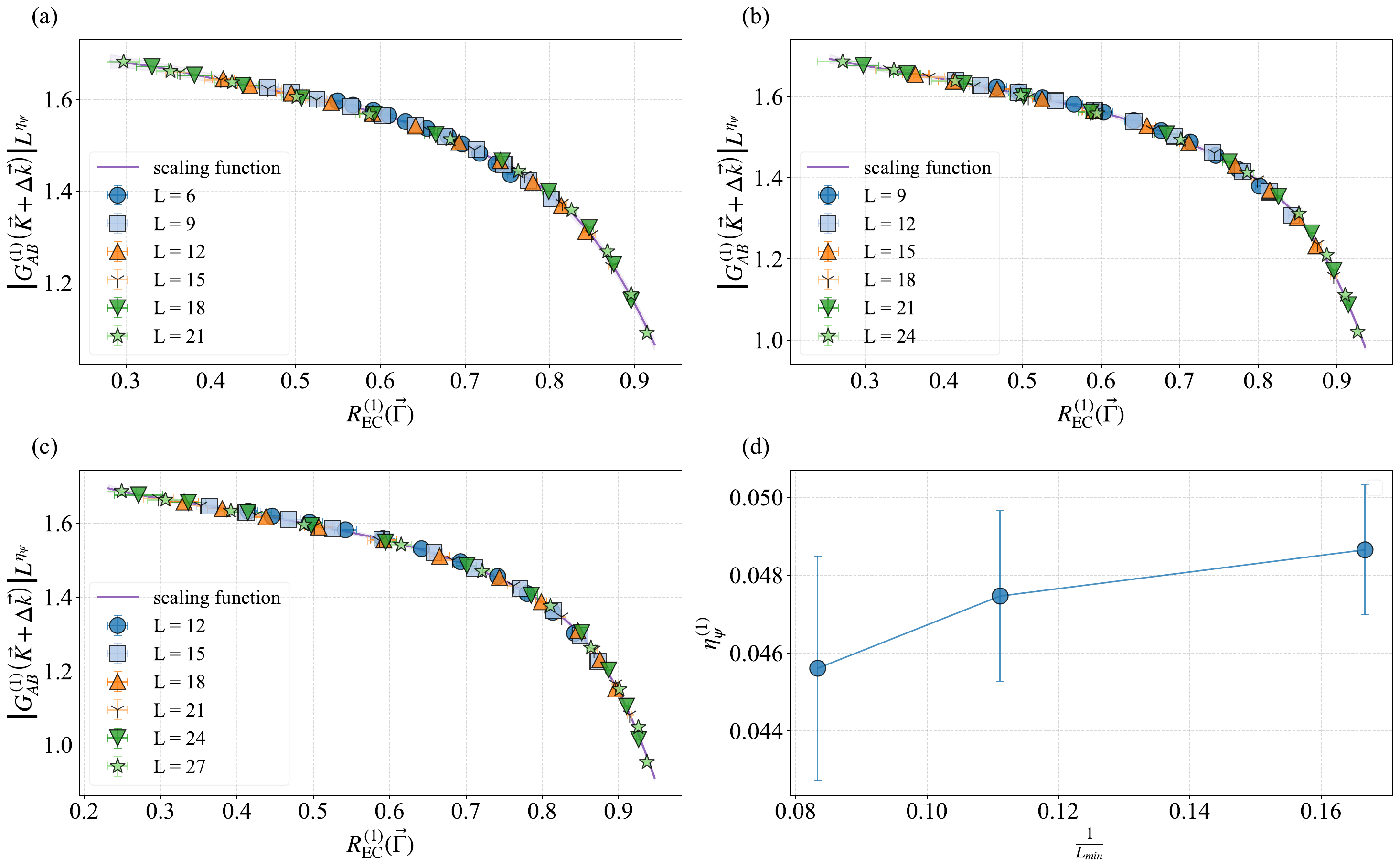}
    \caption{Finite-size scaling of the off-diagonal Green's function at the first phase transition. Panels (a)–(c) show the data collapse of $\left|G_{AB}\left(\vec{K}+\Delta\vec{k}\right)\right|$ for different minimum system sizes $L_{\min}$. (d) The fitted fermionic anomalous dimension $\eta_{\psi}^{\left(1\right)}$ as a function of $1/L_{\min}$. The estimates for $\eta_{\psi}^{\left(1\right)}$ are consistent across all $L_{\min}$ within error bars, demonstrating excellent convergence. The final result from the $L_{\min}=12$ window is $\eta_{\psi}^{\left(1\right)}=0.046\left(3\right)$.}
    \label{fig:ssb-fermion-anomalous-dimension-crossing-1-fss}
\end{figure}

To extract the fermion anomalous dimension, we analyze the off-diagonal Green's function between the two sublattices in momentum space near the Dirac point, which follows the finite-size scaling
\begin{equation} \label{eq:ssb-gab-fss}
    \left|G_{AB}\left(\vec{K}+\Delta\vec{k}\right)\right|\sim L^{-\eta_{\psi}}g_{G}\left(R_{\mathrm{EC}}\left(\vec{\Gamma}\right)\right).
\end{equation}

The first transition involves mass generation of Dirac fermions accompanied by spontaneous $\mathrm{U}(1)$ symmetry breaking. Combining data-collapse analyses of $R_{\mathrm{EC}}$, $S_{\mathrm{EC}}$, and $\left|G_{AB}\right|$, we obtain $J_{c}^{\left(1\right)}=2.45\left(2\right)$, $\nu^{\left(1\right)}=0.98\left(6\right)$, $\eta_{\phi}^{\left(1\right)}=0.74\left(7\right)$, and $\eta_{\psi}^{\left(1\right)}=0.046\left(3\right)$. A comparison of these results with the critical exponents of the $(2+1)D$ $N=4$ Gross-Neveu-XY universality class calculated using different methods is presented in Tab.~\ref{tab:ssb-critical-exponents}. The values of critical exponents are consistent with existing results, confirming that the phase transition belongs to the $(2+1)D$ $N=4$ Gross-Neveu-XY universality class.

\begin{table}[!htp]
    \caption{Comparison of critical exponents for the $(2+1)$D $N=4$ Gross--Neveu--XY universality class obtained by different approaches.}
    \label{tab:ssb-critical-exponents}
    \centering
    \begin{tabular}{cccc}
        \hline
        Method & $\nu$ & $\eta_\phi$ & $\eta_\psi$ \\
        \hline
        QMC (this work) & $0.98\left(6\right)$ & $0.74\left(7\right)$ & $0.046\left(3\right)$ \\
        QMC~\cite{li_fermioninduced_2017} & $1.11\left(3\right)$ & $0.80\left(4\right)$ & -- \\
        QMC~\cite{daliao_valence_2019} & $1.01\left(3\right)$ & $0.80\left(2\right)$ & -- \\
        Interpolation~\cite{hawashin_grossneveuxy_2025} & $1.09\left(6\right)$ & $0.926\left(13\right)$ & $0.0404\left(13\right)$ \\
        $4-\epsilon$ expansion (four loop), $P_{[1/3]}$~\cite{zerf_fourloop_2017,hawashin_grossneveuxy_2025}& $1.06$ & $1.03$ & $0.049$ \\
        $4-\epsilon$ expansion (four loop), $P_{[2/2]}$~\cite{zerf_fourloop_2017,hawashin_grossneveuxy_2025}& $1.130$ & $0.929$ & $0.046$ \\
        $4-\epsilon$ expansion (four loop), $P_{[3/1]}$~\cite{zerf_fourloop_2017,hawashin_grossneveuxy_2025}& $1.130$ & -- & $0.037$ \\
        $4-\epsilon$ expansion (four loop), $P_{[4/0]}$~\cite{zerf_fourloop_2017,hawashin_grossneveuxy_2025}& $0.956$ & $0.995$ & $0.040$ \\
        $2+\epsilon$ expansion (one loop), $P_{[5/1]}$~\cite{hawashin_grossneveuxy_2025}& $1.096$ & $0.938$ & $0.042$ \\
        $2+\epsilon$ expansion (one loop), $P_{[6/0]}$~\cite{hawashin_grossneveuxy_2025}& $1.084$ & $0.938$ & $0.042$ \\
        Large-$N$ expansion~\cite{li_fermioninduced_2017} & $1.25$ & $0.80$ & -- \\
        Large-$N$ expansion, $P_{[1,1]}$~\cite{gracey_critical_2021}& $1.077$ & $0.943$ & $0.038$ \\
        Functional RG~\cite{classen_fluctuationinduced_2017}& $1.08$ & $0.95$ & $0.027$ \\
        \hline
    \end{tabular}
\end{table}

\begin{figure}[!htp]
    \centering
    \includegraphics[width=1\textwidth]{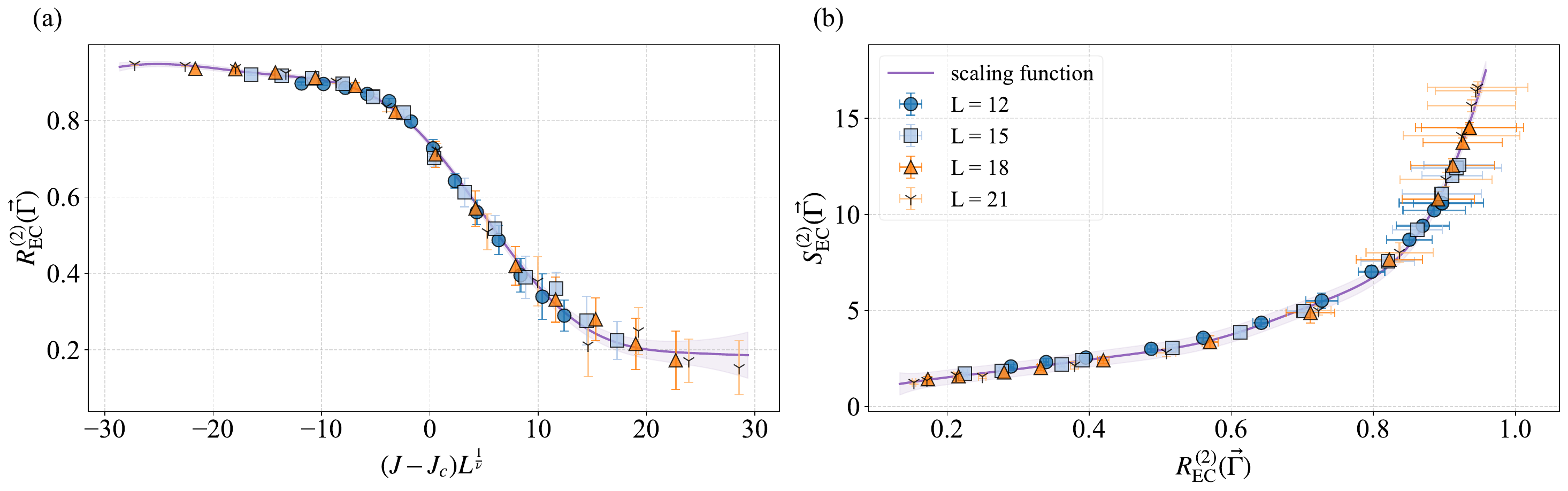}
    \caption{Data collapse analysis for the second phase transition. Panels (a) and (b) show the data collapse for the correlation ratio $R_{\mathrm{EC}}^{\left(2\right)}\left(\vec{\Gamma}\right)$ and structure factor $S_{\mathrm{EC}}^{\left(2\right)}\left(\vec{\Gamma}\right)$, respectively. The critical exponents were fixed to the $(2+1)$D XY universality class values: $1/\nu=1.48864$ and $\eta=0.03816$, resulting in a critical point estimate of $J_{c}^{\left(2\right)}=3.49(2)$. The collapse of data points from various $J$ and $L$ onto a universal curve supports the consistency with the $(2+1)$D XY universality class.}
    \label{fig:ssb-correlation-ratio-and-structure-factor-crossing-2-fss}
\end{figure}

The second transition, from the excitonic insulator to the SMG phase, is driven by the suppression of the excitonic condensate and the restoration of $\mathrm{U}(1)$ symmetry, and is thus expected to fall into the $(2+1)$D XY universality class. High-precision results for the critical exponents of this universality class have already been established. Monte Carlo (MC) methods determined $\nu^{(2)}=0.671718(23)$ and $\eta^{(2)}=0.03816(2)$~\cite{hasenbusch_eliminating_2025}; the conformal bootstrap method yields $\Delta_{\phi}=0.519088(22)$ and $\Delta_{s}=1.51136(22)$, which lead to $\nu^{(2)}=0.67175(10)$ and $\eta^{(2)}=0.038176(44)$~\cite{chester_carving_2020} via the relations $\Delta_{\phi}=\frac{d-2+\eta}{2}$ and $\Delta_{s}=d-\frac{1}{\nu}$. We restrict our analysis to a self-consistency check of the hypothesis. Using the fixed XY critical exponents in the data-collapse analysis, we obtain $J_{c}^{\left(2\right)}=3.49\left(2\right)$. Fig.~\ref{fig:ssb-correlation-ratio-and-structure-factor-crossing-2-fss} shows that both $R_{\mathrm{EC}}$ and $S_{\mathrm{EC}}$ collapse onto universal curves, consistent with the $(2+1)$D XY universality class. Deviations for larger $J$ and larger $L$ are consistent with the increased statistical uncertainties in this regime.

In summary, the $\mathrm{Spin}(5) \times \mathrm{U}(1)/\mathbb{Z}_2$ model, despite satisfying the same anomaly-cancellation condition as the $\mathrm{SU}(2)^3/\mathbb{Z}_2$ model, does not exhibit a direct SMG transition. Instead, the $\mathrm{U}(1)$ symmetry permits the formation of an inter-layer excitonic condensate that intervenes between the Dirac semimetal and the SMG phase. This comparison demonstrates that the pure non-Abelian nature of the $\mathrm{SU}(2)^3/\mathbb{Z}_2$ symmetry---which prohibits all bilinear condensates---is the key ingredient enabling the direct SMG transition observed in the main text.

\end{document}